\def\year{2022}\relax
\documentclass[letterpaper]{article} % DO NOT CHANGE THIS
\usepackage{aaai22}  % DO NOT CHANGE THIS
\usepackage{times}  % DO NOT CHANGE THIS
\usepackage{helvet}  % DO NOT CHANGE THIS
\usepackage{courier}  % DO NOT CHANGE THIS
\usepackage[hyphens]{url}  % DO NOT CHANGE THIS
\usepackage{graphicx} % DO NOT CHANGE THIS
\usepackage{natbib}  % DO NOT CHANGE THIS AND DO NOT ADD ANY OPTIONS TO IT
\usepackage{caption} % DO NOT CHANGE THIS AND DO NOT ADD ANY OPTIONS TO IT
\DeclareCaptionStyle{ruled}{labelfont=normalfont,labelsep=colon,strut=off} % DO NOT CHANGE THIS
\usepackage{booktabs} % For formal tables
\usepackage{enumitem}
\usepackage{multirow}
\usepackage{diagbox}
\usepackage{subcaption}
\usepackage{amsmath}
\usepackage{cleveref}
\usepackage{dsfont}

\usepackage{xspace}

\usepackage[usenames, dvipsnames]{xcolor}
\usepackage{soul}

\usepackage{xargs}
\usepackage[colorinlistoftodos,prependcaption,textsize=tiny]{todonotes}

\newcommand{\header}[1]{{\noindent{\textbf{#1}}}}

\definecolor{ruby}{rgb}{0.6, 0.1, 0.3}
\definecolor{navy}{rgb}{0.1, 0.1, 0.8}
\definecolor{olive}{rgb}{0.1, 0.6, 0.2}

\definecolor{ruby}{rgb}{0.0, 0.0, 0.0}
\definecolor{navy}{rgb}{0.0, 0.0, 0.0}
\definecolor{olive}{rgb}{0.0, 0.0, 0.0}

\newcommand{\rev}[1]{{\color{black}{#1}}}

\newcommandx{\yrnote}[2][1=]{\todo[linecolor=Yellow,backgroundcolor=Yellow!25,bordercolor=Yellow,#1]{#2}}

\newcommandx{\lxnote}[2][1=]{\todo[linecolor=Plum,backgroundcolor=Plum!25,bordercolor=Plum,#1]{#2}}

\usepackage{array} % for better arrays (eg matrices) in maths
\usepackage[]{algorithm2e}
\usepackage{multibib}
\newcites{AP}{Appendix References}

\newcommand{\abortion}{{\tt Abortion}\xspace}
\newcommand{\guncontrol}{{\tt Gun Control}\xspace}
\newcommand{\blm}{{\tt BLM}\xspace}

\title{Linking Online Attention to Offline Opinions: [an iconic finding]}
\title{Linking Cross-Platform Attention to Offline Opinions: When Do Promotions Not Work?}
\title{Linking Cross-Platform Attention to Offline Opinions: Do Groups Tweeting More Get More Video Views?}
\title{Linking Cross-Platform Attention to Offline Attitudes: The Two Tales of Viral Content}
\title{Linking Cross-Platform Online Attention to Offline Attitudes and Behaviors}
\title{Measuring Cross-Platform Attention on Twitter and YouTube:\\ Case Studies on Abortion, Gun Control, and Black Lives Matter}
\title{Measuring Cross-Platform Attention on Twitter and YouTube}
\title{Measuring Cross-Platform Attention across Political Spectrum}
\title{Measuring Cross-Platform Attention on Controversial Topics}
\title{Measuring Attention Across YouTube and Twitter on Controversial Topics}

\title{Whose Advantage? Measuring Attention Dynamics \\ across YouTube and Twitter on Controversial Topics} %% idea for discuss

\author{
    JooYoung Lee,\equalcontrib\textsuperscript{\rm 1}
    Siqi Wu,\equalcontrib\textsuperscript{\rm 3,1}
    Ali Mert Ertugrul,\equalcontrib\textsuperscript{\rm 2}
    Yu-Ru Lin,\textsuperscript{\rm 2}
    Lexing Xie\textsuperscript{\rm 1}
}

\affiliations {
\textsuperscript{\rm 1} School of Computing, Australian National University \\
\textsuperscript{\rm 2} School of Computing and Information, University of Pittsburgh \\
\textsuperscript{\rm 3} School of Information, University of Michigan \\
\{jooyoung.lee,lexing.xie\}@anu.edu.au, \{ertugrul,yurulin\}@pitt.edu, siqiwu@umich.edu
}

\begin{document}
\maketitle

\begin{abstract}
The ideological asymmetries have been recently observed in contested online spaces, where conservative voices seem to be relatively more pronounced even though liberals are known to have the population advantage on digital platforms. Most prior research, however, focused on either one single platform or one single political topic. Whether an ideological group garners more attention across platforms and/or topics, and how the attention dynamics evolve over time, have not been explored. In this work, we present a quantitative study that links collective attention across two social platforms -- YouTube and Twitter, centered on online activities surrounding popular videos of three controversial political topics including Abortion, Gun control, and Black Lives Matter over 16 months. We propose several sets of video-centric metrics to characterize how online attention is accumulated for different ideological groups. We find that neither side is on a winning streak: left-leaning videos are overall more viewed, more engaging, but less tweeted than right-leaning videos. The attention time series unfold quicker for left-leaning videos, but span a longer time for right-leaning videos. Network analysis on the early adopters and tweet cascades show that the information diffusion for left-leaning videos tends to involve centralized actors; while that for right-leaning videos starts earlier in the attention lifecycle. In sum, our findings go beyond the static picture of ideological asymmetries in digital spaces and provide a set of methods to quantify attention dynamics across different social platforms.
\end{abstract}

\section{Introduction}
\label{sec:intro}

Several recent studies have documented the ideological asymmetries between the left-wing and right-wing activism~\cite{brady2019ideological,schradie2019revolution,freelon2020false,waller2021quantifying}. Some highlight the dominance of conservative voices on social media~\cite{brady2019ideological}; others portray the widespread symbolic support for progressive social movements~\cite{jackson2020hashtagactivism}. The term ``conservative advantage'' is coined to describe the strategic dissemination of right-wing users to spread their messages~\cite{schradie2019revolution}. However, most of the existing research bases on the analysis of a single platform or a single political topic. Relatively little is known about how different ideological groups garner attention across platforms, and whether the group advantage of gaining visibility remains across topics and over time. To answer these questions, this work designs several sets of cross-platform measurements on the collective attention dynamics of two different ideological groups across three controversial political topics.

Online platforms, such as Twitter, YouTube, Reddit, and Facebook, are social-technological artifacts that segregate online attention into silos defined by the underlying software and hardware systems. Video views on YouTube are known to be driven by discussions outside the platform~\cite{rizoiu2017expecting}, and to be part of users' broader information diet~\cite{hosseinmardi2021examining}. What is not known, however, is how groups of related content comparatively evolve across different social platforms. Collective attention on political content have been studied on one topic, such as the Occupy Movement~\cite{thorson2013youtube}, Gun Control/Rights~\cite{zhang2019whose}, and Black Lives Matters~\cite{de2016social,stewart2017drawing}. Yet, cross-cutting studies that compare different movements are rare. With data from three long-running controversial topics, this work seeks to provide measures across YouTube and Twitter and paint a nuanced picture about the temporal patterns of attention from left to right.

\begin{figure*}[t]
  \centering
  \includegraphics[width=0.96\linewidth]{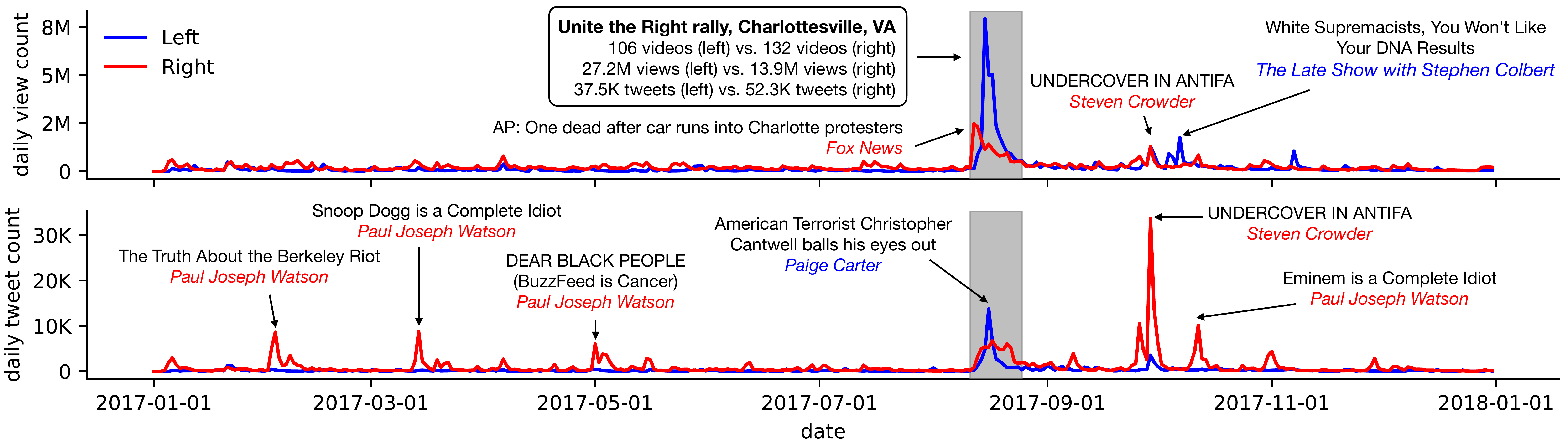}
  \caption{
  Attention time series (top: daily view count; bottom: daily tweet count) related to \blm throughout 2017. The view count dynamics of both left- and right-leaning videos have a handful of sharp peaks, while the tweet count dynamics of right-leaning videos peak more frequently. Notable video releases (with the most views and tweets) are labeled. YouTube channel titles are {\em italicized} and colored by video leanings. Best viewed in colors.
  }
  \label{fig:intro_teaser}
\end{figure*}

We choose three topics: \abortion, \guncontrol, and \texttt{Black Lives Matter} (\blm). We rely on video hyperlinks to connect the content from YouTube to Twitter. A motivating example is given in~\Cref{fig:intro_teaser}. We plot the time series of daily view count for the collected \blm videos from YouTube (top panel) and daily volume of tweets mentioning these \blm videos from Twitter (bottom panel). Both time series are further disaggregated by video uploaders' political leanings. Visually, the view count dynamics of both left- and right-leaning videos are relatively stable in year 2017, except a sharp spike caused by the ``Unite the Right rally''\footnote{\url{https://en.wikipedia.org/wiki/Unite_the_Right_rally}} event in Charlottesville, USA. On the bottom panel, the tweet count dynamics of right-leaning videos have many spikes, which can be attributed to the upload of new videos from far-right YouTube political commentators. The measures on YouTube and Twitter present a contrasting story here: if we focus on the two weeks period after the rally, left-leaning videos attracted more attention on YouTube (measured by views, left: 27.2M, right: 13.9M) while right-leaning videos had higher exposure on Twitter (measured by tweets, left: 37.5K, right: 52.3K). This example demonstrates the need for cross-platform analysis -- findings on one platform may not generalize to another.

We design a set of metrics from publicly available data on YouTube and Twitter, which include total views, video watch engagement, tweet reactions, the evolution of attention over time, and early adopter networks among tweets and Twitter users. On YouTube, we find that left-leaning videos accumulate more views, are more engaging, and have higher viral potential than right-leaning videos. In contrast, right-leaning videos have higher numbers of total tweets and retweets on Twitter. Statistics on the unfolding speed for views and tweets show that the attention on left-leaning videos attenuates faster, while that on right-leaning videos persists for longer. Note that these observations are not generalized unanimously across topics, e.g., for some metrics, we observe significant differences for \abortion and \guncontrol, but not for \blm. These findings expand current wisdom on ideological asymmetries in two ways: the first is exposing the novel facet that left-leaning content attracts more attention in a shorter period of time; the second is the need of contrasting temporal attention statistics between platforms, such as right-leaning tweet cascades tend to start earlier and YouTube views on right-leaning content sustain longer. In sum, our observations paint a richer picture of attention patterns across the political spectrum, provide a basis for further studying political framing and group behavior, and supply fundamental metrics for understanding influences that transcend platforms.

The main contributions of this work include:

\begin{itemize}[leftmargin=*]
  \item a data curation procedure linking content on YouTube and Twitter for longitudinal topic monitoring.\footnote{Our datasets and analysis code are publicly available at \url{https://github.com/picsolab/Measuring-Online-Information-Campaigns}}
  \item several sets of cross-platform metrics that support statistical comparisons for different ideological groups, encompassing the volume and quality of attention, networks of tweets and users, as well as relative temporal evolution.
  \item adding the temporal and cross-platform dimensions to recent observations on ideological asymmetries. We find that polarized content engages users in distinct ways -- more views, more engagement, and faster reactions for videos on the left; comparing to more tweets, more sustained attention for videos on the right.
\end{itemize}

\section{Related Work}
\label{sec:related}

\subsubsection{Online behavior of political groups.} 
Measurement studies have quantified different aspects of users, contents, and their interactions under political polarization on social media. \citet{conover2011political} presented one of the first profiling studies of polarized political groups on Twitter. 
% They studied the communication patterns between two opposing ideological groups and found that the retweet network was highly polarized but the mention network was more mixed. 
There have also been evidences that liberal and conservative groups attract online attention in different manners. \citet{abisheva2014watches} focused on a set of influential Twitter users who promoted YouTube videos, and they found that conservatives tweeted more diverse topics than liberals and that conservatives shared new videos faster. \citet{bakshy2015exposure} quantified the extent to which Facebook users were exposed to politically opposing contents, and they found that conservatives tended to seek out more cross-partisan content. 
\citet{linDynamicsTwitterUsers2020} distinguished online behavioral signals, such as linguistic and narrative characteristics, of two ideology groups in response to mass shooting events.
\citet{garimella2018political} defined several consumption and production metrics and profiled key user behavior patterns. \citet{ottoni2018analyzing} showed that conservatives used more specific language to discuss political topics and showed more negative emotions in the language. On YouTube, a recent study from \citet{wu2021cross} found that left-leaning videos attracted more comments from conservatives than right-leaning videos from liberals. However, all of these works are conducted platform-wide, and are not specialized into particular topics or movements. 

% Online activism, also known as online social movement or digital political campaign, has also been an active topic in computational social science. 
Online activism, also known as online social movement has been actively studied as a form of digital political campaigns.
For example, \citet{de2016social} presented one of the first studies on the \blm movement, measuring geographical differences in participation, and relationships to offline protests. \citet{stewart2017drawing} constructed a shared audience network of users who talked about \blm on Twitter and found the existence of superclusters among liberals and conservatives. \citet{zhang2015modeling} performed policy decision prediction based on tweet texts analysis on \texttt{same-sex marriage}. In a follow-up work, \citet{zhang2016gender} discussed gender disparity by linking tweet texts to the state-level \abortion policy events. \citet{ertugrul2019activism} examined the relation between offline protest events and their social and geographical contexts. \citet{freelon2020false} explained different tactics of liberals and conservatives when approaching audience on social media and articulated the asymmetries of measured behavior between conservatives and liberals. However, all of these works focus on a single controversial topic. In contrast, we examine three political topics and assess consistency of findings across the topics.

\subsubsection{Cross-platform measurement studies.} One early attempt in linking Twitter and YouTube data is from \citet{abisheva2014watches}, in which the authors found that the features of early adopters on Twitter were predictive for the video view counts on YouTube. \citet{rizoiu2017expecting} proposed the Hawkes Intensity Process that linked the time series of tweets and views, which led to a metric called {\it viral potential} for measuring the expected number of views that a video would obtain if mentioned by an average tweet~\cite{rizoiu2017online}. \citet{zannettou2017web} measured the sharing of alternative and mainstream news articles on three different platforms -- Twitter, Reddit, and 4chan. This seminal study characterized the role of fringe communities in spreading news. \citet{hosseinmardi2021examining} used browsing histories to infer video watch behavior, and quantified the consumption and driver of extreme content with respect to users' information diet. However, all of these works cover a breadth of content sources and categories, but none is focused around social movements or consistent political topics. 

This presented work bridges the gap of cross-platform measurement studies on multiple social movements, aiming to provide richer understanding that balances the ideological asymmetry between the left and the right.

\section{Curating Tweeted Video Datasets}
\label{sec:data}

We constructed three new cross-platform datasets by tracking videos on YouTube and posts on Twitter over three controversial topics: \abortion, \guncontrol, and \blm. Those topics have been studied extensively by social and political scientists~\cite{zhang2016gender,de2016social,stewart2017drawing,garimella2018political}. In this section, we first describe the data collection strategy. We then introduce our methods for estimating the political leanings of Twitter users and YouTube videos. \Cref{table:datasets} summarizes the overall statistics of the three topical datasets.

\begin{table*}[tbp]
    \centering
    \small
    \begin{tabular}{rrrrr|rr|rrrrr}
        \toprule
        & \multicolumn{4}{c}{YouTube video} & \multicolumn{1}{c}{Tweet} & \multicolumn{1}{c}{User} & \multicolumn{4}{c}{20\% Early Adopters (Twitter user)} & \multicolumn{1}{c}{(Tweet)} \\
        & L & C & R & Total & Total & Total & Lib. & Neu. & Con. & Total & Total  \\
        \midrule
        \abortion & 58 & 10 & 111 & 179 & 106,776 & 76,337 & 2,534 & 1,708 & 8,202 & 12,444 & 15,843 \\
        \guncontrol & 81 & 33 & 154 & 268 & 270,543 & 156,145 & 10,987 & 3,371 & 13,591 & 27,949 & 37,264 \\
        \blm & 297 & 84 & 396 & 777 & 593,574 & 262,580 & 10,419 & 5,745 & 33,934 & 50,098 & 78,969 \\
        \bottomrule
    \end{tabular}
    \caption{Summary of cross-platform datasets on three controversial topics. L: Left-leaning video; C: Center video; R: Right-leaning video. Lib.: Liberal user; Neu.: Neutral user; Con.: Conservative user. See Section A of~\cite{appendix} for methods of curating tweeted video datasets across YouTube and Twitter, and Section B\&C for methods of estimating the political leanings of Twitter users and YouTube videos. The ratios of early adopters are slightly below 20\% due to banned and protected users.
    }
    \label{table:datasets}
\end{table*}

% ----------------------------------------------------
\subsection{Finding YouTube Videos and Twitter Posts of Controversial Topics}
\label{ssec:data_find}

We are interested in topical YouTube videos and the discussions about them on Twitter. Following the approach used in~\cite{wu2018beyond}, we collected public tweets that mentioned any YouTube URLs via the Twitter filtered streaming API. Our raw Twitter stream spanned 16 months (2017-01-01 to 2018-04-30), and contained more than 1.8 billion tweets. To subsample videos that attracted a reasonable amount of attention before the end of the observation period, we required that the videos must be published in 2017, receive at least 100 tweets and at least 100 views within the first 120 days after upload. We make this filtering choice because (a) analyzing videos with little attention (${<}1$ view per day) is bound to generate noises when comparing different groups; (b) characterizing the timing and structure of a video's tweeting cascades requires a non-trivial number of tweets. This yielded 328,557 videos, which were mentioned in 242M tweets by 29.9M users. For each video, we collected its metadata, daily time series of view count and watch time, and all tweets mentioning it.

To identify topic-relevant videos and tweets, we first curated three separate keyword lists for three controversial topics -- \abortion, \guncontrol, and \blm. We consider a video is potentially relevant if (a) it contains at least one keyword in the video title or description; or (b) it is mentioned in a tweet that contains at least one keyword in the tweet text. Next, we used a mix of manual and semi-automated approaches to annotate the potentially relevant videos. Section A of \cite{appendix} details the topical keywords, their curation, and our video annotation protocol. In total, we obtained 179 \abortion, 268 \guncontrol, and 777 \blm videos, which were mentioned in 970K+ tweets.

\begin{figure}[t!]
  \centering
  \includegraphics[width=0.98\linewidth]{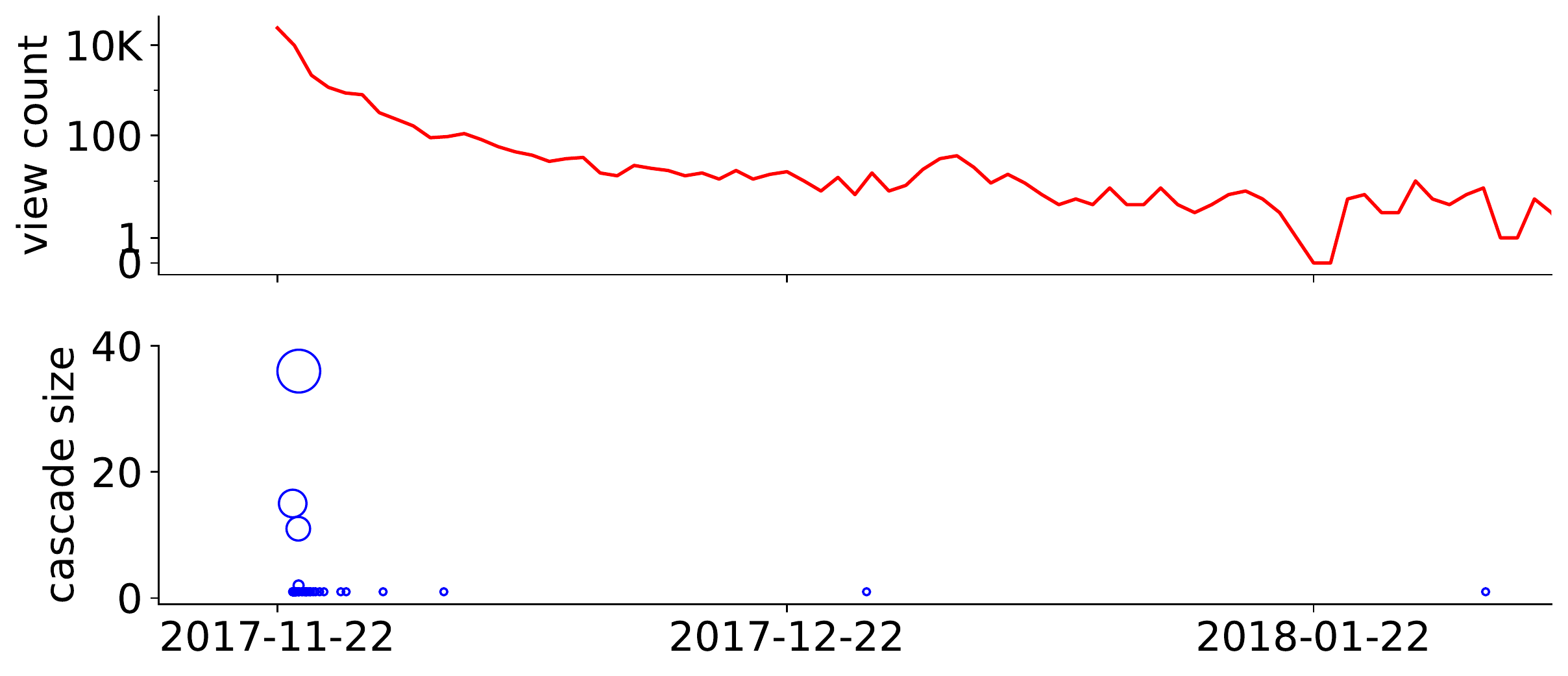}
  \caption{
  Time series of daily view count (top panel, in log scale) and tweet cascades (bottom panel) for an example video (YouTube video id: {\tt PTjMngQwGh8}). A tweet cascade is placed at the time of its first tweet, with y-axis and size proportional to the number of tweets in the cascade (see \Cref{ssec:cascade_metrics}). The viewing dynamic unfolds for months, but tweets only happen sporadically after the first week. 
  }
  \label{fig:data_overtime}
\end{figure}

\Cref{fig:data_overtime} shows the daily view count series and tweet cascades for an example video. A notable contrast between the two sources is that YouTube attention data is only available in daily aggregates, i.e., without individual user logs, while tweets have precise timestamp, user information, and relations between tweets and users. \Cref{sec:metric} is dedicated to designing measures for such cross-platform multi-relational temporal data. We bootstrapped the political information about Twitter users and YouTube videos. In particular, we gathered more information about videos' early adopters, defined as the first 20\% users who tweeted about each video. We collected the follower lists for all early adopters. Network measures of early adopters are found to indicate future popularity~\cite{romero2013interplay}. The threshold (first 20\%) is chosen to balance the need for data and the burden of collecting network information within practical API limits. After filtering out banned and protected users, we extracted 132K early adopter tweets posted by tens of thousands of users across the three topics (see \Cref{table:datasets}).

% ----------------------------------------------------
\subsection{Estimating Political Leanings of Twitter Users and YouTube Videos}
\label{ssec:data_estimate}

We classified the political leanings of early adopters on Twitter into liberal, neutral, and conservative. Meanwhile, we classified the video leanings into left, center, and right. Twitter users and YouTube videos are related in that left-leaning contents (e.g., \guncontrol videos) are generally shared by liberal users while right-leaning contents (e.g., {\tt Gun Rights} videos) are shared by conservative users.

We estimated Twitter users' political leanings by first identifying a group of seed users who included political hashtags in their profile descriptions, and then by using a label propagation algorithm~\cite{zhou2004learning} to propagate the labels of seed users to other users based on the shared follower network. This is a common approach for classifying user leaning on Twitter~\cite{stewart2017drawing} and the {\it follow} relation is found to be the most important in predicting user ideology~\cite{xiao2020timme}. We performed 10-fold cross-validation to evaluate the classification performance. We observed very high scores in precision, recall, and F-score across all three topics ($>95\%$ in all metrics). Section B of~\cite{appendix} describes our classification and evaluation methods for Twitter users in more detail.

We estimated YouTube videos' political leanings by first averaging the leaning scores of videos' early adopters on Twitter. We then used an external YouTube media bias dataset~\cite{ledwich2020algorithmic} to label the video leanings and identified optimal classification thresholds. We were able to find 58/10/111 left, center, and right-leaning videos for \abortion (analogously, 81/33/154 for \guncontrol and 297/84/396 for \blm). To validate our estimation, we performed one round of manual annotation for videos in \guncontrol. We used stratified sampling to sample 50 videos based on the video leaning scores. These videos were annotated independently by three authors. The Fleiss' Kappa was 0.69, suggesting a moderate inter-rater agreement. Section C of~\cite{appendix} details our classification and evaluation methods for YouTube videos.

\section{Measures for Cross-Platform Attention}
\label{sec:metric}

This section designs several sets of metrics for the cross-platform data, in order to compare content across different political ideologies, and examine whether the differences are consistent across topics, across platforms, and over time.

% ----------------------------------------------------
\subsection{Aggregate Attention on YouTube and Twitter}
\label{ssec:aggregate_metrics}

We present four metrics for the total video attention on YouTube.

\header{\em Total view count} sums up a video's view count time series until day 120.

\header{\em Relative engagement} is a metric proposed in~\cite{wu2018beyond} for quantifying the average video watching behavior. Specifically, for each video, we first compute {\em average watch percentage}, defined as the total watch time divided by total number of views (both at 120 days) and then normalized by the video length (in seconds). The relative engagement score is the percentile ranking of average watch percentage among videos of similar lengths. It is a normalized score between 0 and 1. A higher score means more engaging, e.g., a score of 0.8 suggests that this video is on average watched for longer time than 80\% videos of similar length. Note that relative engagement is shown to be stable over time, hence there is no need to examine the temporal variations of watch time, as it would strongly correlate with view counts. In this work, relative engagement is computed based on a publicly available collection of 5.3M YouTube videos~\cite{wu2018beyond}, with details described in Section D of \cite{appendix}.

\header{\em Fraction of likes} measures the video reaction -- provided by YouTube as the total counts of likes and dislikes and collected via the thumb-up and thumb-down icon on the video page. A relatively lower fraction of likes indicates a more diverse audience reaction to the video content. Note that the majority of videos receive a lot more likes than dislikes.\footnote{YouTube announced that the dislike count will no longer be available to the public on Nov 10, 2021. \url{https://blog.youtube/news-and-events/update-to-youtube/}}

\header{\em Viral potential} is a positive number, representing the {\em expected} number of views that a YouTube video will obtain if mentioned by an {\em average} tweet on Twitter~\cite{rizoiu2017online}. More specifically, it is the area under the impulse response function of an integral equation known as Hawkes Intensity Process (HIP)~\cite{rizoiu2017expecting}, which is learned for each video by using the first 120 days of tweeting and viewing history. We choose this quantity rather than simply dividing the number of views by the number of tweets, because the model takes into account views that are yet to unfold due to its sustained circulation via sharing and tweeting. A self-contained summary about HIP and viral potential computation is given in Section E of \cite{appendix}.

On Twitter, tweets can be categorized into four types: original tweets, retweets, quotes, and replies. This leads to five counting metrics: the total number of {\bf\em tweets, original tweets, retweets, quoted tweets}, and {\bf\em replies}. 

% ----------------------------------------------------
\subsection{Views and Tweets over Time}
\label{ssec:temporal_metrics}

\header{\em Viewing half-life} is computed as the number of days to achieve half of its total views at day 120.

\header{\em Tweeting half-life} is computed as the number of days to achieve half of its total tweets at day 120.

\header{\em Tweeting lifetime} is time gap between the first and the last tweets. We do not measure lifetime on viewing because the view count of a video rarely becomes zero even towards the end of the measurement period, but tweets tend to exhaust much sooner.

\header{\em Tweeting inter-arrival time} is the average time difference between every two consecutive tweets about each video.

\header{\em Accumulation of views and tweets.} 
In addition to the summary metrics above, we also compare the attention accumulation on the left- and right-leaning content on a daily basis. On each day $t$, we compute the fraction of the total views that each video has achieved. This leads to two sets of samples $\{v^{(L)}_i\}_{i=1}^n$ and $\{v^{(R)}_j\}_{j=1}^m$, where $n$ is the number of left-leaning videos and $m$ is the number of right-leaning videos. We then compute the normalized Mann-Whitney U (MWU) statistic~\cite{mann1947test}, 
$$\bar U_t = \frac{1}{nm}\sum_i\sum_j \{ I[v_i^{(L)} > v^{(R)}_j] + 0.5\,I[v_i^{(L)} = v^{(R)}_j] \}$$ 

Here $I[\cdot]$ is the indicator function that takes value 1 when the argument is true, 0 otherwise. The U statistic intuitively corresponds to the fraction of sample pairs $(v^{(L)}_i, v^{(R)}_j)$ where the sample from left-leaning distribution is larger, accounting for ties. If the distributions of $v^{(L)}$ and $v^{(R)}$ are indistinguishable, then $\bar U$ would be around 0.5. We compute the statistic $\bar U_t$ on tweets in the same fashion, and both statistics are computed for each day. These two series of statistics allows us to quantify the differences between left- and right-leaning content, and compare the trends on the accumulation of views and tweets over time.

% ----------------------------------------------------
\subsection{Videos' Tweet Cascades}
\label{ssec:cascade_metrics}

\header{\textit{Cascade size.}}
We define that a {\em cascade} consists of a root tweet and all of its retweets, replies, and quotes. It is well-known that the vast majority of cascades in online diffusion networks are very small and only a very small fraction of cascades would become very big~\cite{goel2012structure}. Based on the number of tweets in a cascade, we divide the cascades into isolated (only root tweet), small (2-4 tweets), and large ($\geq 5$ tweets) groups. For videos of each leaning on each topic, we compute the fractions of isolated/small/large cascades and the fraction of tweets in each cascade group. These metrics quantify the structure of online diffusion and allow us to compare behavior on controversial political topics with what was known about tweeted videos in general.

\header{\textit{Cascade start time}} is the percentage of accumulated views of the video when the root tweet of the cascade is posted. It measures how much view attention is accumulated on YouTube before the infusion on Twitter starts. We choose to describe cascade timing relative to the accumulation of view, rather than in absolute number of days since upload, because (a) such relative time more directly correlates the amount of cascades with respect to the views they can potentially drive (rather than through another variable, days); and (b) the percentage of views provides more granularity, since many videos have all views and tweets unfold within a few days after upload.

% ----------------------------------------------------
\subsection{Networks among Early Adopters on Twitter}
\label{ssec:network_metrics}

For each video, we obtain its follower network among the early adopters. If there exists a following relationship between a pair of users, a directed edge is established. This results in one network for each shared video. We compute a set of metrics per video, and then compare their distributions on each topic for left- and right-leaning videos. We describe two key metrics here, and discuss four additional metrics in Section H of~\cite{appendix}.

\header{\textit{Gini coefficient of indegree centrality.}} 
We calculate the indegree centrality for each node in the network. To have a video-level metric, we use the Gini coefficient, which ranges from 0 to 1 and measures the distribution inequality. Specifically, the Gini coefficient of indegree centrality quantifies the degree of inequality of the indegree distribution. A higher value indicates that a few early adopters are followed more by other early adopters, and a lower value indicates that the indegree distribution is more equal. 

\header{\textit{Gini coefficient of closeness centrality}} captures the dispersion in inverse of average shortest path length from one early adopter to all other early adopters of a given video. Higher coefficient implies that a few early adopters can reach the rest of the early adopters within a few hops.

\section{Observations on Cross-Platform Attention}
\label{sec:measure}

We report the results on all metrics described in \Cref{sec:metric}, in mirroring subsections to aid navigation. Many results in this section are presented as violin plots. The outlines are kernel density estimates for the left-leaning (blue) and right-leaning (red) videos, respectively. The center dashed line is the median, whereas the two outer lines denote the inter-quartile range. To compare the distributions of each metric for the left- and right-leaning videos, we adopt the one-sided Mann–Whitney U test. We summarize our results in~\Cref{table:measures} at the end of this paper.

% ----------------------------------------------------
\subsection{Aggregate Attention on YouTube and Twitter}
\label{ssec:aggregate_obs}

\begin{figure}[t!]
  \centering
  \begin{subfigure}[b]{0.48\linewidth}
    \includegraphics[width=\linewidth]{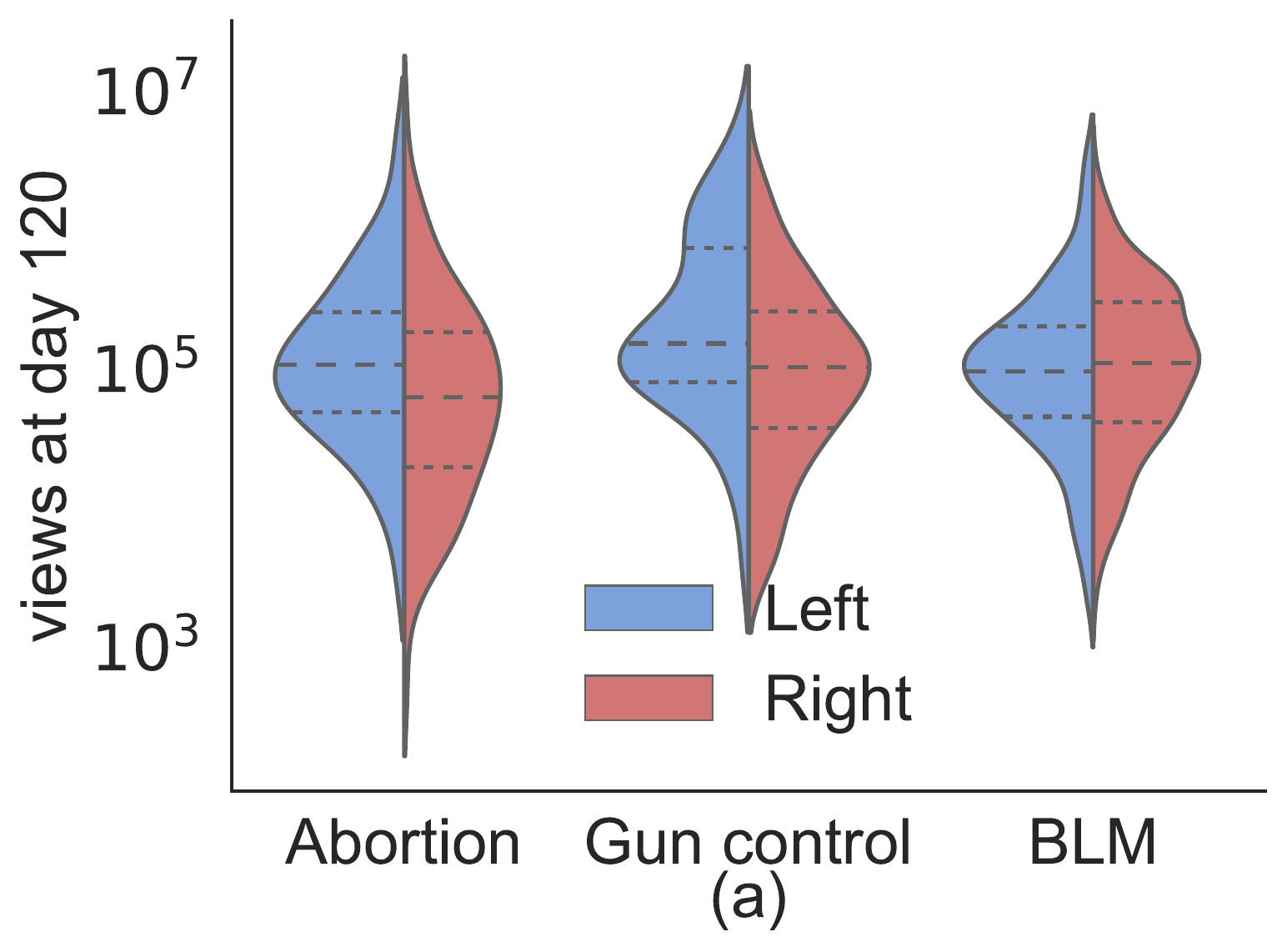}
  \end{subfigure}
  \begin{subfigure}[b]{0.48\linewidth}
    \includegraphics[width=\linewidth]{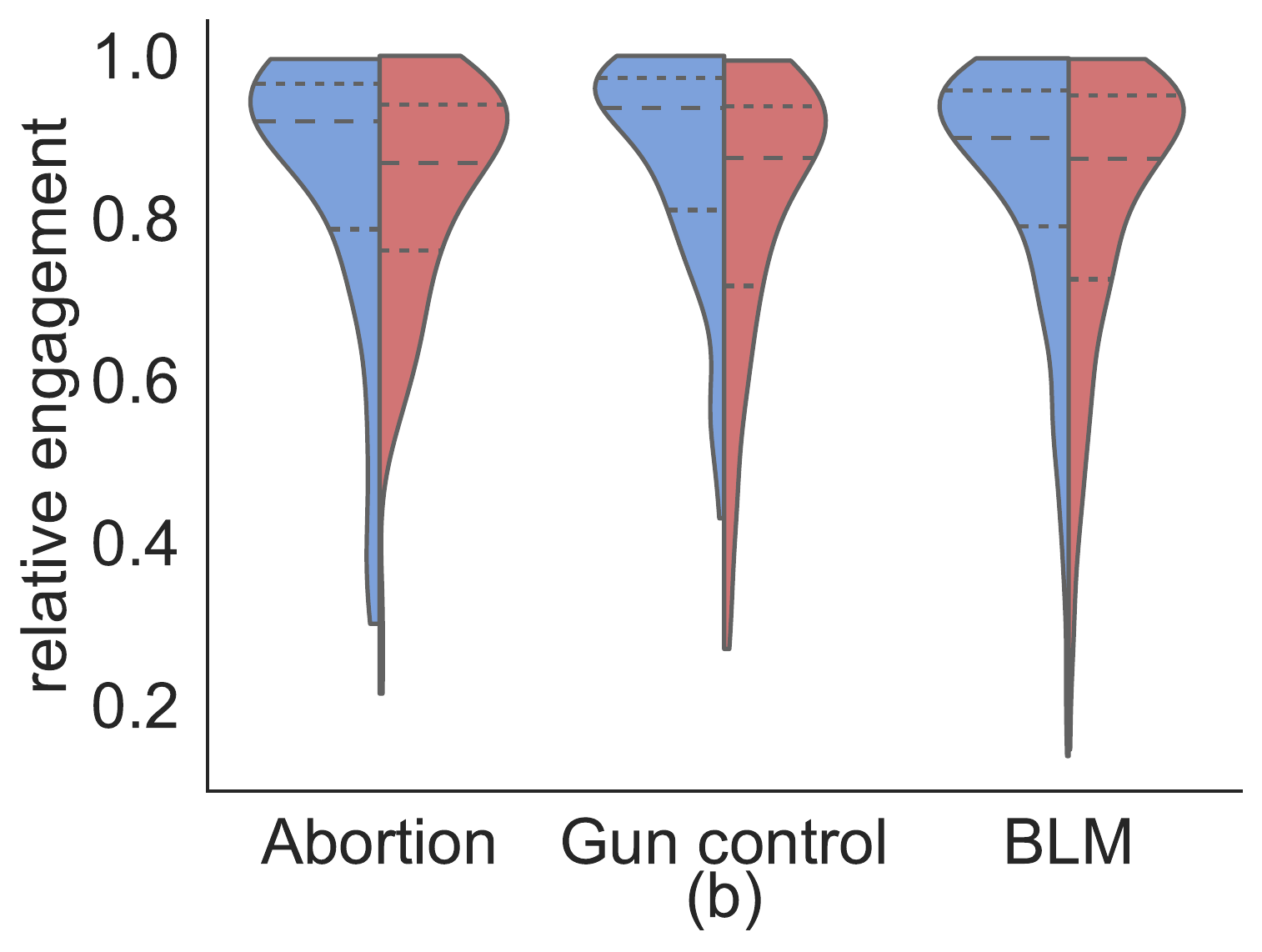}
  \end{subfigure}
  \begin{subfigure}[b]{0.48\linewidth}
    \includegraphics[width=\linewidth]{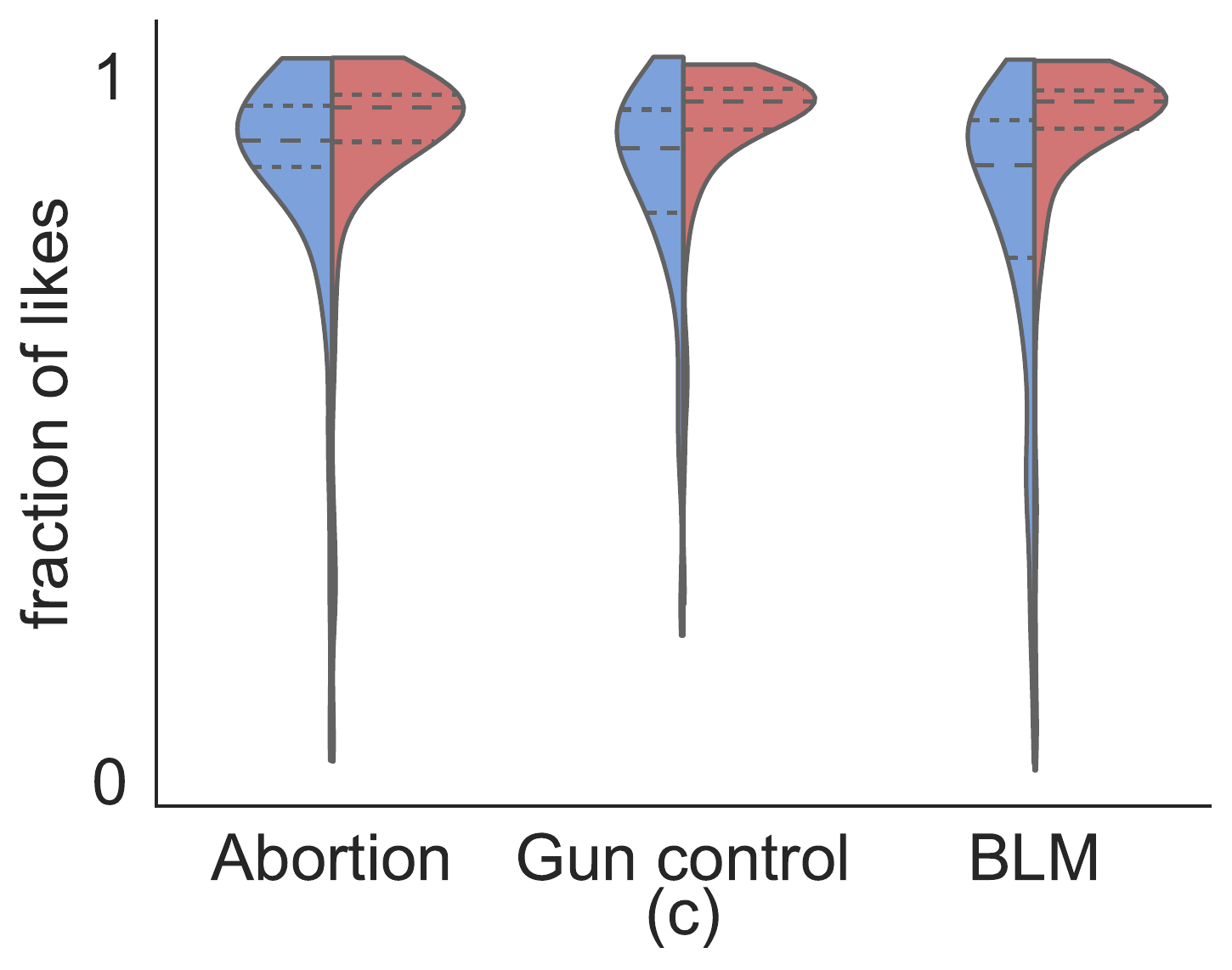}
  \end{subfigure}
  \begin{subfigure}[b]{0.48\linewidth}
    \includegraphics[width=\linewidth]{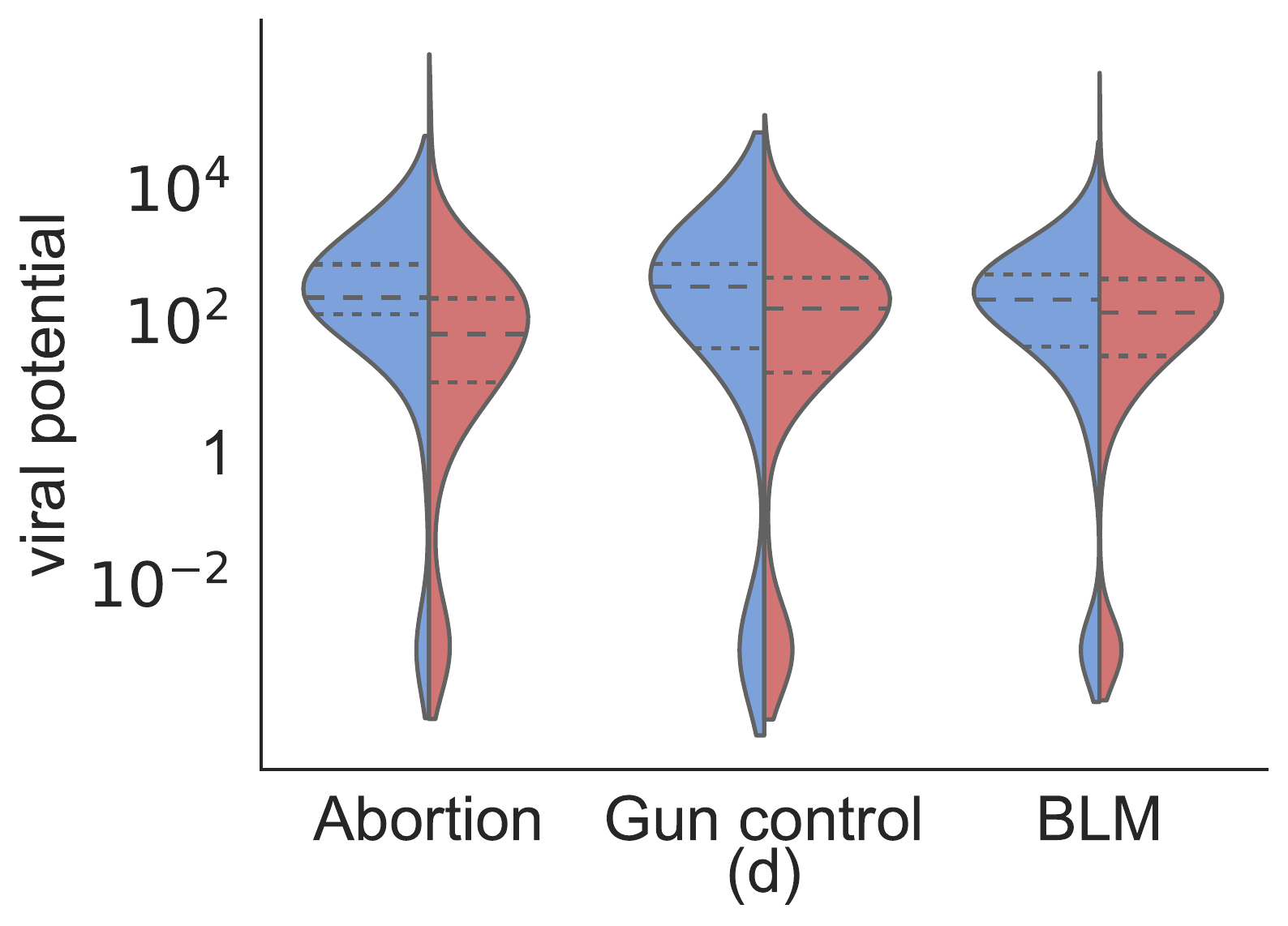}
  \end{subfigure}
  \begin{subfigure}[b]{0.48\linewidth}
    \includegraphics[width=\linewidth]{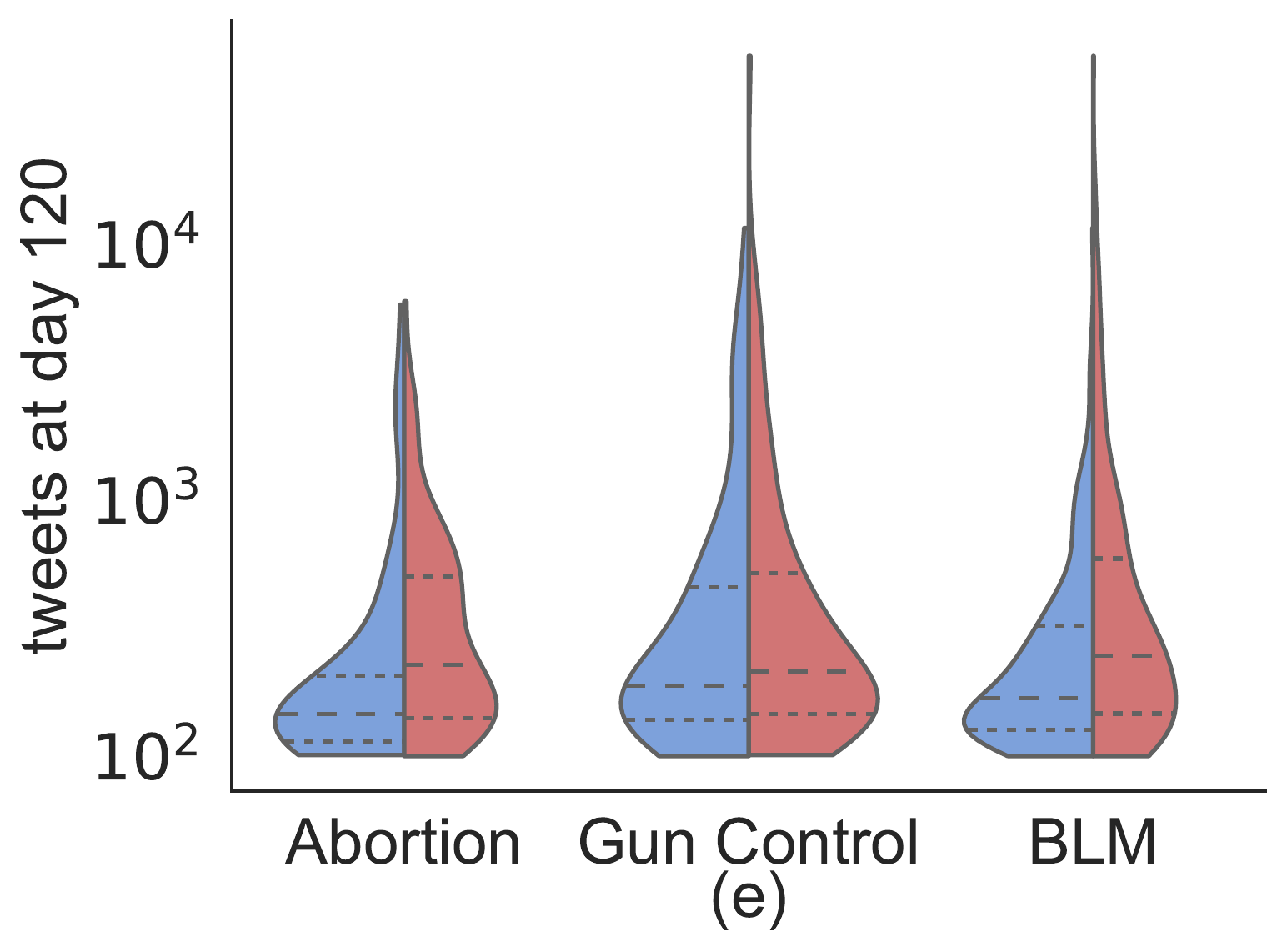}
  \end{subfigure}
  \begin{subfigure}[b]{0.48\linewidth}
    \includegraphics[width=\linewidth]{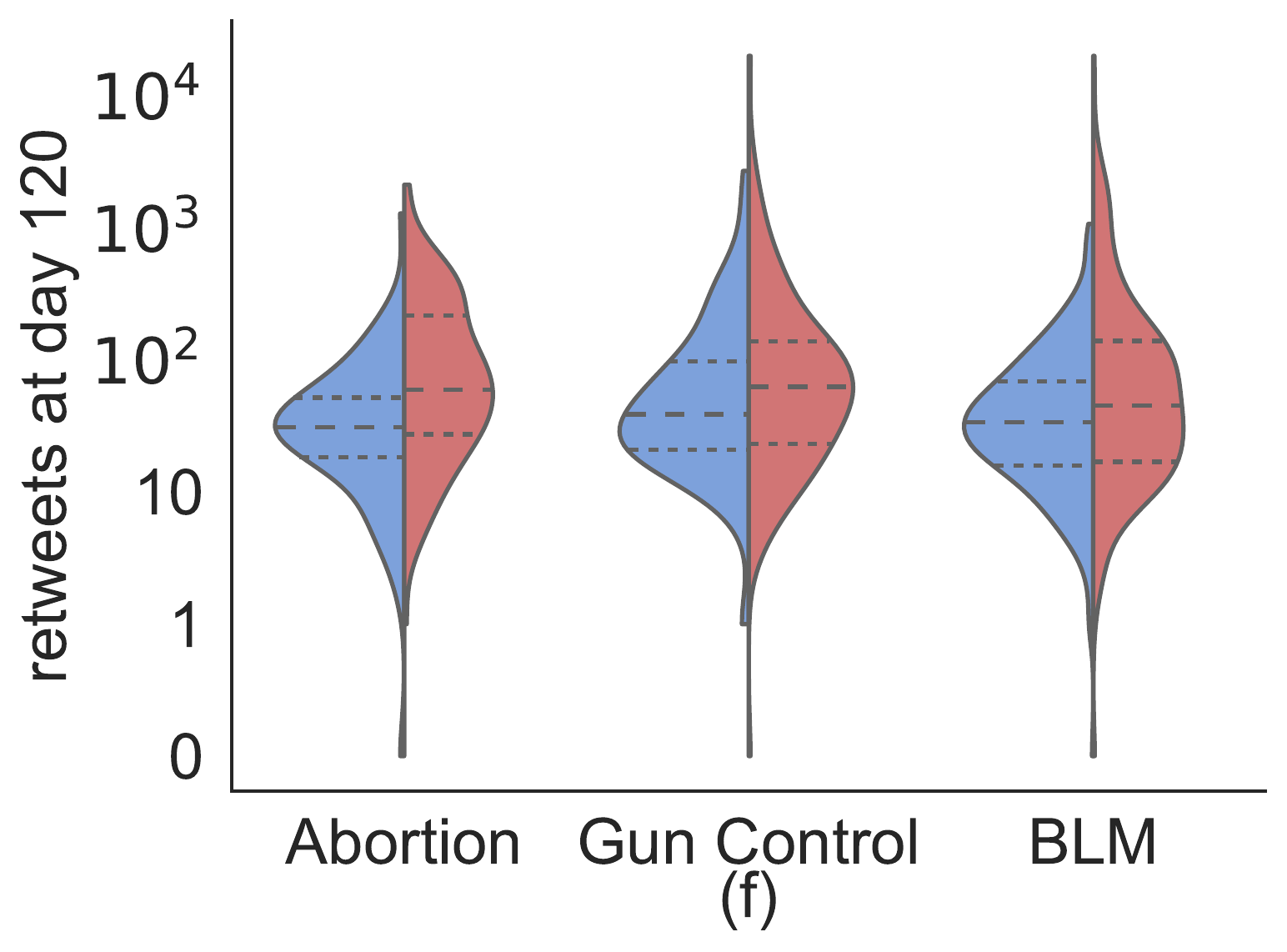}
  \end{subfigure}
  \caption{
  Violin plots comparing the left-leaning (blue) and right-leaning (red) videos in (a) total number of views at 120 days after upload, (b) relative engagement, (c) fraction of likes, (d) viral potential, (e) total number of tweets, and (f) total number of retweets at 120 days after the video upload. Left-leaning videos are more viewed, more engaging, having more diverse reactions, and more viral than right-leaning videos across all three topics (except views in \blm). In \abortion and \blm, right-leaning videos are significantly more tweeted, especially with more retweets.
  }
  \label{fig:obs_aggregate}
\end{figure}

\header{Total view count.}
\Cref{fig:obs_aggregate}(a) shows the distribution of video views at day 120 after upload. Using the view count at the same day removes the effects of video age, so that the videos published for longer time are not taking an unfair advantage. In \abortion and \guncontrol, the median, as well as 25$^{th}$ and 75$^{th}$ percentile of views of left-leaning videos are higher than that of right-leaning videos. The median views for left-leaning videos are 107,346 for \abortion and 153,482 for \guncontrol, versus 62,780 and 103,373 for right-leaning ones. The differences in view distribution are statistically significant ($p < 0.01$, \Cref{table:measures} row 1). For \blm, right-leaning videos have higher median and 75$^{th}$ of views, but the effect is not significant.

\header{Relative engagement.}
From \Cref{fig:obs_aggregate}(b), we can see that videos in all three topics are highly engaging, with mean relative engagement at 0.834 for \abortion, 0.824 for \guncontrol, and 0.831 for \blm. This is because our data processing procedure requires videos to have at least 100 tweets and 100 views, which tends to select videos with significant amount of interests. Left-leaning videos are significantly more engaging than their right-leaning counterparts across all three topics ($p < 0.05$, \Cref{table:measures} row 2).

\header{Fraction of likes.}
\Cref{fig:obs_aggregate}(c) presents the proportion of likes in videos' reactions. Left-leaning videos across all topics have significantly smaller fraction of likes than right-leaning videos ($p < 0.001$, \Cref{table:measures} row 3). This may be explained by the observation that there are far more cross-partisan talks on left-leaning videos~\cite{wu2021cross}.

\header{Viral potential.}
\Cref{fig:obs_aggregate}(d) shows the distributions of viral potential. We find that the left-leaning videos have significantly higher viral scores than the right-leaning videos across all three topics ($p < 0.05$, \Cref{table:measures} row 4), meaning that given the same amount of tweets exposing the video on Twitter, an average left-leaning video can effectively attracts more views than an average right-leaning video. The difference is most notable in \abortion: a typical left-leaning video receives 224 views from an average tweet, whereas a typical right-leaning video receives only 63 views.

\header{Tweet counts.}
\Cref{fig:obs_aggregate}(e) and (f) show the distributions of total tweets and retweets. Contrasting to the observation that left-leaning videos are more viewed, here we find that right-leaning videos are significantly more tweeted, especially with more retweets and more replies ($p < 0.001$, \Cref{table:measures} row 5-7) in \abortion and \blm. On the other hand, we do not observe a significant difference in original tweets and quotes, except for \blm where right-leaning videos have prevailing volume across all tweet types.

To examine the robustness of presented results in this section, we bootstrapped videos for each topic and for each ideological group. Specifically, for each group, we created $1,000$ bootstrapped sets of videos that are of the same size as the original group (shown in~\Cref{table:datasets}). Next, we computed the mean of proposed metrics (shown in~\Cref{fig:obs_aggregate}) for each bootstrapped set. Lastly, we used the {\it independent t-test} to check the statistical significance between left- and right-leaning groups. The results of the {\it t-tests} support all reported relations in~\Cref{table:measures} row 1-6 with $p < 0.001$.

% ----------------------------------------------------
\subsection{Views and Tweets over Time}
\label{ssec:temporal_obs}
We measure how quickly left- and right-leaning videos attract views and tweets. We find that left-leaning videos are reacted on YouTube and Twitter quicker across all topics.

\begin{figure}[t!]
  \centering
  \begin{subfigure}[b]{0.48\linewidth}
    \includegraphics[width=\linewidth]{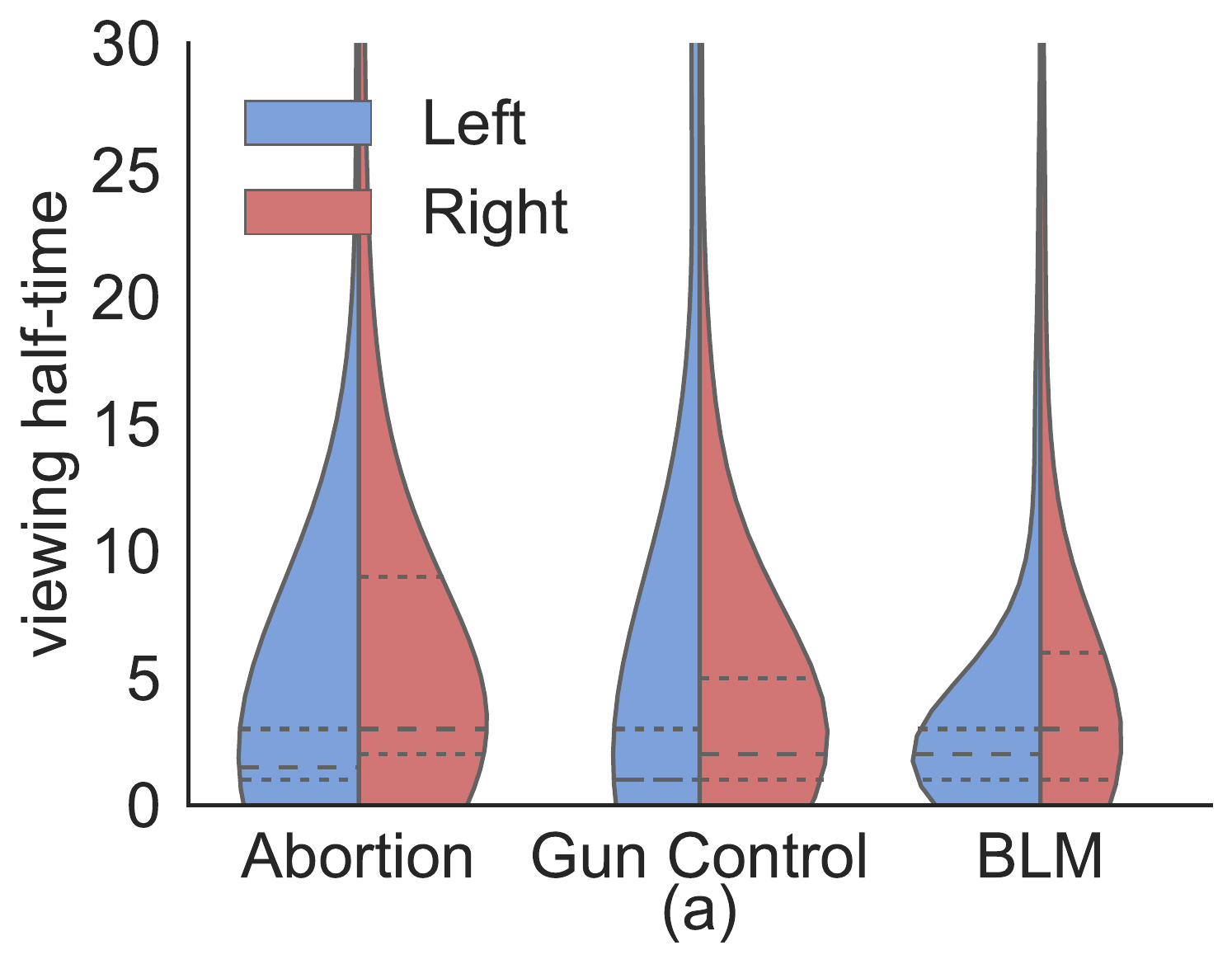}
  \end{subfigure}
  \begin{subfigure}[b]{0.48\linewidth}
    \includegraphics[width=\linewidth]{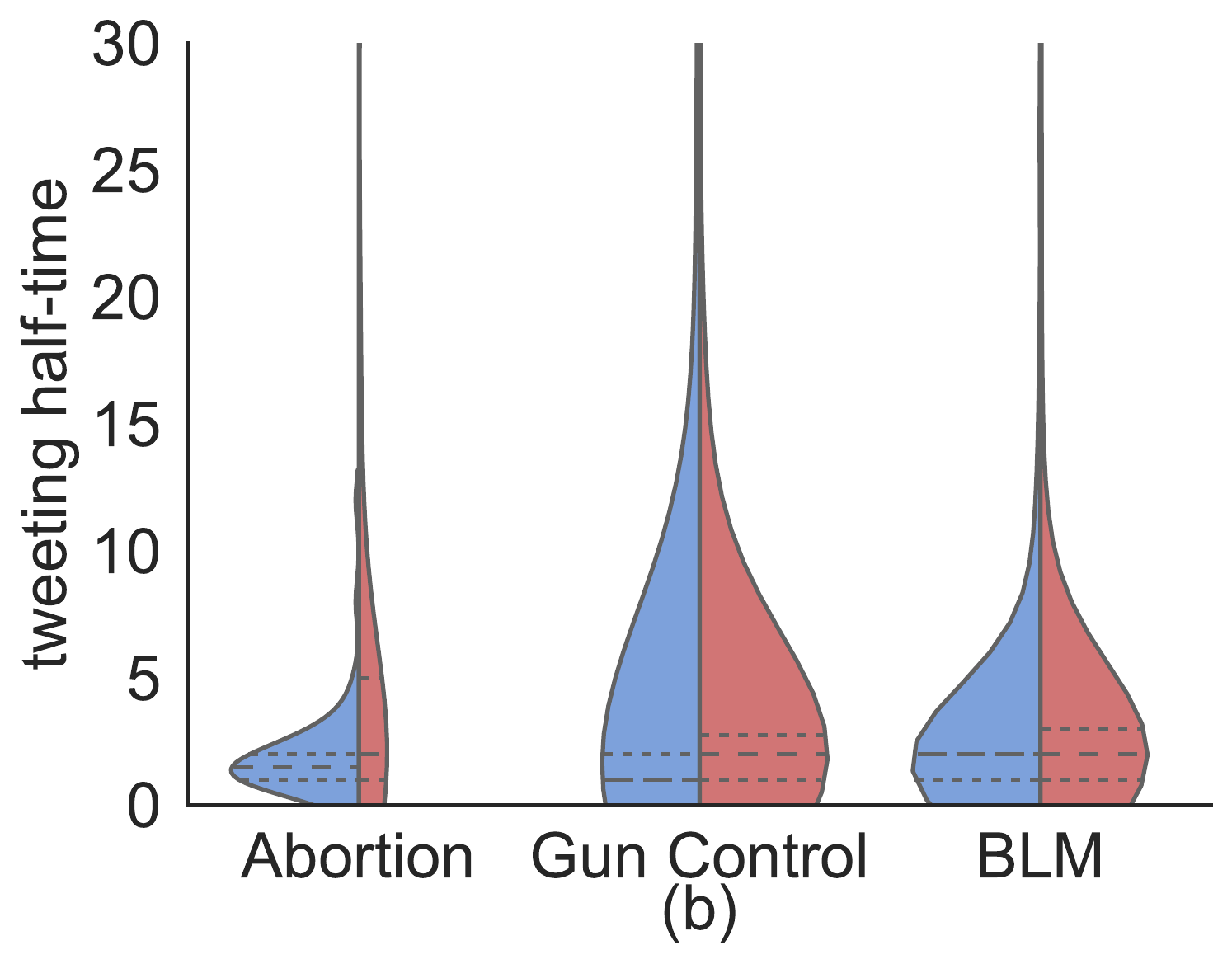}
  \end{subfigure}
  \begin{subfigure}[b]{0.48\linewidth}
    \includegraphics[width=\linewidth]{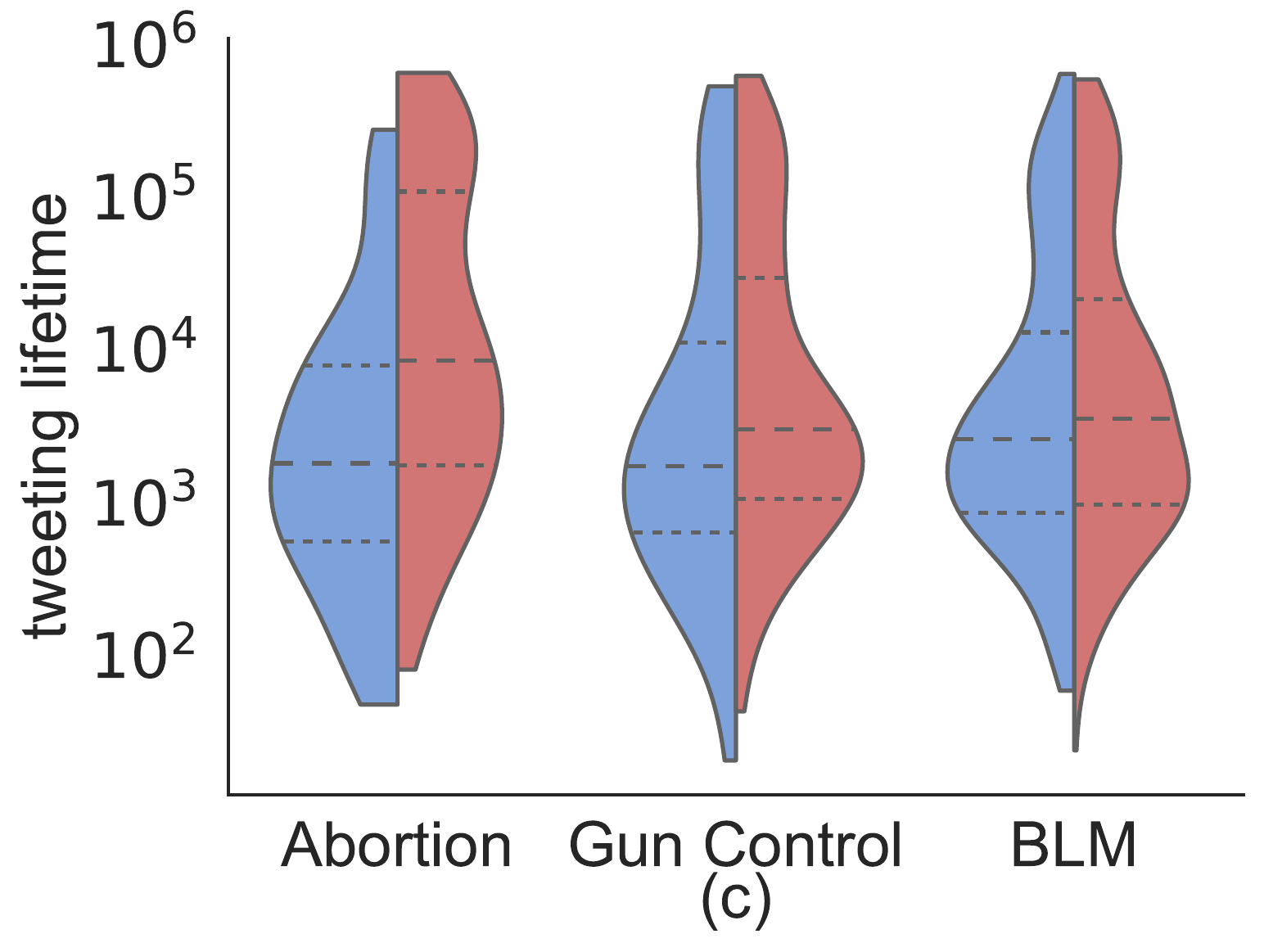}
  \end{subfigure}
  \begin{subfigure}[b]{0.48\linewidth}
    \includegraphics[width=\linewidth]{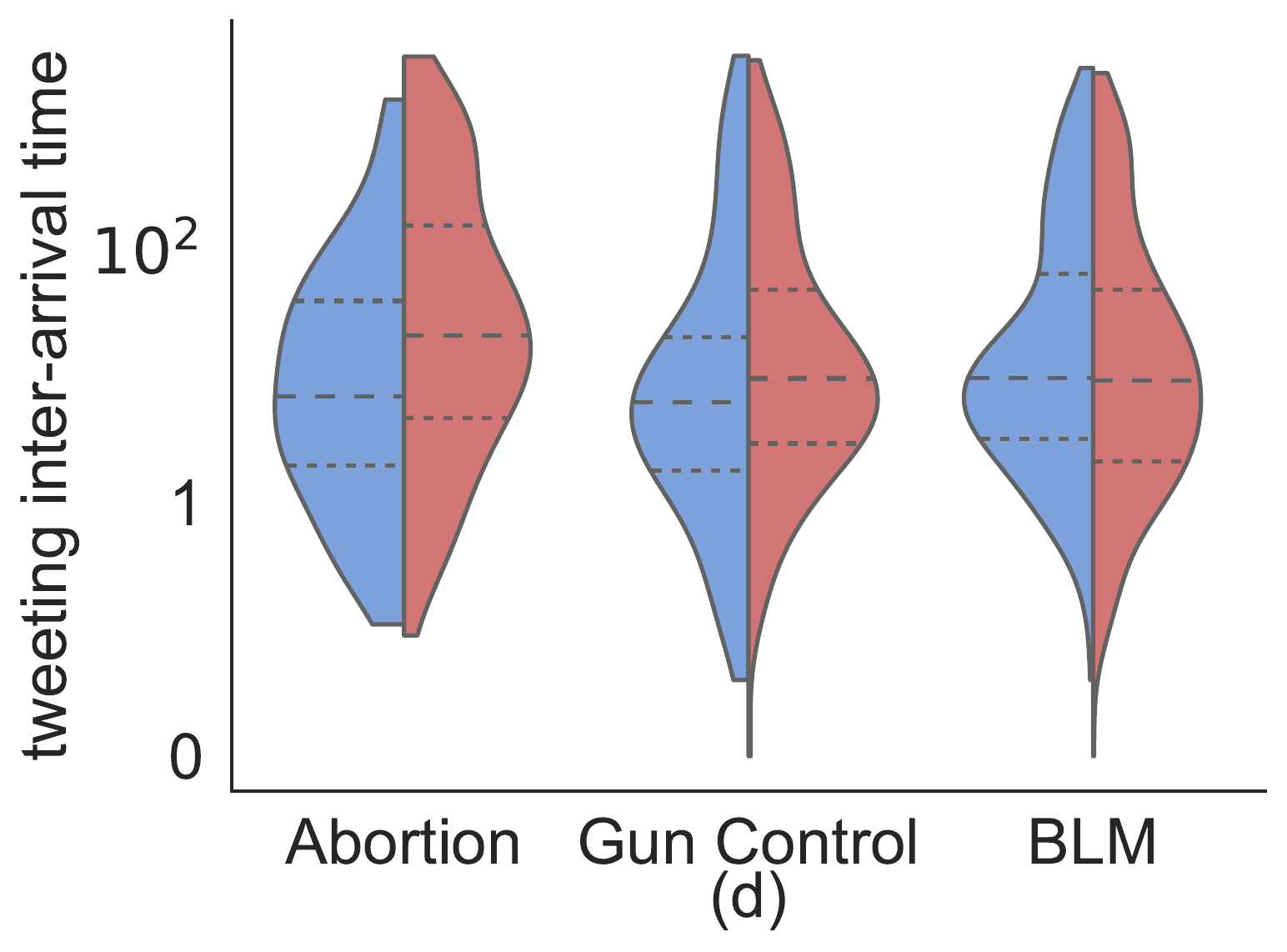}
  \end{subfigure}
  \caption{The number of days to receive 50\% (a) views and (b) tweets. 
  The time gaps (in minutes) between tweets measured as (c) tweeting lifetime and (d) tweeting inter-arrival time.
  Left-leaning videos are mentioned more quickly on Twitter and have shorter circulation duration for all metrics in \abortion and \guncontrol.
  }
  \label{fig:obs_temporal}
\end{figure}

\header{Viewing half-life} and \header{Tweeting half-life.}
We notice that there are significant differences in the attention consumption patterns: right-leaning videos have more prolonged attention spans on YouTube across all topics ($p < 0.01$, \Cref{table:measures} row 10). Right-leaning videos also have longer attention spans on Twitter for \abortion and \guncontrol ($p < 0.05$, \Cref{table:measures} row 11). For example, \Cref{fig:obs_temporal}(a) shows that right-leaning videos for \abortion have the longest attention span -- taking 9 days for 75\% videos to achieve viewing  half-life, while left-leaning videos only take 3 days. Comparing \Cref{fig:obs_temporal}(a) to \Cref{fig:obs_temporal}(b), we find that attention spans on Twitter are shorter than that on YouTube. In \abortion, left-leaning videos take 2 days for 75\% of videos to reach tweeting half-life (vs. 3 days for views) and right-leaning videos take 5 days for 75\% of videos to reach tweeting half-life (vs. 9 days for views).

\header{Tweeting lifetime} and \header{Tweeting inter-arrival time.} 
\Cref{fig:obs_temporal} (c) and (d) shows the distributions of tweeting lifetime and inter-arrival time. The results are mixed across topics. For \abortion and \guncontrol, the MWU test results show that left-leaning videos have significantly less scores in both metrics than right-leaning videos ($p < 0.05$, \Cref{table:measures} row 12-13). But for \blm, both metrics show similar distributions between left- and right-leaning videos.

These results suggest that left-leaning videos have shorter circulation duration and are mentioned more quickly (except \blm). The most notable difference is in \abortion where the median of tweeting lifetime and inter-arrival time for right-leaning videos are more than three times of those for left-leaning videos. For instance, the median tweeting lifetime is 1811.6 minutes for left-leaning videos, while the median is 8453.8 minutes for right-leaning videos.

\begin{figure}[t!]
  \centering
  \includegraphics[width=0.97\linewidth]{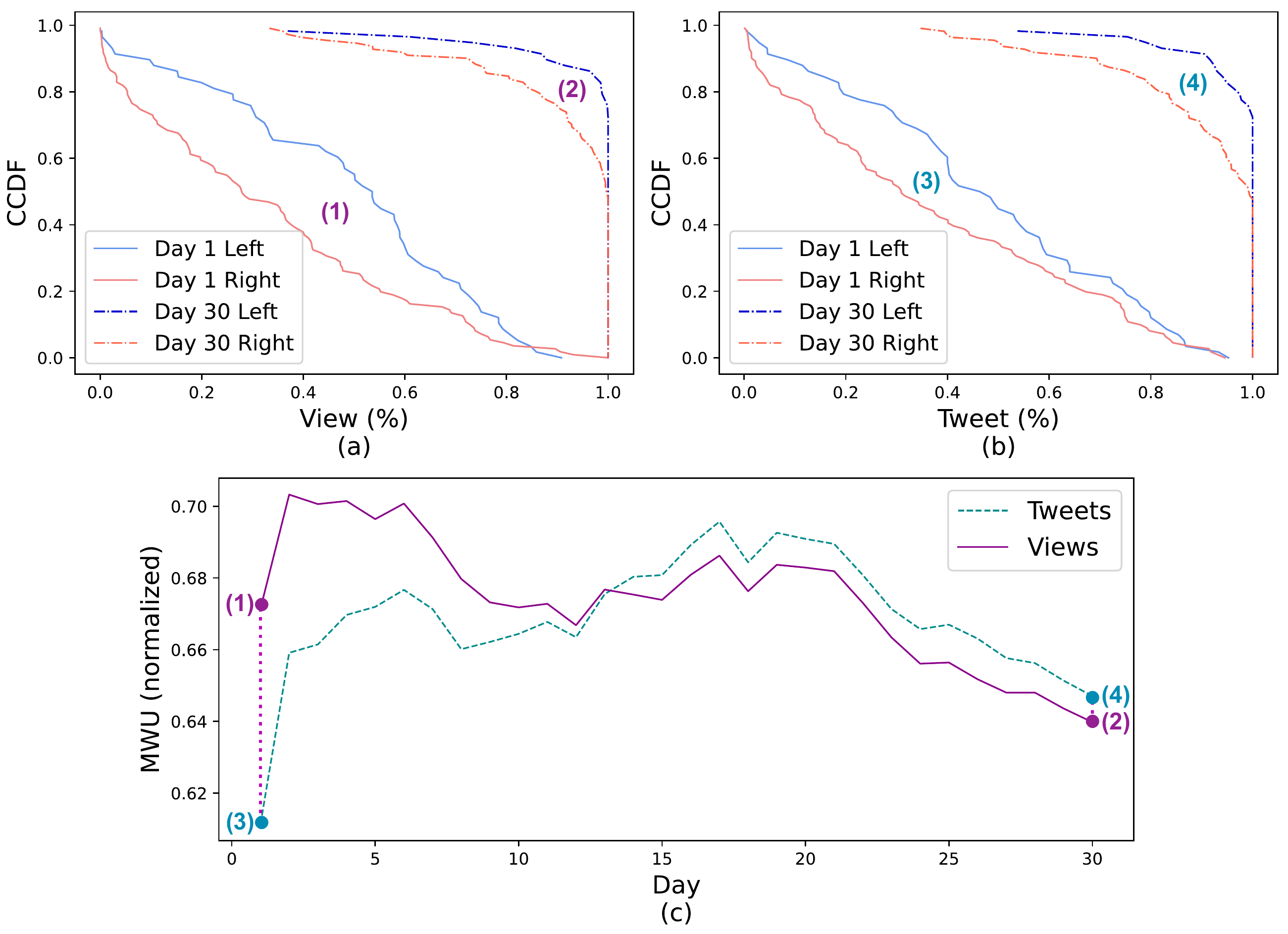}
  \caption{The distribution of accumulated views and tweets in \abortion videos on the first day of video upload and at $30^{th}$ day: (a) percentage of views achieved, (b) percentage of tweets achieved, (c) MWU statistic values for each day from the first day to day 30. The area marked with (1,2,3,4) in (a) and (b) are shown in (c) as a normalized MWU statistic. For example, (1) shows difference between left- and right-leaning videos in terms of view accumulation. This difference is measured using MWU test and shown in (c) where (1) is marked. The difference between left- and right-leaning videos, in both views and tweets accumulation, tend to become smaller in later days. Section F of~\cite{appendix} includes the same set of plots for \guncontrol and \blm.
  }
  \label{fig:obs_accumulation}
\end{figure}

\header{Accumulation of views and tweets.}
We examine how much views and tweets are accumulated each day. \Cref{fig:obs_accumulation}(a) and (b) show the Complementary Cumulative Distribution Function (CCDF) of views and tweets percentages accumulated for the first day (video published date) and $30^{th}$ day for \abortion videos. We observe that left-leaning videos tend to achieve views/tweets faster than right-leaning videos, which is consistent with \Cref{fig:obs_temporal}. For example, after day 1, 56.9\% left-leaning videos have achieved viewing half-life, but only 26.1\% right-leaning videos achieved the same. For tweet accumulation, after 1 day, the gap of tweeting half-life between left- and right- leaning videos is 11.4\% (46.6\% and 35.1\%, respectively). By day 30, only 3 left-leaning videos have not accumulated 80\% of views.

\Cref{fig:obs_accumulation}(c) compares the normalized MWU statistic values of left- and right-leaning videos in views $\bar U_v$ and tweets $\bar U_t$ on each of the 120 days since upload. It shows that the difference between left- and right-leaning videos is larger in the beginning and decreases towards 0.5 over time. At the $120^{th}$ day (as the observation period ends), both will be 0.5 by definition, we thus truncate the plot at day 30. We also observe that the differences in views is larger than that of tweets across left- and right- leaning videos initially, whereas the discrepancy between views and tweets narrows as videos get older. This is because more videos have already fully achieved all the views and tweets. 

% ----------------------------------------------------
\subsection{Videos' Tweet Cascades}
\label{ssec:cascade_obs}

\begin{figure}[t!]
  \centering
  \begin{subfigure}[b]{0.49\linewidth}
    \includegraphics[width=\linewidth]{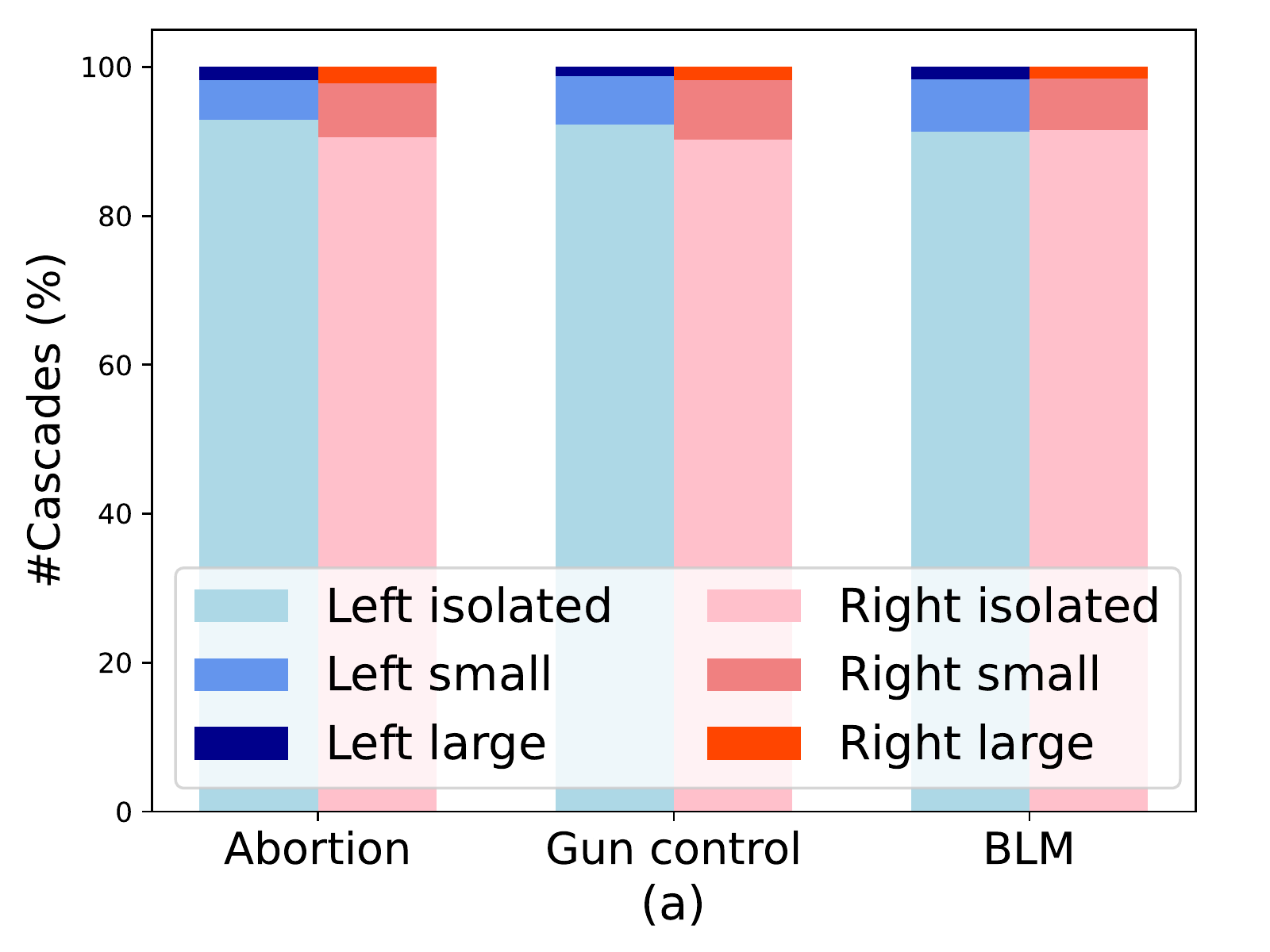}
  \end{subfigure}
  \begin{subfigure}[b]{0.49\linewidth}
    \includegraphics[width=\linewidth]{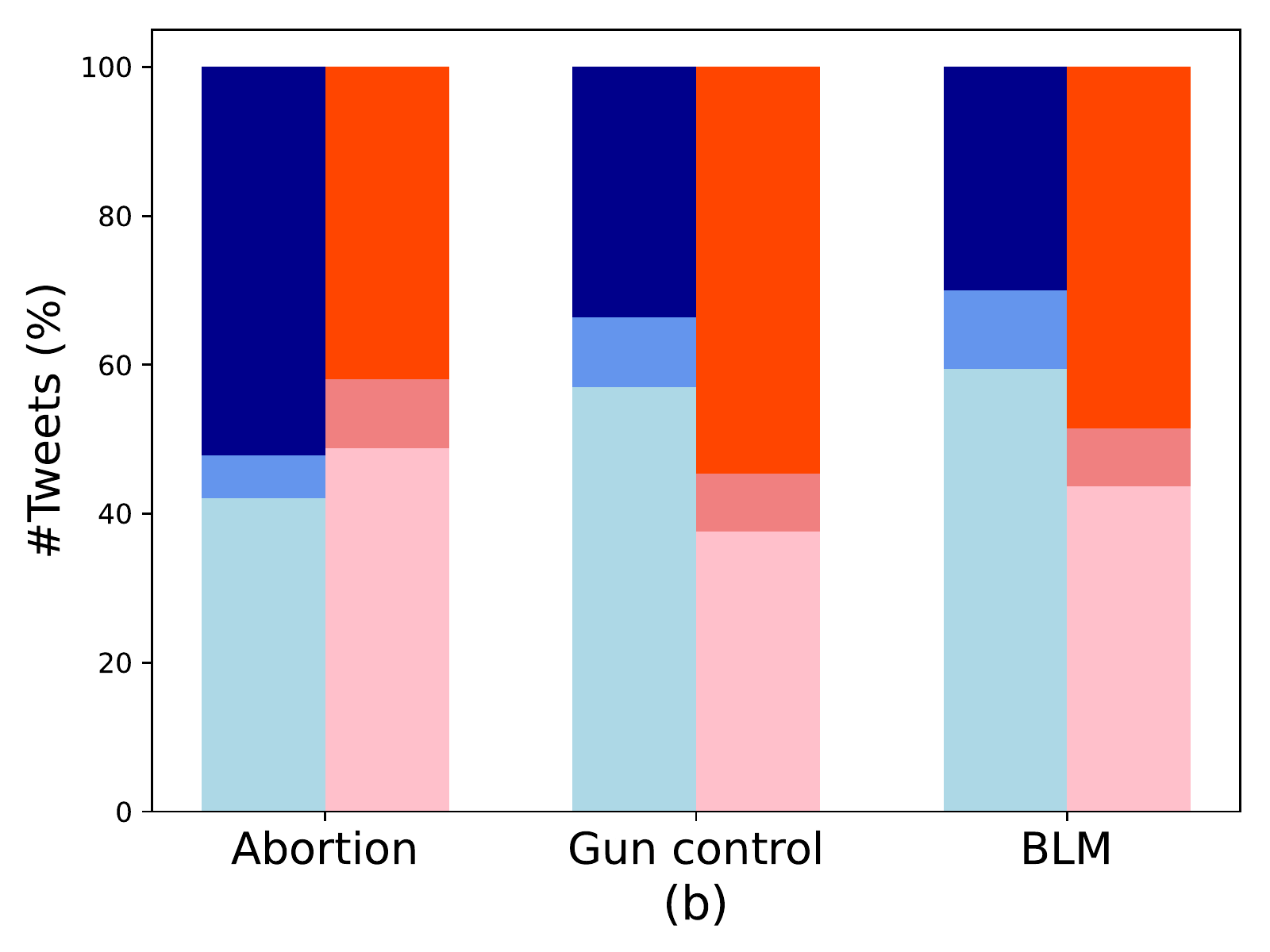}
  \end{subfigure}
   \begin{subfigure}[b]{0.49\linewidth}
    \includegraphics[width=\linewidth]{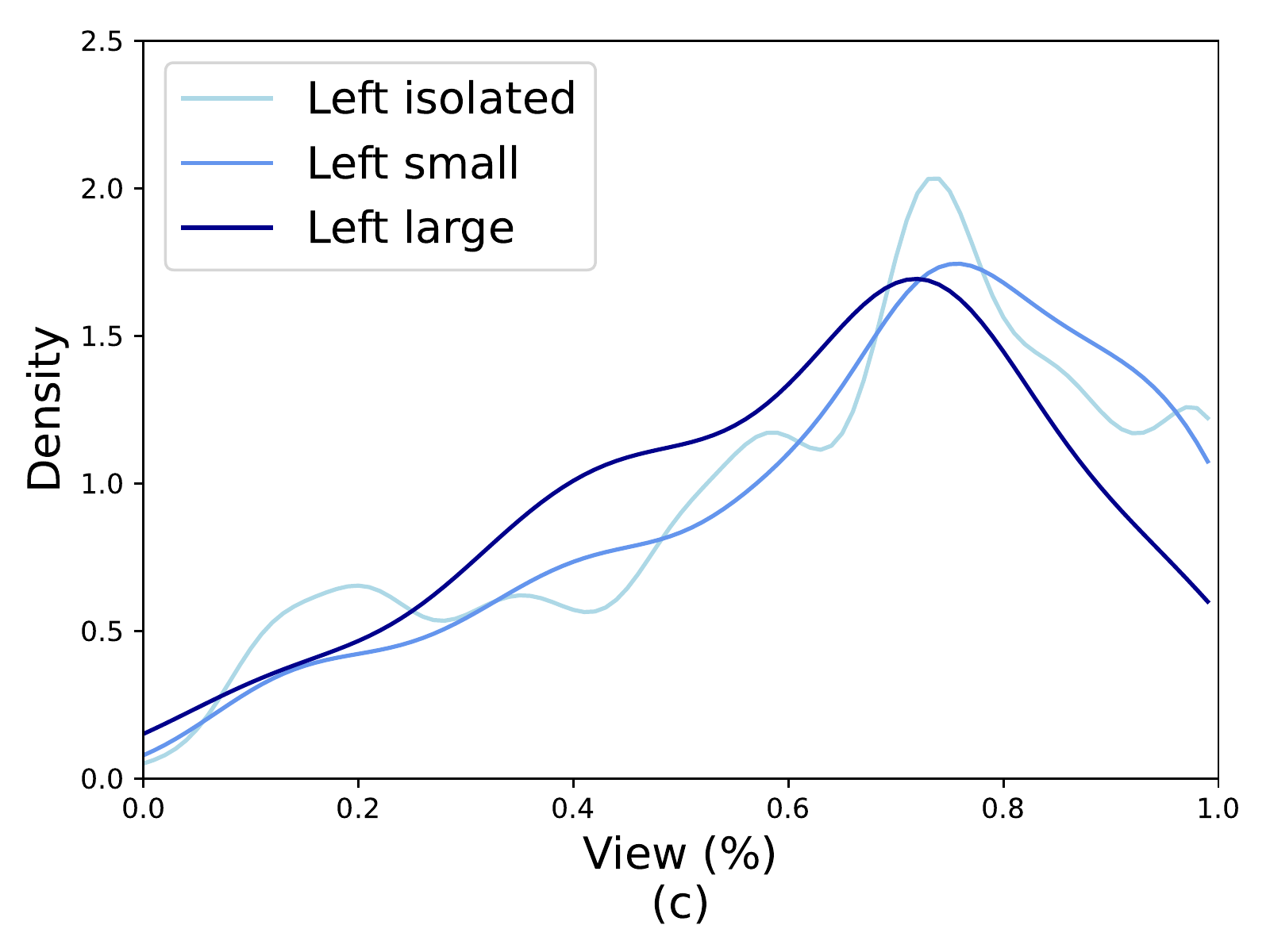}
  \end{subfigure}
  \begin{subfigure}[b]{0.49\linewidth}
    \includegraphics[width=\linewidth]{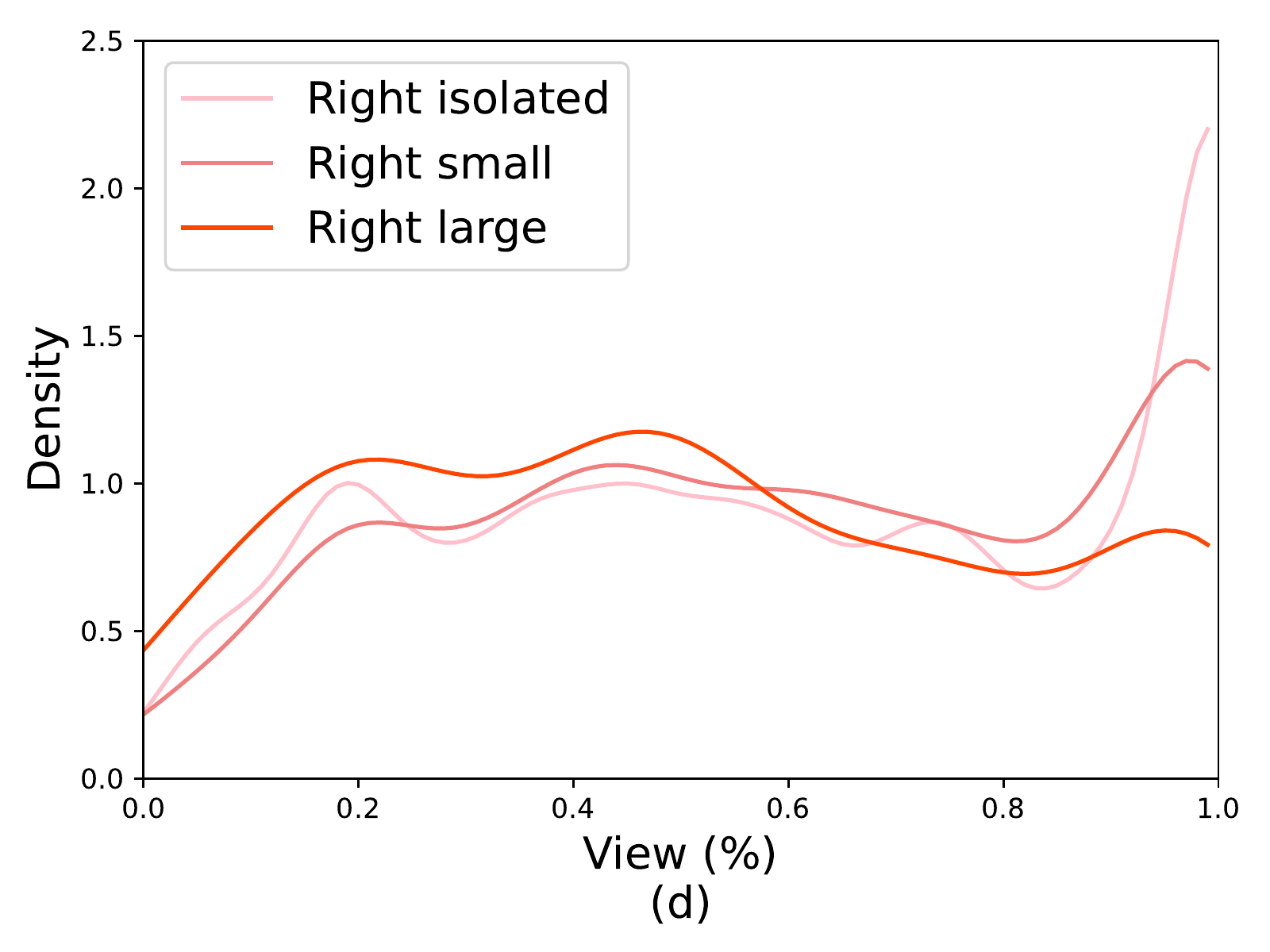}
  \end{subfigure}
  \caption{
  Comparing the left- and right-leaning videos in (a) proportions of cascade sizes -- isolated (1 tweet), small (2-4 tweets) and large (5+ tweets); (b) volume of tweets in each cascade type. (c) and (d) show the density of cascade start time in relation to  accumulated view percentage for \abortion videos. Because the left-leaning videos accumulate views quicker (90\% of left-leaning videos reach viewing half-life within 3 days after upload while 59.4\% of right-leaning videos do), more left-leaning tweet cascades are shown to start after viewing half-life. Section G of~\cite{appendix} includes the same set of plots for \guncontrol and \blm.
  }
  \label{fig:obs_cascades}
\end{figure}

\header{Cascade size.} 
\citet{goel2012structure} found that one-node-cascades (isolated cascades) account for $96\%$ of all cascades in their {\it Twitter Videos} dataset. \Cref{fig:obs_cascades}(a) shows that the proportions of isolated cascades in our datasets are lower, measured at $91.5\%$, $91.2\%$, and $91.9\%$ respectively for \abortion, \guncontrol, and \blm. Notwithstanding a confounder that the dynamics on Twitter has changed significantly since \cite{goel2012structure}, this may still suggest that tweets on controversial topics are less isolated than tweets of videos about any topics. \Cref{fig:obs_cascades}(b) shows the volume of tweets involved in each cascade group. It is interesting to find that most tweets belong to either isolated cascades or large cascades. Aggregated over left and right-leaning videos in \abortion, \guncontrol, and \blm, 47\%, 43.4\%, and 47.3\% of tweets are isolated, whereas 44.7\%, 48.4\%, and 44.3\% are in large cascades of size 5 and above, dominated by a handful of cascades over 1,000 tweets.

\header{Cascade start time.} 
We compare when tweet cascades start in the process of view accumulation, grouped by different cascade sizes. \Cref{fig:obs_cascades}(c) and (d) show the distribution of cascade start times of left- and right-leaning videos in \abortion relative to percentages of views. For right-leaning videos, there is a peak of isolated cascades started at the end of the videos' viewing lifetime. We also observe that for left-leaning videos, the peaks for isolated, small and large cascades are concentrated near $70\%$ of view accumulation while peaks for right-leaning videos are more distributed over different stages of view accumulation. Moreover, in all topics, more right-leaning tweet cascades start before viewing half-life regardless of cascades size. For example, 41\% of right-leaning isolated cascades started before viewing half-life while 25\% of left-leaning isolated cascades started before viewing half-life in \abortion. This is consistent with the observation that views of right-leaning videos unfold much slower (See \Cref{fig:obs_temporal}(a) and \Cref{fig:obs_accumulation}(a)), allowing tweet cascades to start at earlier stages of view accumulation process. The difference in cascade start time is significant between left- and right-leaning videos in all topics for isolated and small cascades ($p < 0.001$, \Cref{table:measures} row 14-15) and is significant for \abortion and \guncontrol for large cascades ($p < 0.001$, \Cref{table:measures} row 16).

% ----------------------------------------------------
\subsection{Networks among Early Adopters on Twitter}
\label{ssec:network_obs}

\begin{figure}[t!]
  \centering
  
  \begin{subfigure}[b]{0.45\linewidth}
      \includegraphics[width=\linewidth]{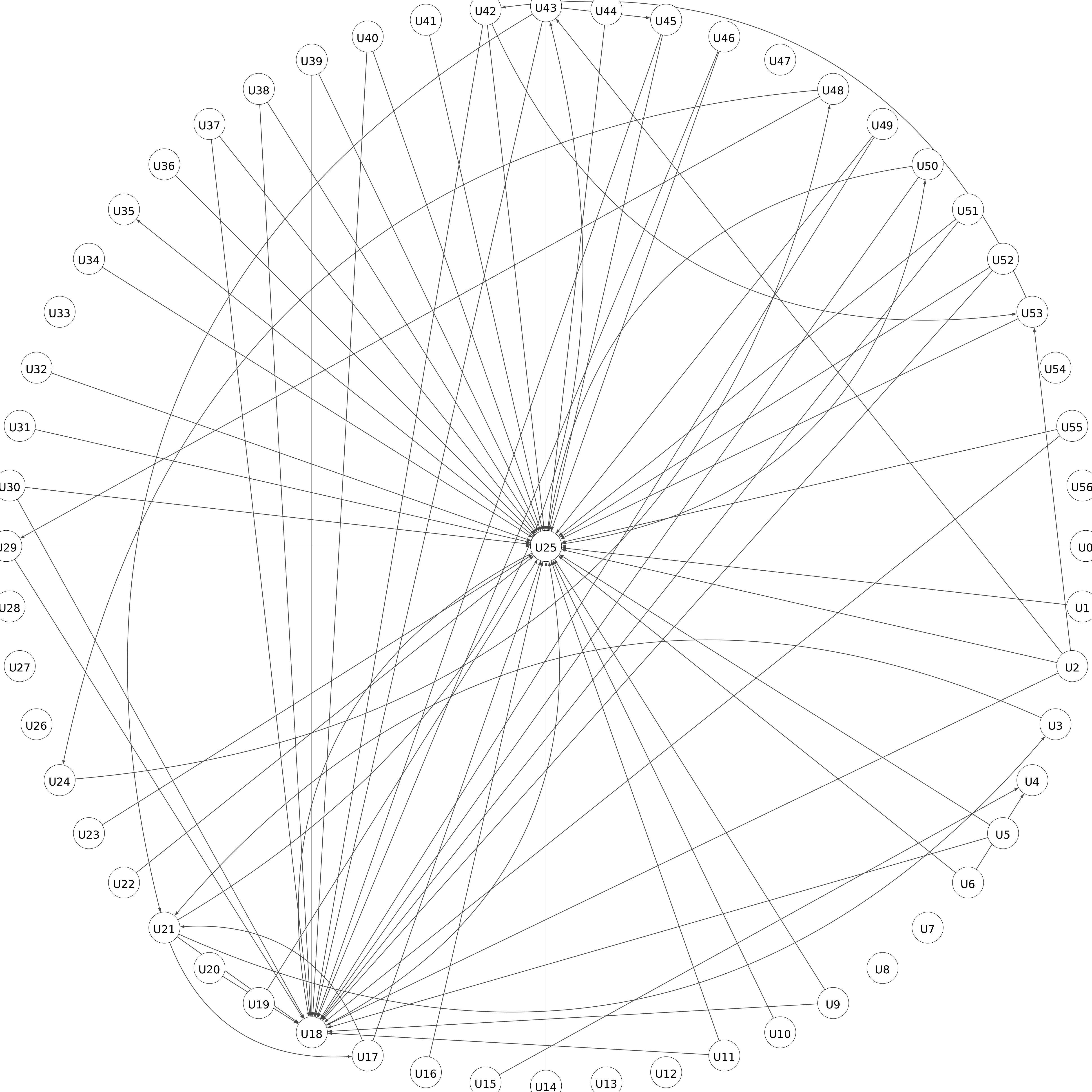}
      \caption{\textit{aeFjqdiVAuY} (Left-leaning)}
  \end{subfigure}
  \hfill
  \begin{subfigure}[b]{0.45\linewidth}
      \includegraphics[width=\linewidth]{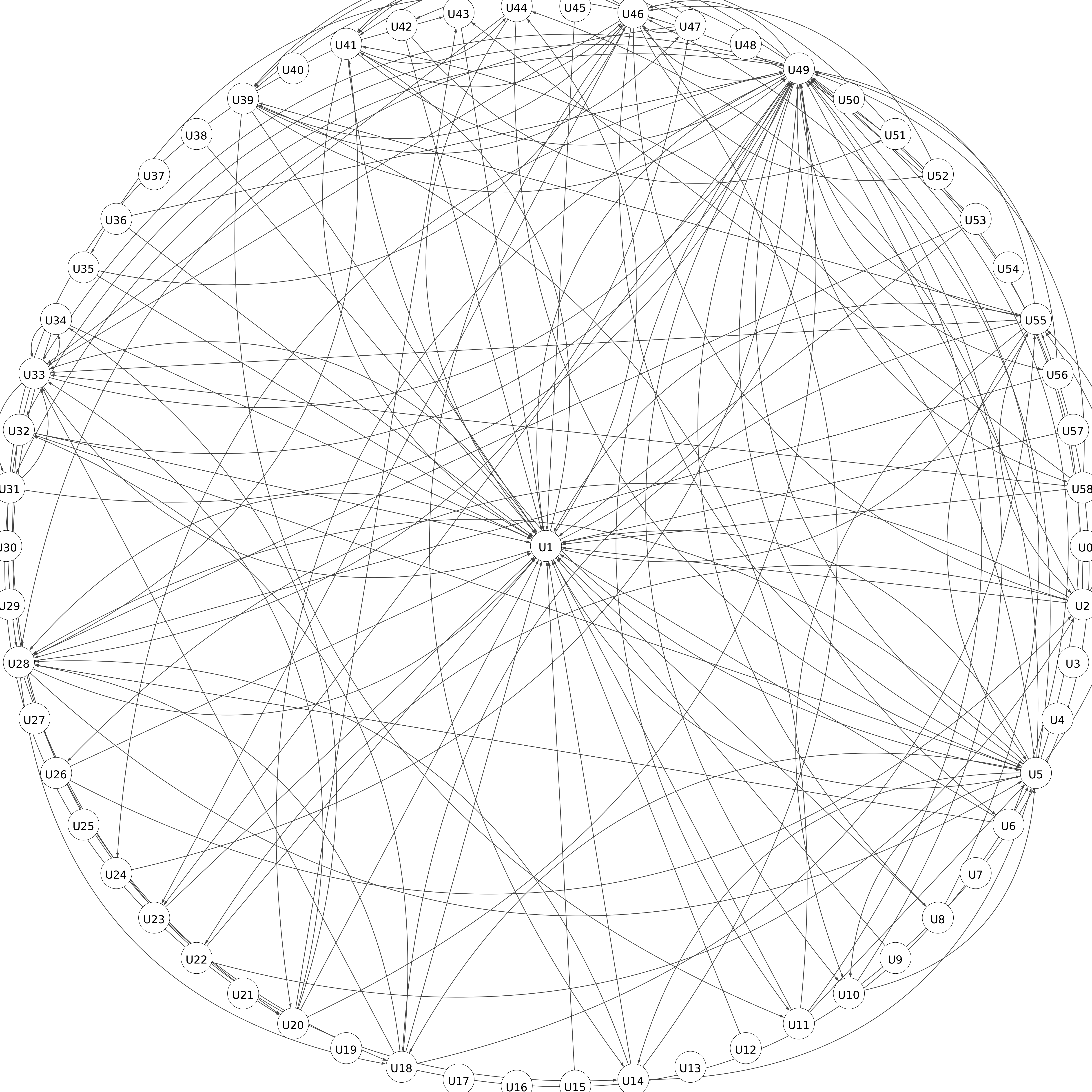}
      \caption{\textit{w5zN1Af7JXQ} (Right-leaning)}
  \end{subfigure}

  \begin{subfigure}[b]{0.48\linewidth}
    \includegraphics[width=\linewidth]{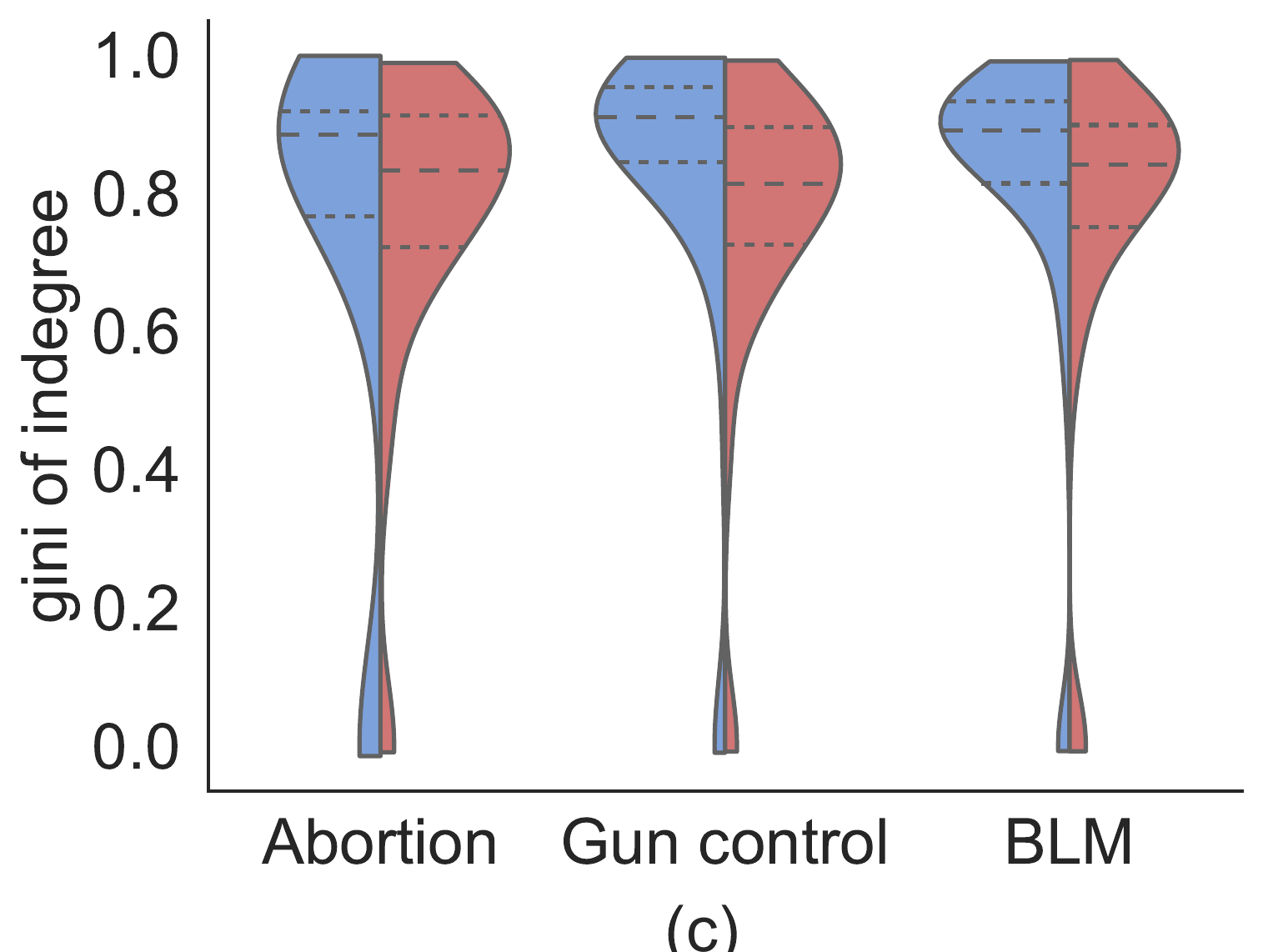}
  \end{subfigure}
  \begin{subfigure}[b]{0.48\linewidth}
      \includegraphics[width=\linewidth]{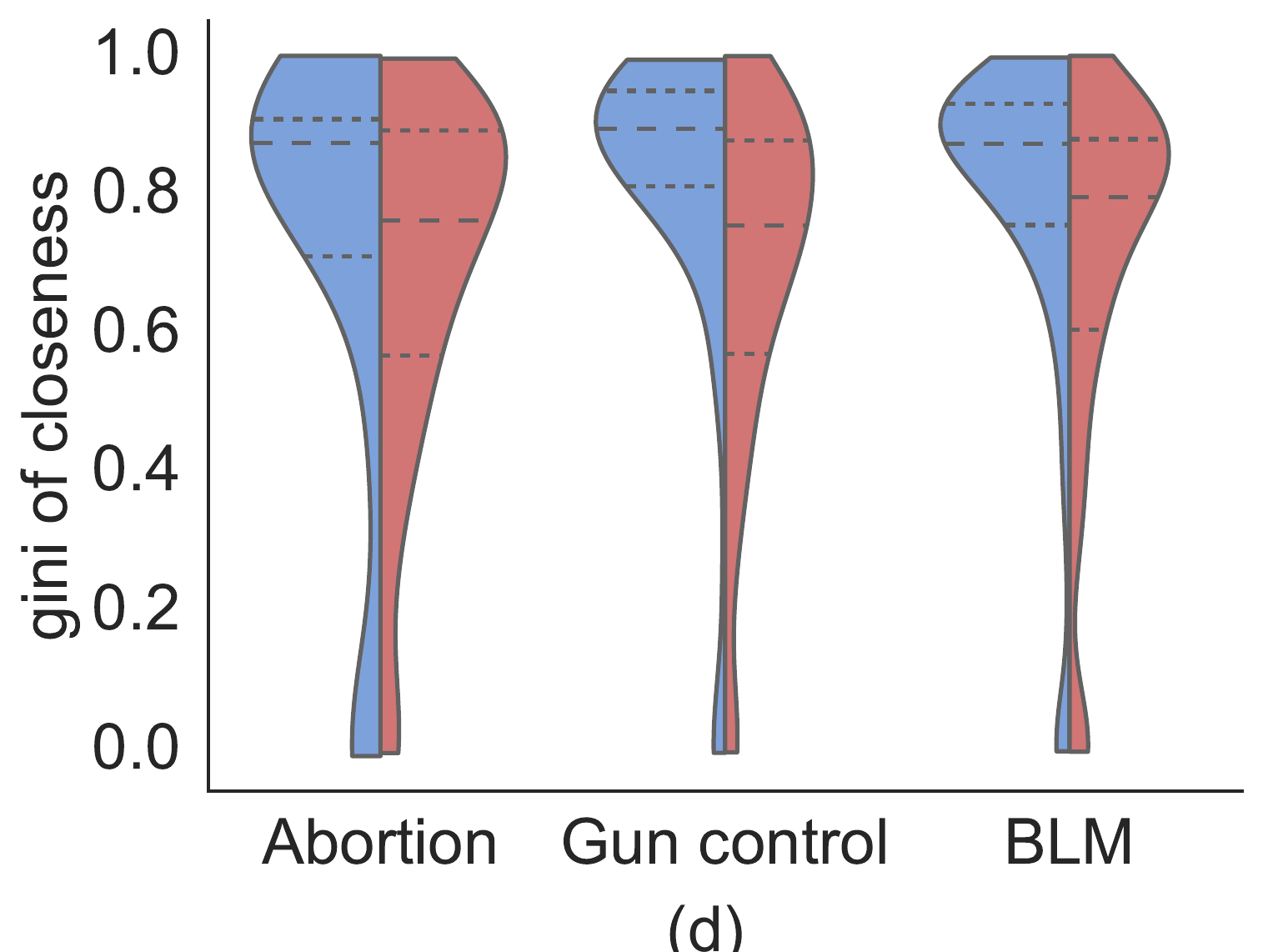}
  \end{subfigure}
  \caption{Follower networks of early adopters for (a) a left-leaning and (b) a right-leaning video with similar network size in \abortion. 
  We compare the left- and right-leaning videos in (c) Gini coefficient of indegree centrality, and (d) Gini coefficient of closeness centrality.
  %Right-leaning videos have significantly greater network size.
  %But the sharing of left-leaning videos relies more on early adopters who are followed by more early adopters, and who have shorter path to other early adopters.
  }  
  \label{fig:obs_network}
\end{figure}

\Cref{fig:obs_network}(c) and (d) show the distributions of the Gini coefficient of indegree centrality and the Gini coefficient of closeness centrality. Gini coef. of indegree centrality of left-leaning videos' networks is significantly larger in \guncontrol and \blm ($p < 0.001$, \Cref{table:measures} row 17). For Gini coefficient of closeness centrality, the MWU test results indicate that left-leaning videos' networks have significantly greater Gini index than those of right-leaning videos' networks across all topics ($p < 0.05$, \Cref{table:measures} row 18). 
 
This suggests that the networks of early adopters for left-leaning videos have more users serving as hubs, i.e., who are followed by more early adopters and have shorter path to other early adopters. This also suggests that in the networks of early adopters for right-leaning videos, users are more equally facilitating dissemination of political information which is consistent with the findings shown in~\cite{conover2012partisan}. As an example of this, we present the follower networks of early adopters of one left-leaning video and one right-leaning video in \abortion in \Cref{fig:obs_network}(a,b). To have a fair comparison we sample two videos having similar network size (57 and 59 for left and right-leaning videos, respectively). The left-leaning video has Gini coef. of indegree centrality: 0.918, Gini coef. of closeness centrality: 0.90. The right-leaning video has Gini coef. of indegree centrality: 0.748, Gini coef. of closeness centrality: 0.536. It can be observed that the sharing of this left-leaning video relies more on central users who are followed by more early adopters and have shorter path to others. On the other hand, in the follower network of the early adopters of this right-leaning video, indegree and closeness centrality distributions are more equal.

Apart from the reported metrics, we have also performed preliminary examination on the correlation and trends between two and more metrics. An example on linking relative engagement to the view and tweet counts is presented in Section I of~\cite{appendix}. We have not seen consistent and salient patterns that are not already captured by individual measures.

\section{Conclusion and Discussion}
\label{sec:conclusion}

This work presents a quantitative study that links collective attention towards online videos across YouTube and Twitter over three political topics: \abortion, \guncontrol, and \blm. For each topic, we curated a cross-platform datatset that contained hundreds of videos and hundreds of thousands of tweets spanning 16 months. The extracted videos all have a non-trivial amount of views and tweets. The key contributions include several sets of video-centric metrics for comparing attention consumption patterns between left-leaning and right-leaning videos across two platforms. We find that left-leaning videos are more viewed and more engaging, while right-leaning videos are more tweeted and have longer attention spans. We also found that the follower networks of early adopters on left-leaning videos are of higher centrality, whereas tweet cascades for right-leaning videos start earlier in the attention lifecycle. This study enriches the current understanding of ideological asymmetries by adding a set of temporal and cross-platform analyses.

\header{Limitations.} 
Extensive discussions about social data biases are presented in~\cite{olteanu2019social}. The biases can be introduced due to the choice of social platforms, data (un)availability, sampling methods, etc. Here we discuss three limitations in our data collection process.

A recent study found that Twitter filtered streaming API subsamples high-volume data streams that consist of more than 1\% of all tweets~\cite{wu2020variation}. The authors proposed a method of using Twitter rate limit messages to quantify the data loss. Based on this method, we find our 16-month Twitter stream has a sampling rate of 79.4 \% -- we collected 1,802,230,572 out of 2,270,223,254 estimated total tweets. Under a Bernoulli process assumption, the chance of collecting a video tweeted more than once in our tweet stream is 95.8\%. Since most missing videos are tweeted sporadically, the sampling loss from Twitter APIs is small, which minimally affects the measures on tweeting activities, including attention volumes, timing, and cascade sizes. Confidence intervals for simple measures such as volume can be derived~\cite{wu2020variation}.

In this paper, we present various measurements focused on YouTube videos, which are the main entities that link the two platforms. YouTube viewers are unknowable (via publicly available data) and Twitter users are hard to track consistently over time. Therefore, we track videos which attract views and tweets. All the presented metrics are video-centric and we do not assume that the viewers or tweeters of the videos represent specific groups of users. We believe that each set of videos (\abortion, \guncontrol, and \blm) represent YouTube videos that are relevant to the topic, curated by keyword queries and semi-manual coding. The number of videos belonging to each topic is not large but we attempted to include all relevant videos shared on Twitter which have non-trivial activities. Thus our results intend to explain attention gathering behavior of topic-relevant videos. Nevertheless, it is unclear that our observations about ideological asymmetries can generalize to videos with less attention and/or videos about other topics. We leave this generalization validation as future work.

One data integrity limitation is the time gap between tweet collection in 2017--2018 and early adopters' follower networks collected in early 2020.
Our data collection is limited by the Twitter search API quota, which restrains collecting tweets and Twitter user followers simultaneously. The tweeted videos stream has on average 3.7M tweets in each day. Collecting the follower network for all these tweets far exceeds the capacity of Twitter API, focusing on the early-adopter network is a practical trade-off between still having informative results and making data collection feasible. A related issue comes from unavailable YouTube videos and Twitter users since content publicly available in 2017 may be deleted, banned, or protected in 2020. We found that between 17\% to 19\% candidate videos become unavailable in our dataset.

\header{Practical implications and future work.}
We believe this work adds a new dimension to the understanding of online political behavior and discourse -- cross platform links. Further examination in this direction could bear theoretical and empirical fruits. The measurements presented in this work are mostly quantitative. One direction of future work is to complement qualitative analysis. For example, to gain deeper insight into our observations about how the user attention to left-leaning YouTube videos was driven by a group of elite early adopters, one can examine typical diffusion networks from both left- and right-leaning groups and study the diffusion process of the video spreading. One could also examine the framing of left- and right-leaning content in both video descriptions and tweets about them. For example, \citet{linDynamicsTwitterUsers2020} used mixed-methods approaches to identify the primary framing and rhetorics in online conversations related to gun control, which can be expanded to enrich the quantitative analyses, such as investigating the linguistic features of YouTube descriptions and tweet cascades, and their relationships to the changes in collective attitudes. Finally, understanding the collective attention across multiple social platforms is important for content producers, who could devise better strategies in promoting their content in another domain.

\section*{Ethical Statement}
All data that we obtained was publicly available at the time of data collection. We discarded deleted, protected, and private content at the time of analysis. In our released dataset, we anonymized user identities. Therefore, the analyses reported in this work do not compromise any user privacy.

\begin{table*}
    \centering
    \resizebox{1\textwidth}{!}{
    \begin{tabular}{rrr>{\raggedright\arraybackslash}p{5.6cm}lll}
        \toprule
        row & crossref & metric & definition & \multicolumn{3}{c}{significance} \\
        & & & & \abortion & \guncontrol & \blm \\
        \midrule
        \multicolumn{7}{c}{\textbf{YouTube and Twitter aggregate attention metrics (\Cref{ssec:aggregate_obs})}}\\
        1 & \cref{fig:obs_aggregate}a & view\_{\em x} & Total number of views at day {\em x} & \textbf{L} $>$ R** & \textbf{L} $>$ R*** & $\times$ \\
        2 & \cref{fig:obs_aggregate}b & relative\_engagement & Rank percentile of watch percentage among videos of similar lengths & \textbf{L} $>$ R* & \textbf{L} $>$ R*** & \textbf{L} $>$ R** \\
        3 & \cref{fig:obs_aggregate}c & fraction of likes & Number of likes divided by total number of reactions  & L $<$ \textbf{R}*** & L $<$ \textbf{R}*** & L $<$ \textbf{R}*** \\
        4 & \cref{fig:obs_aggregate}d & viral\_potential & Number of views potentially excited by one tweet & \textbf{L} $>$ R*** & \textbf{L} $>$ R** & \textbf{L} $>$ R* \\
        5 & \cref{fig:obs_aggregate}e & tweet\_{\em x} & Total number of tweets at day {\em x} & L $<$ \textbf{R}*** & $\times$ & L $<$ \textbf{R}*** \\
        6 & \cref{fig:obs_aggregate}f & retweet\_{\em x} & Total number of retweets at day {\em x} & L $<$ \textbf{R}*** & $\times$ & L $<$ \textbf{R}*** \\
        7 & -- & reply\_{\em x} & Total number of replies at day {\em x} & L $<$ \textbf{R}*** & $\times$ & L $<$ \textbf{R}*** \\
        8 & -- & original\_tweet\_{\em x} & Total number of original tweets at day {\em x} & $\times$ & $\times$ & L $<$ \textbf{R}*** \\
        9 & -- & quote\_{\em x} & Total number of quotes at day {\em x} & $\times$ & $\times$ & L $<$ \textbf{R}*** \\
        \midrule
        \multicolumn{7}{c}{\textbf{Temporal metrics of views and tweets (\Cref{ssec:temporal_obs})}}\\
        10 & \cref{fig:obs_temporal}a & viewing half-life & Number of days to reach 50\% views & L $<$ \textbf{R}*** & L $<$ \textbf{R}** & L $<$ \textbf{R}*** \\
        11 & \cref{fig:obs_temporal}b & tweeting half-life & Number of days to reach 50\% tweets & L $<$ \textbf{R}** & L $<$ \textbf{R}* & $\times$ \\
        12 & \cref{fig:obs_temporal}c & tweeting lifetime & Time gap between the first and the last tweets in minutes & L $<$ \textbf{R}*** & L $<$ \textbf{R}* & $\times$ \\
        13 & \cref{fig:obs_temporal}d & tweeting inter-arrival time & Average time gap between every two consecutive tweets in minutes & L $<$ \textbf{R}** & L $<$ \textbf{R}* & $\times$ \\
        \midrule
        \multicolumn{7}{c}{\textbf{Tweet cascades measures (\Cref{ssec:cascade_obs})}}\\
        &  & \underline{cascade start time} & Percentage of accumulated views of the video when the root of the cascade is tweeted & & & \\
        14 & \cref{fig:obs_cascades}c,d & (isolated cascades) & & \textbf{L} $>$ R*** & \textbf{L} $>$ R*** & \textbf{L} $>$ R***\\
        15 & \cref{fig:obs_cascades}c,d & (small cascades) & & \textbf{L} $>$ R*** & \textbf{L} $>$ R*** & \textbf{L} $>$ R*** \\
        16 & \cref{fig:obs_cascades}c,d & (large cascades) & & \textbf{L} $>$ R*** & \textbf{L} $>$ R*** & $\times$ \\
        \midrule
        \multicolumn{7}{c}{\textbf{Network metrics of early adopters (\Cref{ssec:network_obs})}}\\
        17 & \cref{fig:obs_network}c & Gini\_indegree & Gini coefficient of indegree centrality & $\times$ & \textbf{L} $>$ R*** & \textbf{L} $>$ R*** \\
        18 & \cref{fig:obs_network}d & Gini\_closeness & Gini coefficient of closeness centrality & \textbf{L} $>$ R* & \textbf{L} $>$ R*** & \textbf{L} $>$ R*** \\
        19 & -- & Gini\_betweenness & Gini coefficient of betweenness centrality & $\times$ & \textbf{L} $>$ R* & $\times$ \\
        20 & -- & network\_density & Density of early adopters' follower network & $\times$ & $\times$ & \textbf{L} $>$ R*** \\
        21 & -- & max\_indegree & Max indegree value in early adopters' follower network & L $<$ \textbf{R}** & $\times$ & $\times$ \\
        22 & -- & global\_efficiency & Average efficiency over all pairs of distinct early adopters & L $<$ \textbf{R}* & L $<$ \textbf{R}** & $\times$ \\
        \bottomrule
    \end{tabular}
    }
    \caption{
    Summary of all metrics and comparison across political leanings. \textbf{L} $>$ R* means that the metric of a randomly selected left-leaning video is significantly larger than that of a randomly selected right-leaning video. Significance is measured by one-sided Mann–Whitney U test. *$p < 0.05$, **$p < 0.01$, ***$p < 0.001$. The significantly larger leaning is boldfaced.  ``$\times$'' sign indicates non-significant relation. Sample size: \abortion (L: 58; R: 111), \guncontrol (L: 81; R: 154), \blm (L: 297; R: 396). See the corresponding cross-referenced figures and discussions in \Cref{sec:measure} for more details, ``--'' sign means the metric is not presented in a figure.
    }
    \label{table:measures}
\end{table*}

\section*{Acknowledgments}
We thank Xian Teng and Muheng Yan for their help in data annotation. We would like to acknowledge the support from NSF \#2027713, ARC DP180101985, and AFOSR projects 19IOA078, 19RT0797, 20IOA064, and 20IOA065. Any opinions, findings, and conclusions or recommendations expressed in this material do not necessarily reflect the views of the funding sources.

% \small
\bibliography{xplatform-main-ref}

% ====================================================
% comment everything in this box for camera-ready
\clearpage
\appendix
\onecolumn
\normalsize

Supplementary materials accompanying the paper \textit{Whose Advantage? Measuring Attention Dynamics across YouTube and Twitter on Controversial Topics}.

% ====================================================
\section{Identifying Topical Videos and Tweets}
\label{sec:app-data}

We curated a keyword list for each of the three controversial topics -- \abortion, \guncontrol, and \blm. Keywords for \abortion and \guncontrol came from another recent work \cite{guo2020inflating} and were expanded using contextual expansion with sample tweets. Keywords for \blm were manually curated, separating into pro- and anti-\blm keywords respectively. We expanded the latter with white-supremacy related keywords. \Cref{table:app-keyword} lists all keywords.

We consider a video is potentially relevant if (a) it contains at least one keyword in the video title or description; or (b) it is mentioned in a tweet that contains the keywords in the tweet text. We obtained 815 candidate videos for \abortion, 974 candidate videos for \guncontrol, and 8,571 candidate videos for \blm.

For \abortion and \guncontrol, we manually annotated all candidate videos. All videos were randomly divided into five buckets (with 163 or 195 videos in each) and each bucket was labeled by an annotator. The five authors of this paper, along with two postgraduate students who had extensive research experience in computational social science, consisted of the annotator team. Annotators were instructed to watch a video for at least five minutes before making a decision of whether or not this video is relevant to the target topic. To measure the inter-rater reliability, we randomly selected a bucket, and had two more annotators labeling the same bucket. We computed the Fleiss' Kappa. \abortion had Fleiss' Kappa of 0.830 and \guncontrol had that of 0.826, suggesting a high agreement among annotators.

For \blm, we used a semi-automatic approach because the number of candidate videos (8,571) is much larger. Based on our observations on \abortion and \guncontrol, we designed the following protocol:

\begin{itemize}[leftmargin=*]
  \item For videos that contained any topic-relevant keywords in the video titles or descriptions,

  \begin{itemize}[leftmargin=*]
    \item if the videos were also mentioned by tweets with topic-relevant keywords, 89.0\%/86.7\% of the videos were labeled \textit{relevant} (in \abortion and \guncontrol respectively, same below). These videos consisted of 16.1\%/12.5\% of all candidate videos. We labeled such videos as relevant, there were 418 in \blm.
    \item if the videos were not mentioned by tweets with topic-relevant keywords, 77.8\%/89.4\% of the videos were labeled \textit{irrelevant}. These videos consisted of 6.7\%/11.9\% of all candidate videos. We labeled such videos as irrelevant.
  \end{itemize}

  \item For videos that did not contain any topic-relevant keywords in the titles nor descriptions, but had been mentioned by some topic-relevant tweets,

  \begin{itemize}[leftmargin=*]
    \item if the videos were tagged as ``Music'', or did not have English captions, or did not contain any topic-relevant keywords in the captions, 95.2\%/80.6\% of the videos were labeled \textit{irrelevant}. We labeled such videos as irrelevant.
    \item we used the manual annotation steps for \abortion and \guncontrol to label the remaining 749 videos, out of which 359 videos were labeled relevant. The Fleiss' Kappa was 0.711.
  \end{itemize}
\end{itemize}

In total, we identified 179 \abortion videos, 268 \guncontrol videos, and 777 \blm videos. We extracted all tweets and Twitter users that mentioned any of these topic-relevant videos from our 16 months data stream.

\begin{table*}[!htbp]
    \centering
    \small
    \begin{tabular}{rp{12cm}}
        \toprule
        Topic & Keywords \\
        \midrule
        \abortion & abortion, pro choice, pro life, ru486, planned parenthood, unborn baby, unborn babies, unborn child, roe v wade, roe versus wade, reproductive health care, birth control \\
        \guncontrol & gun violence, gun reform, gun safety, gun regulation, gun death, gun law, gun right, gun grab, gun owner, gun control, gun free, gun sense, guns kill people, pro gun, firearm, no bill no break, disarm hate, second amendment, moms demand, wear orange, molon labe \\
        \blm & blm, black lives matter, all lives matter, blue lives matter, white lives matter, michael brown, mike brown, eric garner, freddie gray, walter scott, tamir rice, john crafford, tony robinson, eric harris, ezell ford, akai gurley, kajimeme powell, tanisha anderson, victor white, jordan baker, jerame reid, yvette smith, philip white, dante parker, mckenzie cochran, tyree woodson, jocques clemmons, fire bannon, charlottesville, kkk, ku klux klan, klansman, klansmen, domestic terrorism, white supremacist \\
        \bottomrule
    \end{tabular}
    \caption{Keyword list for the three controversial topics.}
    \label{table:app-keyword}
\end{table*}

% ====================================================
\clearpage

\section{Estimating and Validating the Leanings of Twitter Users}
\label{sec:app-estimate-twitter}

We estimated the political leanings of early adopters on Twitter in four steps with increasing coverage.

\header{1) Curating seed hashtags.}
Twitter users are known to use political hashtags in their user profiles as a marker of identity and a tool of network gatekeeping, such as marking participation in an ongoing activism, facilitating the establishment of social network ties around specific topics, or expressing community identities~\cite{kulshrestha2017quantifying}. Based on this insight, we started by identifying users who clearly signaled political ideologies in their Twitter profile descriptions. We curated two sets of ideologically descriptive hashtags that were typically used by \rev{liberal} and \rev{conservative} users. We described a hashtag as ideologically descriptive if it either supported or criticized political leaders, their statements, and their campaigns (e.g., \textit{\#trump}, \textit{\#america1st}, \textit{\#maga}). We called these hashtags as political seed hashtags. The whole list can be found in \Cref{table:seed_hashtag}. There are 25 left-leaning hashtags used by 260, 552, 817 users, and 35 right-leaning hashtags used by 1023, 2065, 3798 users in \abortion, \guncontrol, and \blm, respectively.

\begin{table*}[!htbp]
    \centering
    \small
    \begin{tabular}{rp{11cm}}
        \toprule
        Leaning & Hashtags \\
        \midrule
        Left-leaning & theresistance, resist, resistance, fbr, impeachtrump, impeachtrumpnow, stillwithher, bluewave, uniteblue, bluewave2018, resisttrump, nevertrump, fucktrump, imwithher, demforce, voteblue, followbackresistance, bernie2020, resistanceunited, impeach45, imstillwithher, democrat, democrats, alwayswithher, flipitblue. \\
        Right-leaning & maga, trump, americafirst, draintheswamp, trumptrain, trump2020, kag, makeamericagreatagain, lockherup, presidenttrump, cruzcrew, trumppence2020, trumparmy, womenfortrump, istandwithtrump, keepamericagreat, neverhillary, trump2016, trumpworld, republican, republicans, hillaryforprison, kag2020, gop, trumpsupporter, trumpville, trumppence16, alwaystrump, donaldtrump, america1st, trumpismypresident, trump45, women4trump, teamtrump. \\
        \bottomrule
    \end{tabular}
    \caption{Political seed hashtags.}
    \label{table:seed_hashtag}
\end{table*}

\header{2) Expanding seed hashtags.}
To obtain new political hashtags, we followed an expansion protocol based on co-occurrence with the political seed hashtags in the user profile descriptions. Firstly, we required that a candidate hashtag must co-occur with the seed hashtags in at least 0.1\% of all users' profile descriptions. Secondly, the candidate hashtag should be relevant to only one ideology (either left or right). To achieve this, we computed the Shannon entropy based on the co-occurrence of the candidate hashtag with the seed left- and right-leaning hashtags. We selected the hashtags with entropy less than or equal to 0.1. \Cref{table:expanded_hashtag} presents the expanded political hashtags. The number of users using left- and right-leaning hashtags are 264 and 1250 for \abortion, 677 and 2375 for \guncontrol, 970 and 4412 for \blm, respectively. We noticed that the number of \rev{conservative} users are about four times that of \rev{liberal} users across the three topics.

\begin{table*}[!htbp]
    \centering
    \small
    \begin{tabular}{@{}lll@{}}
    \toprule
    & \multicolumn{1}{l}{Left-leaning hashtags} & \multicolumn{1}{l}{Right-leaning hashtags} \\ 
    \midrule
    \abortion & notmypresident & \begin{tabular}[c]{@{}l@{}}2a, nra, conservative, 1a, patriot, prolife, \\ buildthewall, tcot, ccot, military, bluelivesmatter, \\ usa, qanon, deplorable\end{tabular} \\
    \guncontrol & blacklivesmatter, blm, notmypresident & \begin{tabular}[c]{@{}l@{}}2a, nra, 1a, conservative, patriot, prolife, christian, \\ buildthewall, qanon, deplorable, backtheblue, vets, \\ bluelivesmatter, usa\end{tabular} \\
    \blm & \begin{tabular}[c]{@{}l@{}}blacklivesmatter, blm, notmypresident\\ progressive, metoo\end{tabular} & \begin{tabular}[c]{@{}l@{}}2a, nra, 1a, conservative, prolife, buildthewall, \\ christian, bluelivesmatter, tcot, ccot, qanon, \\ deplorable, israel, constitution, usa, backtheblue, \\ vets, covfefe, banislam, codeofvets, bansharia\end{tabular} \\ \bottomrule
    \end{tabular}
    \caption{Expanded political hashtags from political seed hashtags.}
    \label{table:expanded_hashtag}
\end{table*}

\header{3) Identifying seed users.}
After curating and expanding the left- and right-leaning hashtags (jointly called {\em political hashtags}), we assigned the political ideologies to the Twitter users. For users whose profile descriptions included at least one political hashtag, we counted the numbers of left- and right-leaning hashtags in their profile descriptions. The average numbers of political hashtags per such user were 1.96, 1.97, and 2.04 for \abortion, \guncontrol and \blm, respectively. If the ratio of one leaning was greater than or equal to 0.9, we assigned the corresponding ideology to the user. We called the users {\em seed users}, since their labels were later used to estimate the ideologies of other users. The numbers of liberal and conservative seed users are given in~\Cref{table:seed_users}. 

\begin{table*}[!t]
    \centering
    \small
    \begin{tabular}{@{}lccclccc@{}}
    \toprule
     & \multicolumn{3}{c}{\#Seed users} &  & \multicolumn{3}{c}{\rev{Ideology} propagation (10-fold CV performance)} \\ 
     %\midrule
     \cmidrule{2-4} \cmidrule{6-8}
     & \rev{Liberal} & \rev{Conservative} & Total &  & Precision (Lib-Con) & Recall (Lib-Con) & F-score (Lib-Con) \\
    \abortion & 258 (2.07\%) & 1244 (9.99\%) & 1502 (12.06\%) &  & 0.99 - 0.99 & 0.96 - 1.00 & 0.98 - 1.00 \\
    \guncontrol & 664 (2.38\%) & 2362 (8.45\%) & 3026 (10.83\%) &  & 0.98 - 0.99 & 0.97 - 0.99 & 0.98 - 0.99 \\
    \blm & 950 (1.90\%) & 4392 (8.76\%) & 5342 (10.66\%) &  & 0.98 - 0.99 & 0.96 - 1.00 & 0.97 - 0.99 \\ \bottomrule
    \end{tabular}
    \caption{Political seed users and \rev{ideology} propagation results. See \Cref{sec:app-estimate-twitter} for methods.}
    \label{table:seed_users}
\end{table*}

\header{4) Propagating labels to other users.}
We aimed to infer the ideologies of users who had no explicit political hashtags in their profile descriptions. We first created a user-to-user network where the edge weight between two users $(u, v)$ represented their {\em shared audience}, calculated using the Jaccard similarity of their follower lists:

\begin{equation}
  \label{eq:jaccard}
  J(u, v) = \dfrac{\text{followers}(u) \cap \text{followers}(v) }{\text{followers}(u) \cup \text{followers}(v)}
\end{equation}

The shared audience network is shown to be an effective method for splitting users with different political behaviors~\cite{stewart2017drawing}. When estimating the user ideologies, our assumption is that users with the same ideology share similar audience. The shared audience network was a dense network since we created edges between any two users regardless of whether they had a direct follower-following relation. The initial networks consisted of approximately $28M$, $119M$, and $394M$ edges for \abortion, \guncontrol, and \blm, respectively. The corresponding network densities were 18.1\%, 15.2\%, and 15.7\%. To reduce the network complexity, we used a disparity filtering algorithm~\cite{serrano2009extracting} to extract the network backbone. The disparity filtering algorithm can identify important edges in a network. It has a hyperparameter $\alpha$ to control the degree of edge importance. We used $\alpha = 0.05$ as suggested in the initial work~\cite{serrano2009extracting}. After disparity filtering, the resulting networks contained approximately $1.9M$, $7.8M$, and $25.5M$ edges for \abortion, \guncontrol, and \blm, respectively. The densities of these networks ranged between 1\% and 1.2\%.

Next, we applied a label propagation algorithm~\cite{zhou2004learning} on the filtered network to infer the political leanings of the unknown users from those of the seed users. Label propagation is a semi-supervised algorithm that builds on two intuitions: (a) the seed labels should not change too much after propagation; (b) the inferred labels should not be too different with their neighbours. A teleport factor $\beta$ controls the trade-off between these two intuitions. We used $\beta = 0.85$ as suggested in the prior work~\cite{yan2020mimicprop}. The label propagation algorithm returned a liberal score and a conservative score for each user in the network, which we then re-normalized into a score between 0 and 1, where 0 indicated strong liberal and 1 indicated strong conservative. Re-normalization was done by computing the ratio of conservative score to the summation of liberal and conservative scores.

We stratified one-tenth of seed users in each fold and hided their labels. Next, we compared the inferred political leanings to actual leanings of these users. Note that we employed whole network during the label propagation validation. \Cref{table:seed_users} shows the evaluation results for the label propagation.

% ====================================================
\clearpage

\section{Estimating and Validating the Leanings of YouTube Videos}
\label{sec:app-estimate-youtube}

Once we obtained the ideology probabilities of early adopters on Twitter, we computed the leaning score for each YouTube video by averaging its promoters' ideology scores. To assign discrete leaning labels (i.e. \textit{Left}, \textit{Right} and \textit{Center}) to the videos, we leveraged the Recfluence dataset~\cite{ledwich2020algorithmic} as an external source to annotate YouTube videos' leanings. This dataset includes 816 YouTube channels where each channel has 10k+ subscribers and more than 30\% of its content is relevant to US politics, or cultural news, or cultural commentary. It consists of channels of YouTubers in addition to those of traditional news outlets. In this dataset, each channel is assigned to \textit{Left (L)}, \textit{Center (C)} or \textit{Right (R)} leaning. The annotations were conducted based on two well known media bias resources\footnote{Media Bias/Fact Check: \url{https://mediabiasfactcheck.com/}; Ad Fontes Media: \url{https://www.adfontesmedia.com/}} as well as manual labeling.

The videos whose channels' leanings are available in the Recfluence dataset are assigned to the same leaning. We were able to label 109 videos (L: 38, R: 66, C:5) from 38 channels for \abortion, 153 videos (L: 52, R: 88, C: 13) from 62 channels for \guncontrol, and 521 videos (L: 194, R: 285, C:42) from 120 channels for \blm. To label the videos whose channels are not available in the Recfluence dataset, we identified two thresholds for each topic -- $thr_{(L, C)}$ and $thr_{(C, R)}$ where $thr_{(L, C)} < thr_{(C, R)}$. For a given video $\nu$, we assigned a discrete leaning label, $leaning\_label(\nu)$, based on its leaning score, $leaning\_score(\nu)$, as follows:

\[
leaning\_label(\nu) = 
\begin{cases}
  L & \text{if $leaning\_score(\nu) < thr_{(L, C)}$} \\
  R & \text{if $leaning\_score(\nu) > thr_{(C, R)}$} \\
  C & \text{otherwise.}
\end{cases}
\]

\begin{figure*}[!htb]
  \centering
  \begin{subfigure}[b]{0.32\linewidth}
    \includegraphics[width=\linewidth]{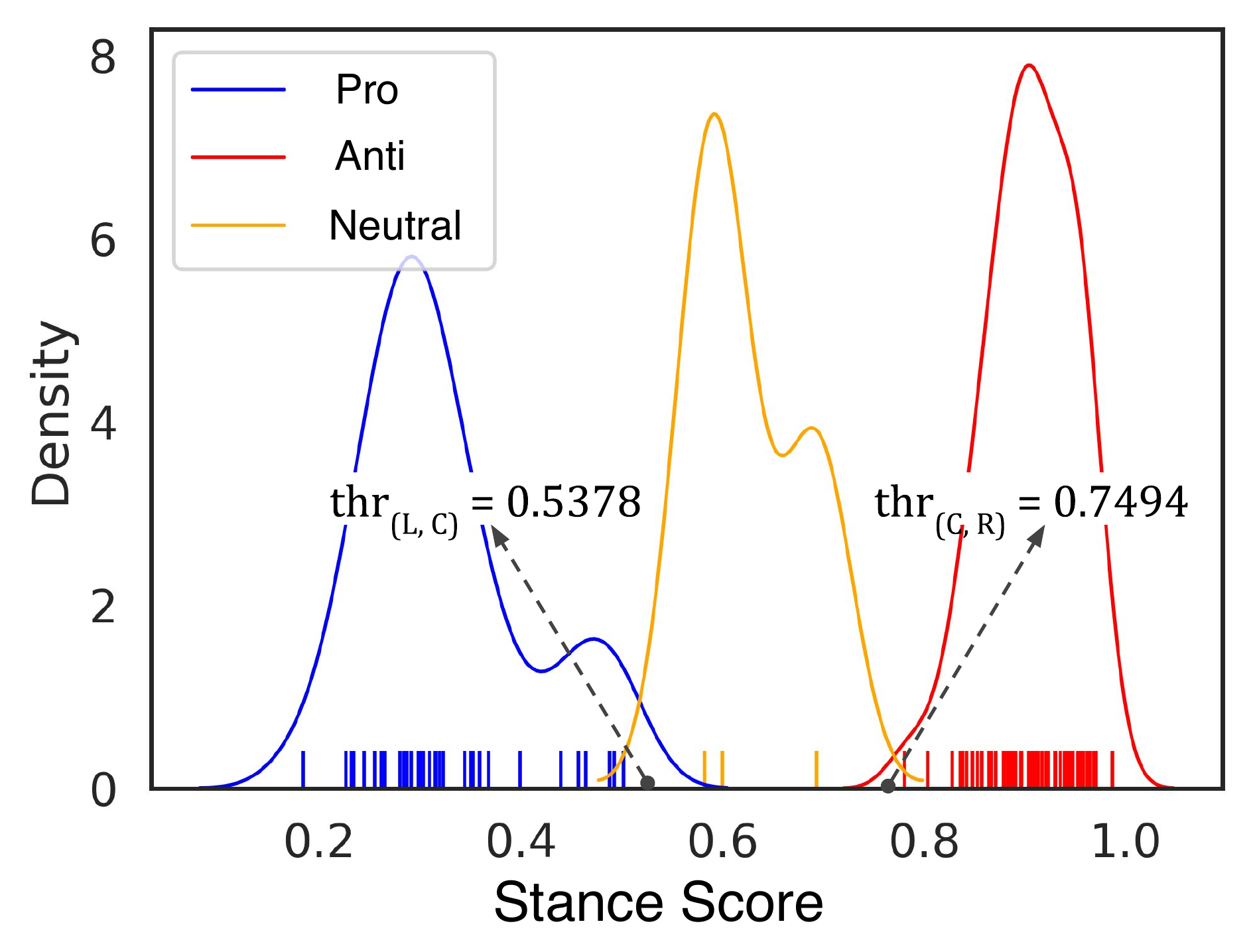}
    \caption{\abortion}
  \end{subfigure}
  \begin{subfigure}[b]{0.32\linewidth}
    \includegraphics[width=\linewidth]{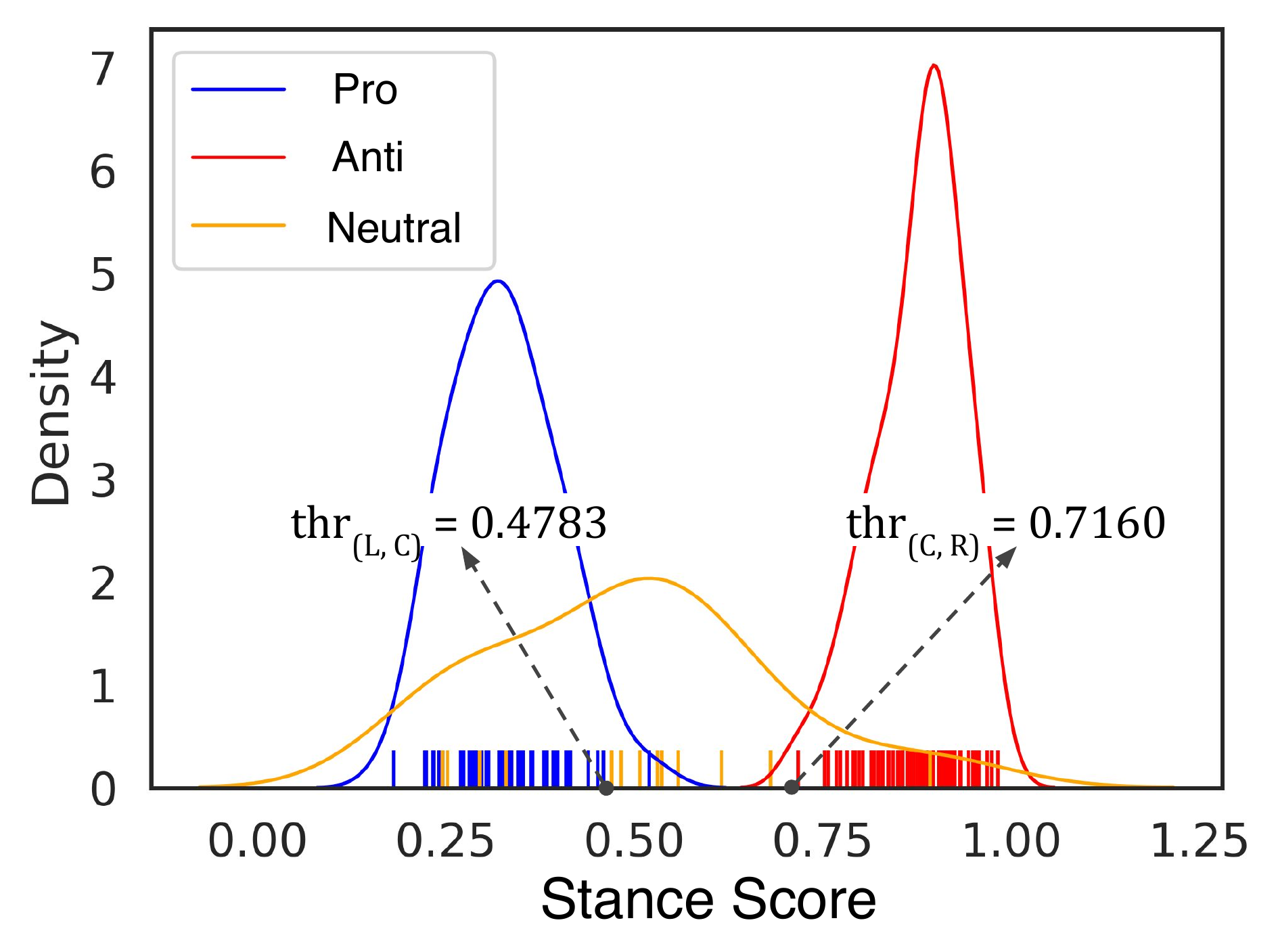}
    \caption{\guncontrol}
  \end{subfigure}
  \begin{subfigure}[b]{0.32\linewidth}
  \includegraphics[width=\linewidth]{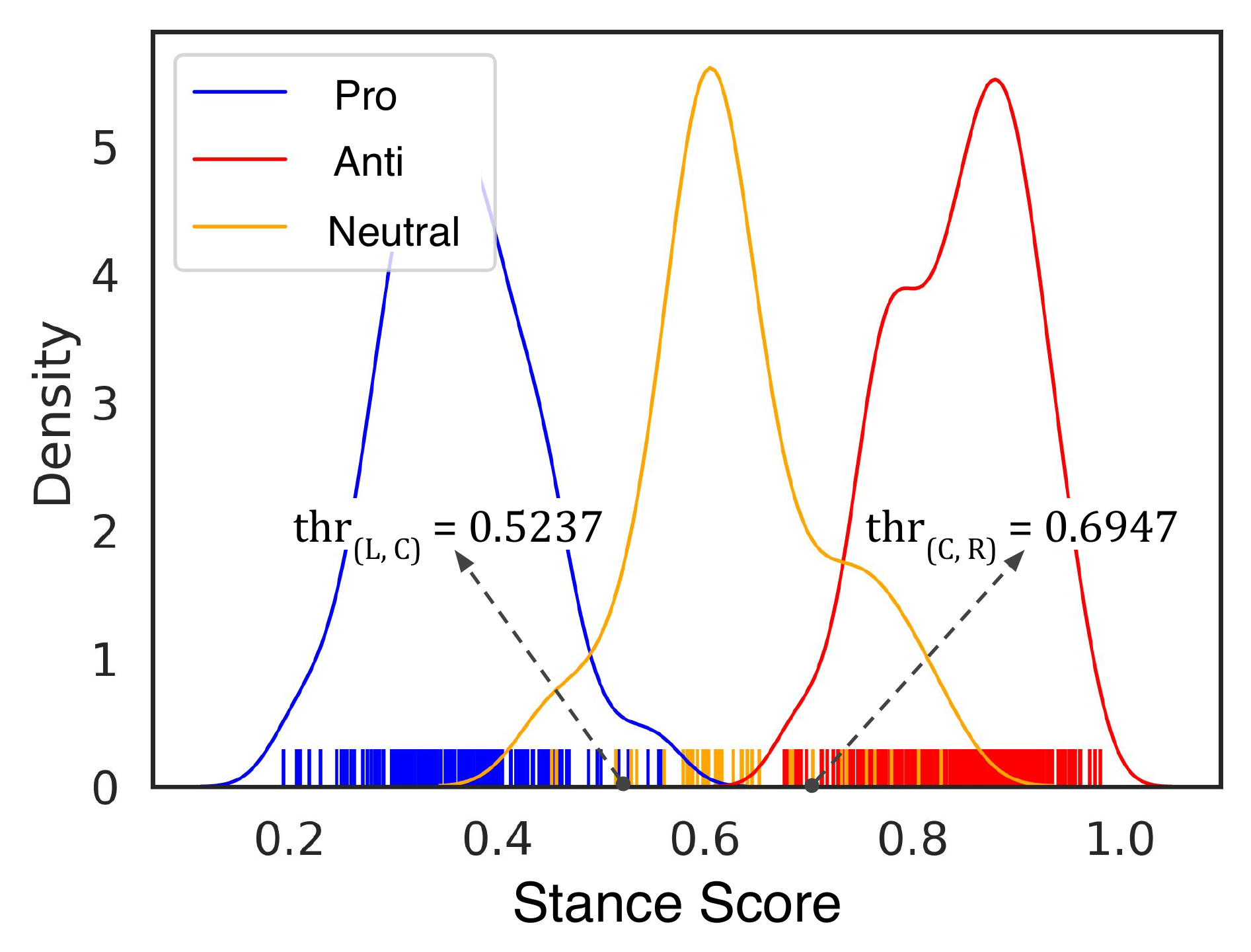}
  \caption{\blm}
  \end{subfigure}
  \caption{Distribution of leaning scores of videos (generated with kernel density estimation after removal of outliers) whose channels are available in Recfluence dataset \cite{ledwich2020algorithmic}. Each rod in the figures corresponds to a video, and its color represents the leaning label of the video obtained from Recfluence dataset.}
  \label{fig:recfluence_thr_plots}
\end{figure*}

We used videos with labels from Recfluence to find optimum thresholds $thr_{(L, C)}$ and $thr_{(C, R)}$. We obtained a leaning score distribution for each group (i.e., L, R and C). We identified outliers for each group using inter-quartile rule and filtered them out. The leaning score distributions by groups for each topic are given in \Cref{fig:recfluence_thr_plots}. Next, we found the thresholds $thr_{(L, C)}$ and $thr_{(C, R)}$ using posterior probabilities as follows:
\begin{eqnarray}
     P(L|leaning\_score(\nu)) = P(C|leaning\_score(\nu)),&\text{where}~ leaning\_score(\nu) = thr_{(L, C)} \\
    P(C|leaning\_score(\nu)) = P(R|leaning\_score(\nu)),&\text{where}~ leaning\_score(\nu) = thr_{(C, R)}   
\end{eqnarray}

The resulting threshold pairs ($thr_{(L, C)}$, $thr_{(C, R)}$) are $(0.538, 0.749)$ for \abortion, $(0.478, 0.716)$ for \guncontrol and $(0.524, 0.695)$ for \blm, respectively. Based on these thresholds, we inferred the leaning labels of 70 videos (L: 20, R: 45, C: 5) from 58 channels for \abortion, 115 videos (L: 29, R: 66, C: 20) from 98 channels for \guncontrol, 256 videos (L: 103, R: 111, C: 42) from 197 channels for \blm. 

To validate the video leaning estimation, we conducted a manual leaning annotation for videos in \guncontrol. We used stratified sampling where we sampled 5 videos from each 10 percentile bin of their leaning scores. The leaning of these 50 videos were annotated independently by three authors considering three categories (\textit{left-leaning}, \textit{right-leaning}, \textit{unknown}). The Fleiss' Kappa is 0.691, suggesting a moderate agreement among annotators. We used the majority vote to assign a video label. Table \ref{table:gun_stance_manual_annotation} summarizes the annotation results. 

We observed that the only decile bin with both manually annotated left- and right-leaning video is the 30th-40th percentile of leaning scores $(0.465 - 0.665)$. Suggesting that the annotator had high agreement with the estimated leaning scores, and that the threshold should be in this interval -- which is indeed the case for \guncontrol with $thr_{(L, C)}$ lies within its left boundary, and $thr_{(C, R)}$ only 0.05 away from its right boundary. In addition, most of the videos which were assigned \textit{unknown} by the raters are the footage of shooting of Philando Castile and mass shooting events including Las Vegas shooting and TX massacre. These videos do not include any stance on their contents. However, we observed that the leaning scores of these videos align well with the leaning of channels. One example is ``Comments on TX church shooting'' where the video leaning score belongs to the second bin $(80th - 90th)$ percentile. The channel owner is a far-right YouTuber. The video leaning score matches with the channel owner's leaning, our approach is thus successful for solving such ambiguities.

\Cref{table:gun_stance_manual_annotation} shows manual leaning annotation results of 50 Gun Control videos. Compared to the leaning score range in the second column (which is estimated), the first three rows of 14 videos (which are labeled as left-leaning) are consistently labeled (the result of manual labeling and estimated leanings agree) as well as the last five rows of 19 videos (which are labeled as right-leaning). The computed thresholds for Gun Control videos, as presented in~\Cref{fig:recfluence_thr_plots}, are $0.478$ and $0.716$.

\begin{table*}[!htbp]
    \centering
    \small
    \begin{tabular}{lccccc}
        \toprule
        %\multicolumn{1}{l}{Score propagation} & \multicolumn{4}{l}{Manual annotation} \cmidrule{1-6}\\
        \multicolumn{1}{l}{Percentiles} & \begin{tabular}[c]{@{}c@{}}Leaning score\\ range\end{tabular} & \begin{tabular}[c]{@{}c@{}}\#left-\\leaning\\ videos\end{tabular} & \begin{tabular}[c]{@{}c@{}}\#right-\\leaning\\ videos\end{tabular} & \begin{tabular}[c]{@{}c@{}}\#unknown\\ videos\end{tabular} & Videos annotated as unknown \\ \midrule
        0th - 10th & (0.138 - 0.289) & 4 & 0 & 1 & \begin{tabular}[c]{@{}c@{}} (1) Shooting footage of Philando Castile\\ (by CBC News (left-center)) \end{tabular} \\ \cmidrule{1-6}
        10th - 20th & (0.289 - 0.364) & 5 & 0 & 0 & - \\ \cmidrule{1-6}
        20th - 30th & (0.364 - 0.465) & 5 & 0 & 0 & - \\ \cmidrule{1-6}
        30th - 40th & (0.465 - 0.665) & 2 & 1 & 2 & \begin{tabular}[c]{@{}c@{}}(1) Shooting footage of Philando Castile\\ (2) Las Vegas shooting footage\end{tabular} \\ \cmidrule{1-6}
        40th - 50th & (0.665 - 0.789) & 0 & 4 & 1 & \begin{tabular}[c]{@{}c@{}}(1) Discussion on gun violence in Chicago\\ (by Ron Paul)\end{tabular} \\ \cmidrule{1-6}
        50th - 60th & (0.789 - 0.837) & 0 & 3 & 2 & \begin{tabular}[c]{@{}c@{}}(1) Las Vegas shooting footage.\\ (2) Las Vegas shooting footage\\ (multiple shooters claim).\end{tabular} \\ \cmidrule{1-6}
        60th - 70th & (0.837 - 0.875) & 0 & 4 & 1 & \begin{tabular}[c]{@{}c@{}}(1) Comments on Las Vegas shooting\\ (by a Trump supporter.)\end{tabular} \\ \cmidrule{1-6}
        70th - 80th & (0.875 - 0.895) & 0 & 4 & 1 & (1) Las Vegas shooting footage. \\ \cmidrule{1-6}
        80th - 90th & (0.895 - 0.924) & 0 & 3 & 2 & \begin{tabular}[c]{@{}c@{}}(1) Comments on TX church shooting \\ (by a far-right YouTuber).\\ (2) Comments on Las Vegas shooting \\ (by Milo).\end{tabular} \\ \cmidrule{1-6}
        90th - 100th & (0.924 - 0.982) & 0 & 5 & 0 & - \\ \cmidrule{1-6}
        {\bf Total} & - & 16 & 24 & 10 & - \\
        \bottomrule
    \end{tabular}
    \caption{Manual leaning annotation of \guncontrol videos.}
    \label{table:gun_stance_manual_annotation}
\end{table*}

% ====================================================
\clearpage

\section{Computing Relative Engagement of a Video}
\label{sec:app_relengagement}

From the collected videos we extracted topic (\abortion, \guncontrol and \blm) relevant videos for our topical analysis, but all collected videos were used to compute relative engagement scores. %We filter videos based on recency and their level of attention. We remove videos that are published prior to this two-month period to avoid older videos, since being tweeted a while after being uploaded may indicate higher engagement~\cite{wu2018beyond}.
We filter out videos that receive less than 100 views within their first 30 days after upload, which is the same filter used by~\citet{brodersen2012youtube}.
We summarize how relative engagement score is computed. For a given video, first we compute two aggregate metrics:
\begin{itemize}
    \item average watch time ($\bar{\omega}_t$): the total watch time $x_{w}[1 : t]$ divided by the total view count $x_v[1 : t]$ up to day $t$
    \[\bar{\omega}_t = \frac{\Sigma^t_{i=1}x_{w}[i]}{\Sigma^t_{i=1}x_{v}[i]}\]
    \item average watch percentage ($\bar{\mu}_t$): the average watch time $\bar{\omega}_t$ normalized by video duration D
    \[\bar{\mu}_t = \frac{\bar{\omega}_t}{D}\]
\end{itemize}

\begin{figure*}[!htb]
  \centering
  \begin{subfigure}[b]{0.4\linewidth}
    \includegraphics[width=\linewidth]{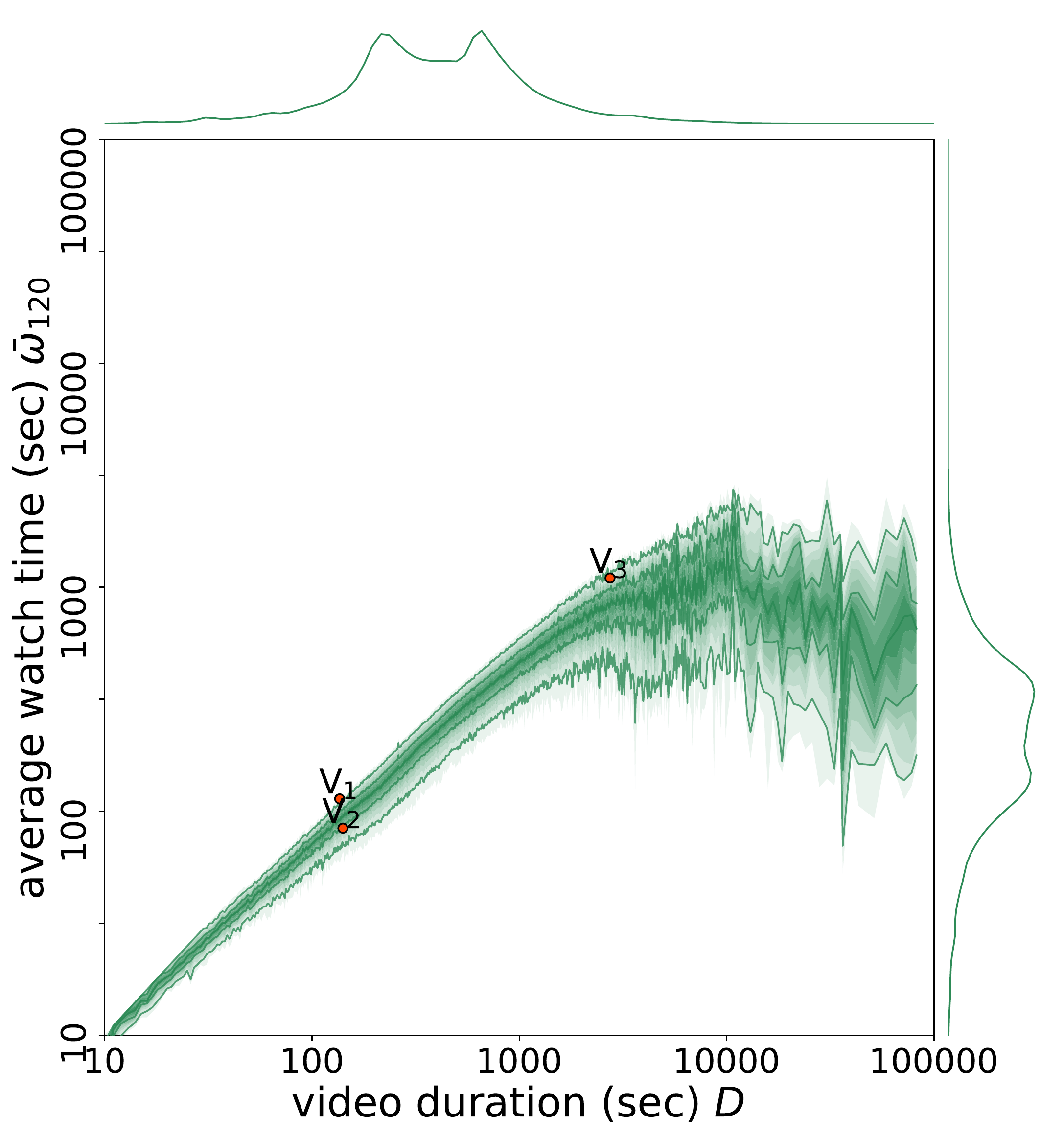}
    \caption{Duration vs watch time}
  \end{subfigure}
  \begin{subfigure}[b]{0.4\linewidth}
      \includegraphics[width=\linewidth]{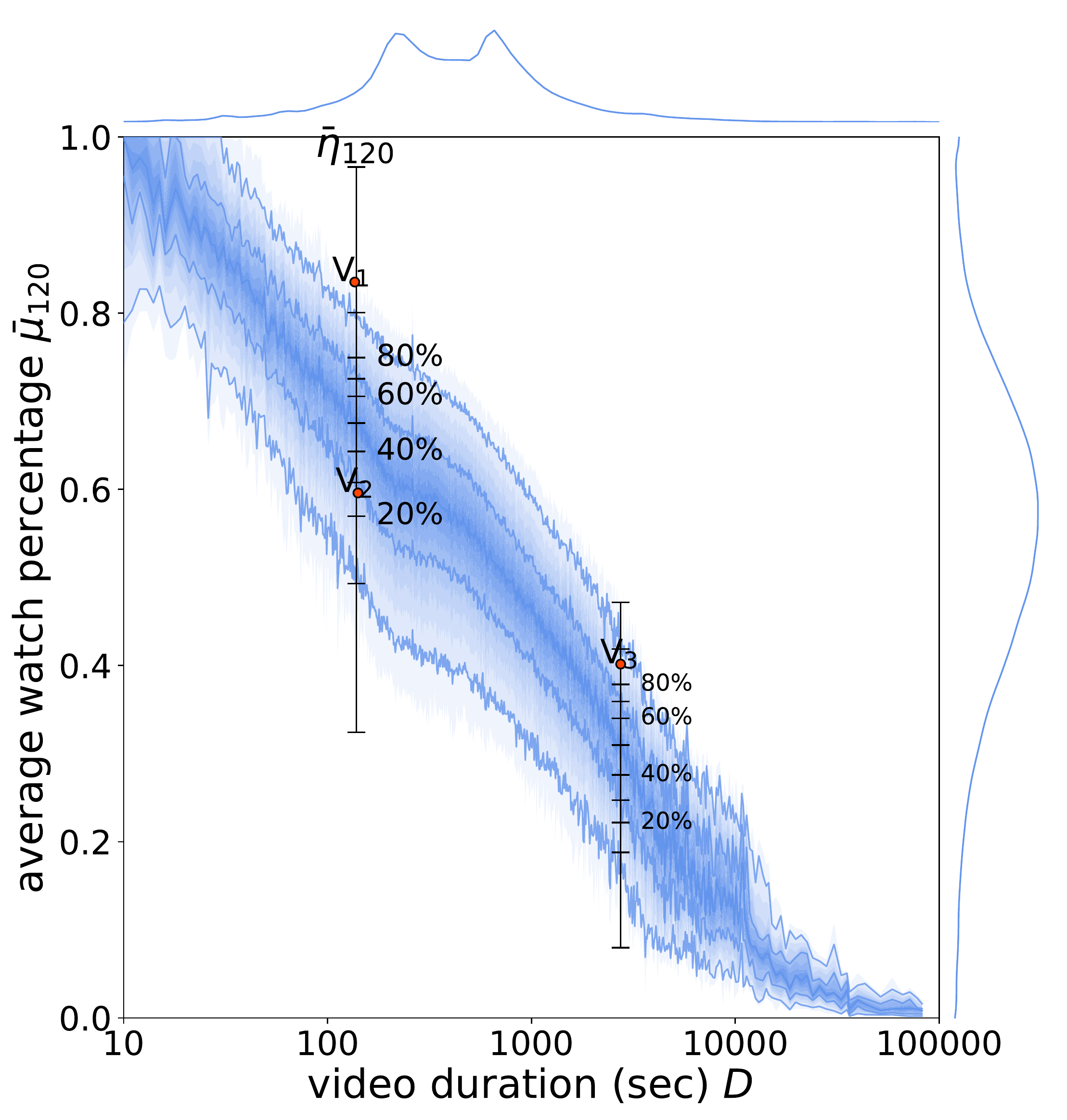}
      \caption{Engagement map}
  \end{subfigure}
  \caption{Video engagement in the tweeted videos dataset at the age of 120 days.  (a) video duration D vs average watch time $\bar{\omega}_{120}$ (b) the engagement map: video duration D
vs average watch percentage $\bar{\eta}_{120}$.
This figure originally appeared in \cite{wu2018beyond}.}
  \label{fig:appendix_relative_engagement}
\end{figure*}

Then {\it Engagement map} of tweeted videos is constructed as follows. Two maps are created; the first where x-axis
shows video duration D, and the y-axis shows average
watch time over the first 120 days ($\bar{\omega}_{120}$) (\Cref{fig:appendix_relative_engagement}(a)) and the second with the same axis with average watch percentage over the first 120 days ($\bar{\mu}_{120}$) (\Cref{fig:appendix_relative_engagement}(b)) as y-axis. All videos in the tweeted videos dataset are projected onto both maps. The x-axis is split into 1,000 equally wide bins in log scale. These 2 maps are logically identical but \Cref{fig:appendix_relative_engagement}(b) is easier to read as y-axis is bounded between [0,1]. This second map is denoted as {\it Engagement map}.

Based on this {\it Engagement map},  relative engagement $\bar{\eta}_t\in [0,1]$ defined
as the rank percentile of video in its duration bin.  This is an average engagement measure in the first t days. \Cref{fig:appendix_relative_engagement}(b) illustrates the relation between video duration D, watch percentage $\bar{\mu}_{120}$ and relative engagement $\bar{\eta}_{120}$ for three example videos. $V_1$ is a left-leaning video in \abortion topic with duration of 136 seconds. $V_2$ is a right-leaning video in \abortion topic with duration of 141 seconds. $V_1$ and $V_2$ have similar lengths but different watch percentages, $\bar{\mu}_{120}(V_1) = 0.83$ and $\bar{\mu}_{120}(V_2) = 0.59$. The difference becomes more apparent with relative engagement scores, $\bar{\eta}_{120}(V_1) = 0.94$ and $\bar{\eta}_{120}(V_2) = 0.27$ as shown in \Cref{fig:appendix_relative_engagement}(b).
$V_3$ is also a right-leaning video in \abortion which is significantly long than the others. Although $V_3$ is not as much watched as the previous two, $\bar{\mu}_{120}(V_3) = 0.40$, relative engagement of $V_3$ is much higher than $V_2$, $\bar{\eta}_{120}(V_3) = 0.86$ as longer videos tend to have lower average watch percentages. 

% ====================================================
\clearpage

\section{Computing Viral Potential of a Video via the Hawkes Intensity Process}
\label{sec:app_hip}

\header{Viral potential}
estimates the expected number of views a video will obtain if mentioned by an  {\em average} tweet ~\cite{rizoiu2017online}.
More specifically, it is the area under the impulse response function of an integral equation known as Hawkes Intensity Process (HIP)~\cite{rizoiu2017expecting}, which are learned for each video by using the first 120 days of tweeting and viewing history. The HIP model is designed to describe the phenomena of tweets driving or attracting video views including factors such as network effect (such as views beget views), system memory (such as interests in news videos wane within a week, but interests on music videos sustain for several years), and video quality (including factors such as sensitivity to tweet mentions and inherent attractiveness). 
We estimate the number of views per tweet via the HIP model rather than simply dividing the number of views by the number of tweets, because the model takes into account future views that are expected to unfold after the 120-day cut-off. 
% This effect is significant for videos for which the word-of-mouth unfolds over many days, or with high re-watch value.
% , such as \TODO{}{do we have an example? I remember a music video, or maybe summaries of BLM events}.
The same reason motivates the name viral {\em potential} rather than viral score.
One limitation of HIP and viral potential is the notion of an  ``average'' tweet being implemented by marginalizing over power-law distributed number of followers. This means
% is an over-simplification of the relation of the diverse users tweeting behavior, and 
network size variations across different topics are not taken into account.
A self-contained summary of HIP and viral potential computation is included below. The viral potential is a positive number, since views cannot be negative (i.e., no video loses views). However, it can take values less than one, corresponding to the effect of tweets being dampened (one view per several tweets) rather than amplified, as often seen in videos linked by spam tweets. 

Hawkes Intensity Process (HIP)~\cite{rizoiu2017expecting} extends the well-known Hawkes (self-exciting) process~\cite{hawkes1971spectra} to describe the volume of activities within a fixed time interval (e.g. daily).
This is done by taking expectations over the stochastic event history.
Specifically, this model describes a self-exciting phenomenon that is commonly observed in online social network~\cite{rizoiu2017online}.
It models the target quantity $x[t]$ as a self-consistent equation into three parts: the unobserved external influence, the effects of external promotions, and the influence from historical events.
Formally, it can be written as

\begin{equation}
  \label{eq:hip}
  x[t] = \gamma \mathds{1}[t = 0] + \eta \mathds{1}[t > 0] + \alpha s[t] + C\sum_{\tau=1}^{t} x[t - \tau](\tau+c)^{-(1+\theta)}
\end{equation}

The first two terms represent unobserved external influences.
$\gamma$ and $\eta$ model the strengths of an initial impulse and a constant background rate, respectively.
In the middle component, $\alpha$ is the sensitivity to external promotion, $s[t]$ is the volume of promotion, and $\alpha s(t)$ is the instantaneous response to promotion.
In the last component, $\theta$ is the exponent of a power-law memory kernel $(\tau+c)^{-(1+\theta)}$.
$c$ is a nuisance parameter for keeping the kernel bounded, and $C$ accounts for the latent content quality.
Overall, this last component models the impact over its own event history $x[\tau]$ for $\tau{=}1:t{-}1$.

In our case, $x[t]$ is the time series of daily views $x_v[t]$ on YouTube.
$s[t]$ is the daily number of tweets on Twitter.
The parameter set $\{\gamma, \eta, \alpha, C, c, \theta\}$ is estimated from the first 90-day interval of each video using the constrained L-BFGS algorithm in \texttt{SciPy} Python package.\footnote{\url{https://github.com/andrei-rizoiu/hip-popularity}}

% ====================================================
\clearpage

\section{View and Tweets over time: additional plots}
\label{sec:app-overtime_additional}

\begin{figure*}[!htb]
  \centering
 \includegraphics[width=0.66\linewidth]{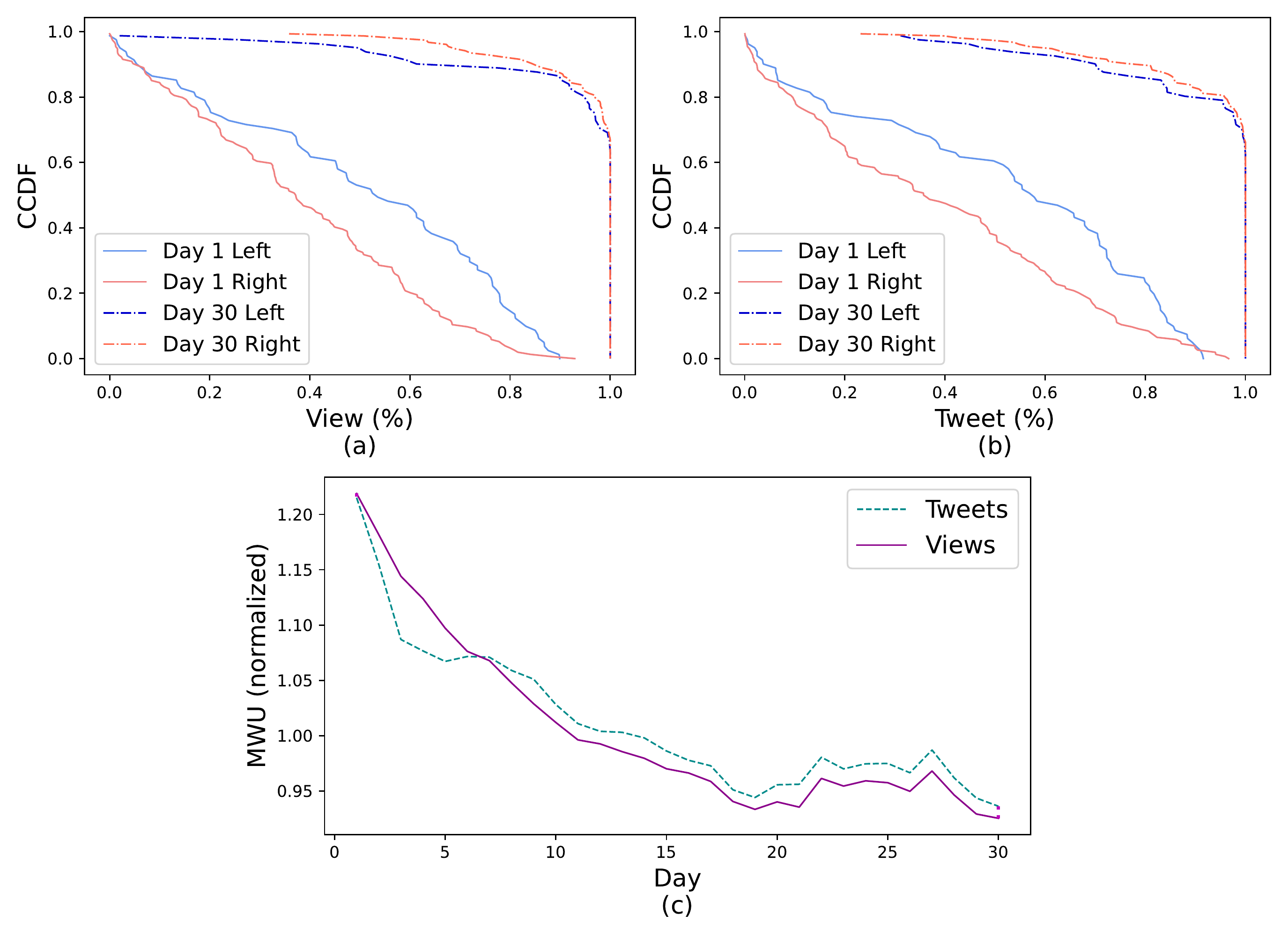}
  \caption{Comparing view and tweet accumulations of left- and right-leaning videos in \guncontrol. (a) and (b) show CCDF of views and tweets accumulated for Day 1 and Day 30,  respectively. As seen in (c), the differences between left- and right-leaning videos are larger than the differences of \abortion videos, but the differences between accumulation of views and tweets are smaller than \abortion videos.}
  \label{fig:app_gun_ccdf}
\end{figure*}

\begin{figure*}[!htb]
  \centering
 \includegraphics[width=0.66\linewidth]{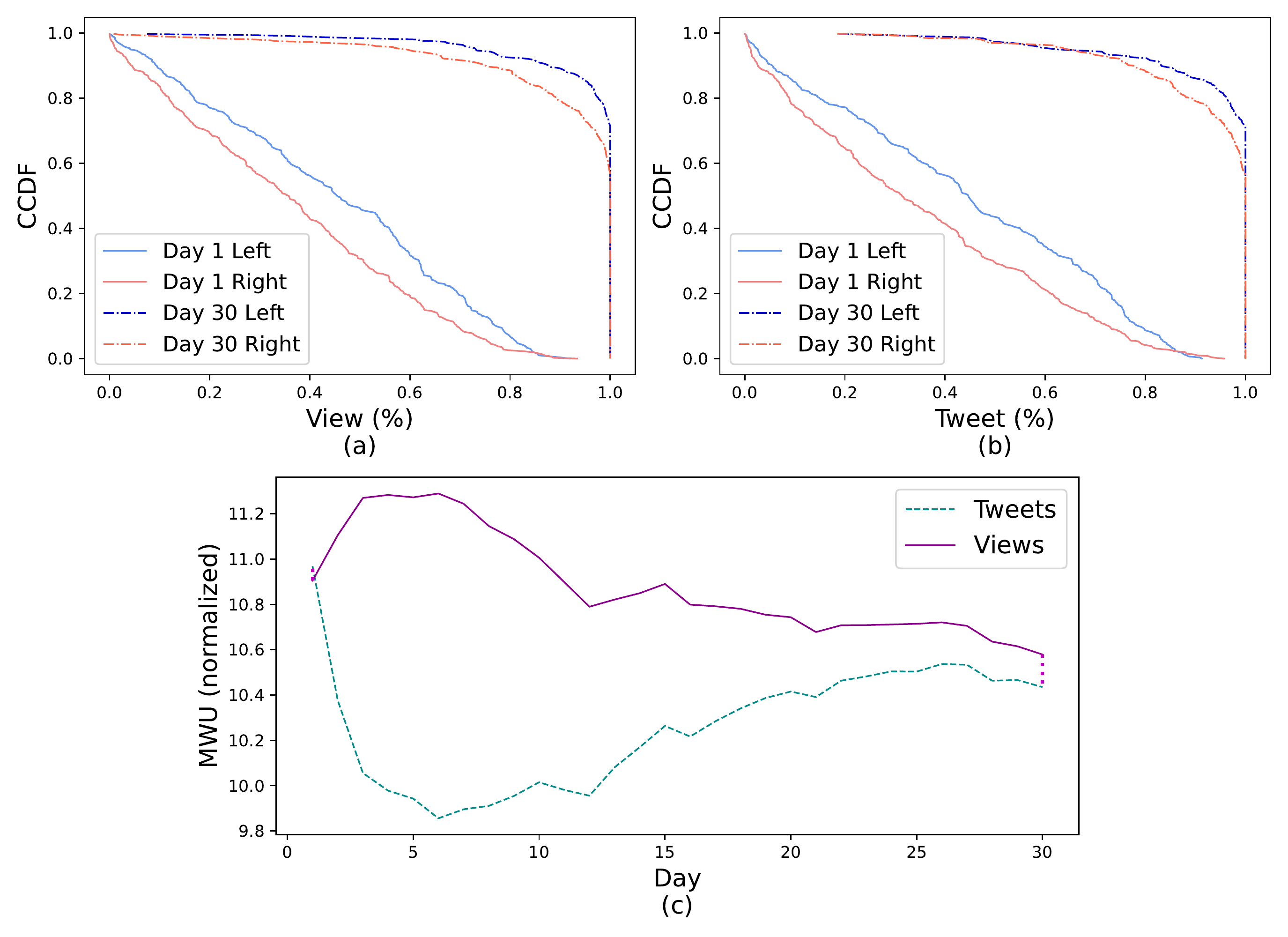}
  \caption{Comparing view and tweet accumulations of left- and right-leaning videos in \blm. (a) and (b) show CCDF of views and tweets accumulated for Day 1 and Day 30,  respectively. (c) shows that the differences between left- and right-leaning videos are the largest among all topics.}
  \label{fig:app_blm_ccdf}
\end{figure*}

\Cref{fig:app_gun_ccdf} and \Cref{fig:app_blm_ccdf} visualize the attention accumulation for \guncontrol and \blm topic.

% ====================================================
\clearpage

\section{Tweet cascades: additional plots}
\label{sec:app-cascade_additional}

\begin{figure}[!htb]
  \centering
  \begin{subfigure}[b]{0.45\linewidth}
    \includegraphics[width=\linewidth]{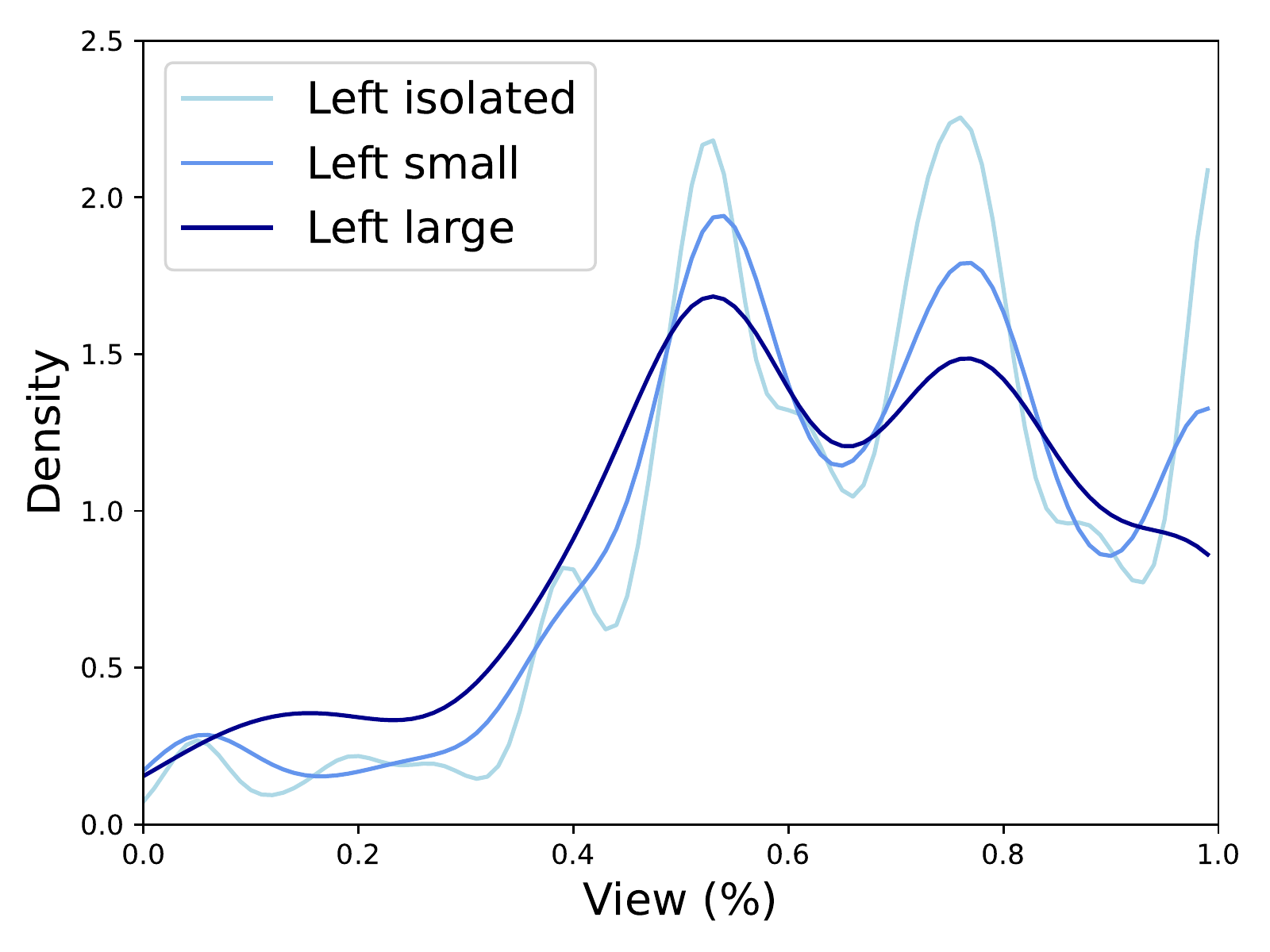}
    \caption{Tweet cascades in \guncontrol left-leaning videos.}
  \end{subfigure}
  \begin{subfigure}[b]{0.45\linewidth}
      \includegraphics[width=\linewidth]{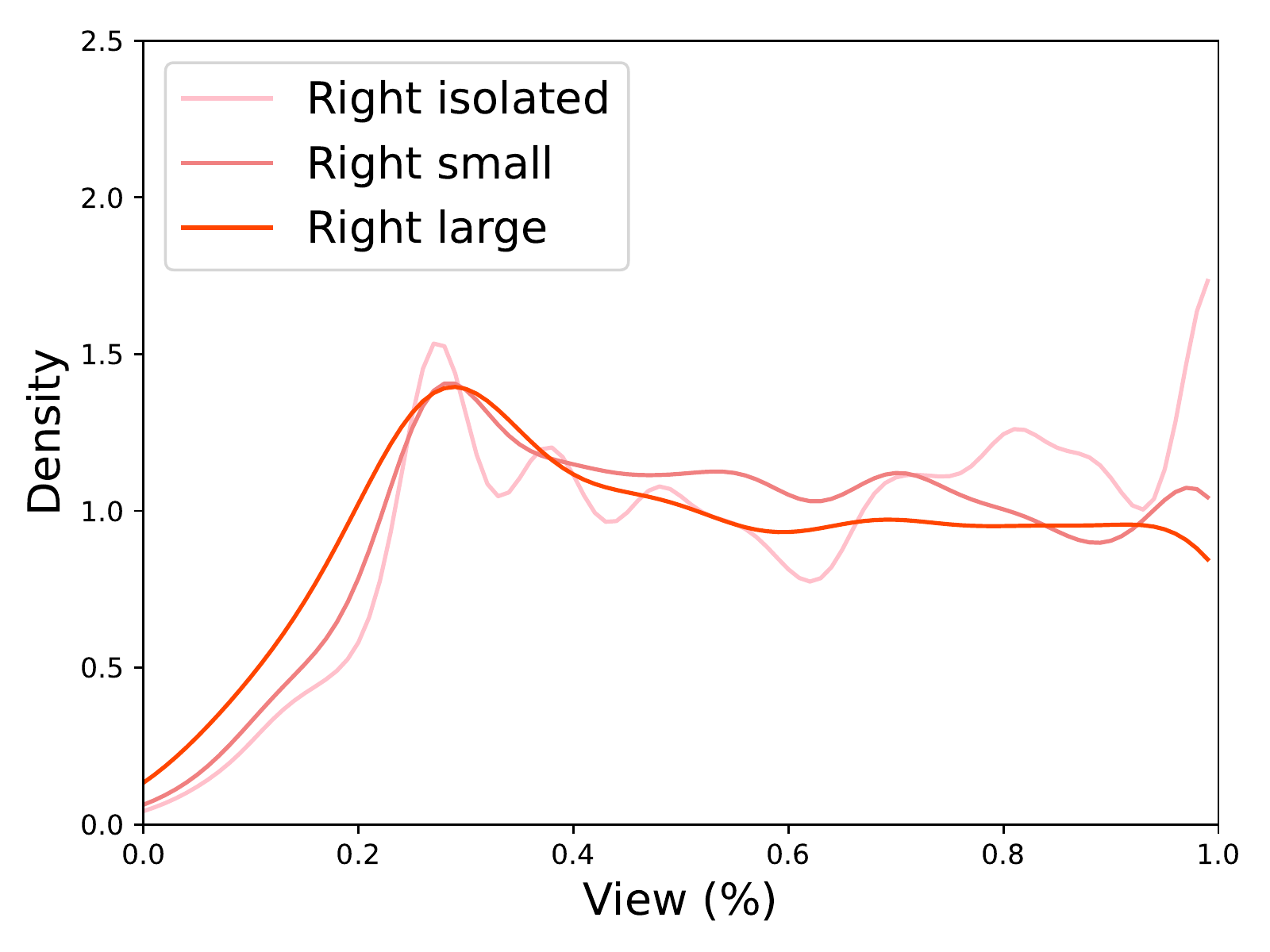}
      \caption{Tweet cascades in \guncontrol right-leaning videos.}
  \end{subfigure}
  \begin{subfigure}[b]{0.45\linewidth}
    \includegraphics[width=\linewidth]{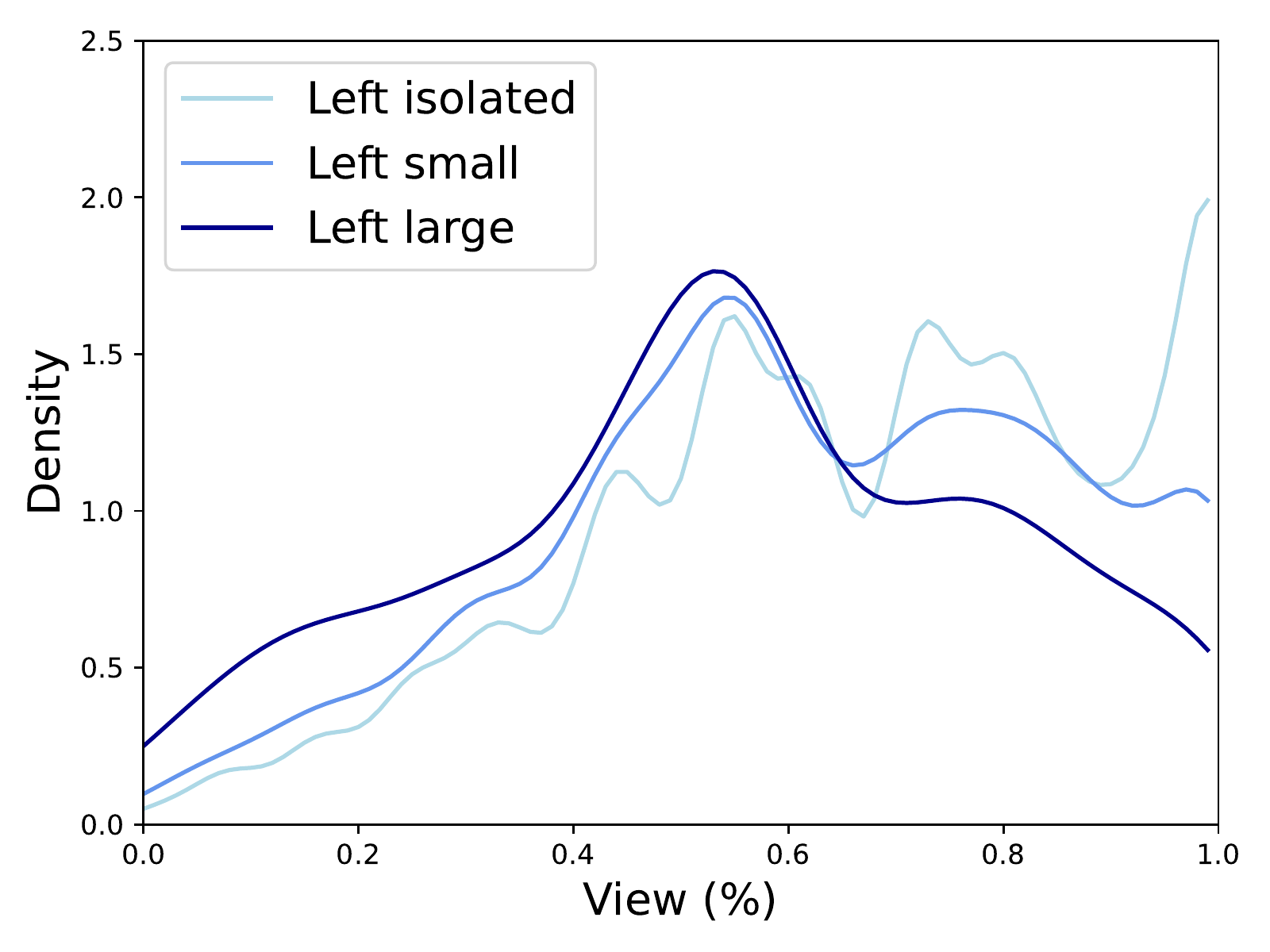}
    \caption{Tweet cascades in \blm left-leaning videos.}
  \end{subfigure}
  \begin{subfigure}[b]{0.45\linewidth}
      \includegraphics[width=\linewidth]{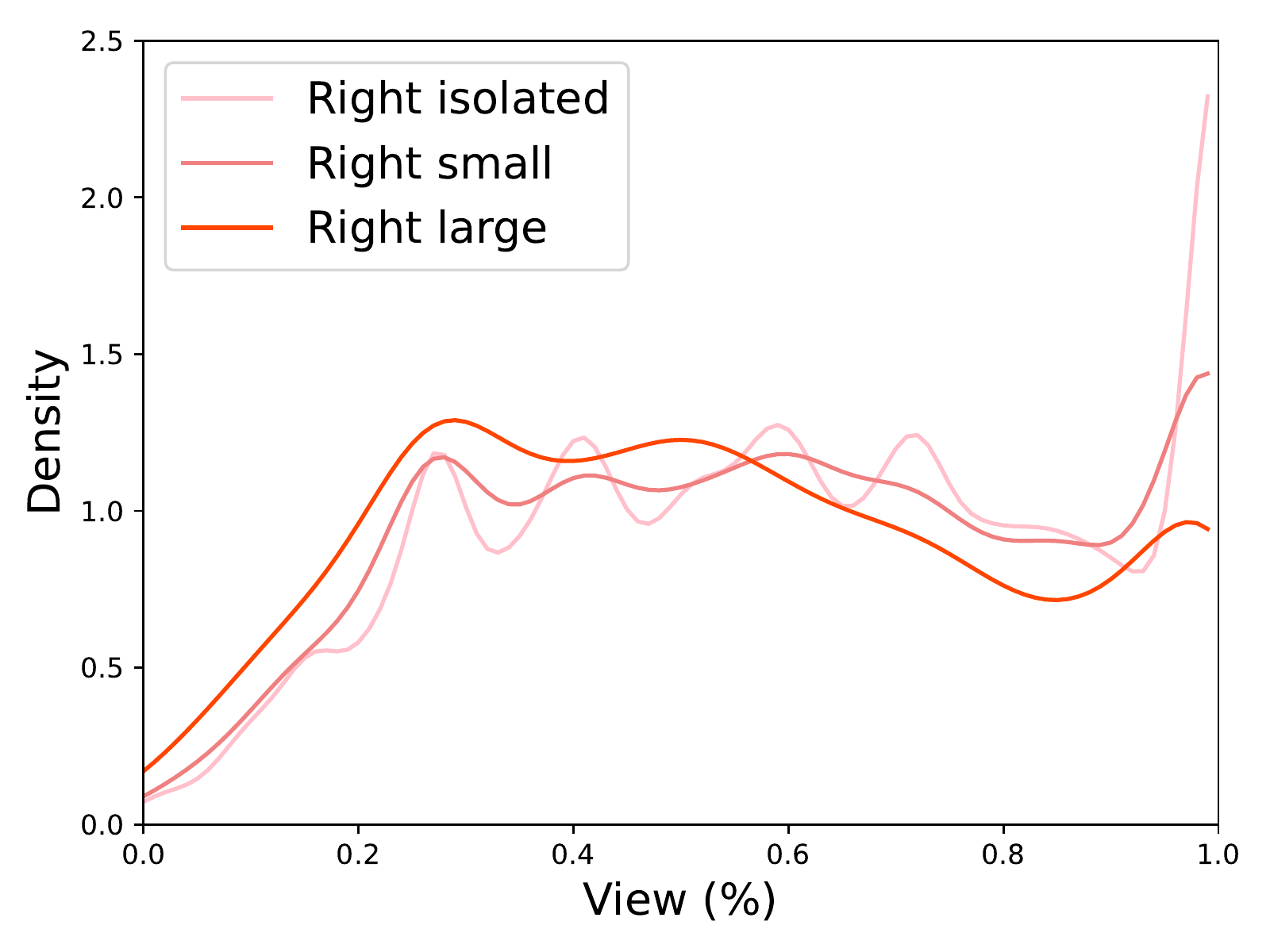}
      \caption{Tweet cascades in \blm right-leaning videos.}
  \end{subfigure}
  \caption{Cascade start times measured at view accumulation process. \Cref{table:measures} reports that retweet cascades for left-leaning videos start significantly later in view accumulation process for all topics.}
  \label{fig:app_cascade_additional}
\end{figure}

\Cref{fig:app_cascade_additional} visualizes attention distribution of the isolated, small, and large cascades for \guncontrol and \blm topic.

% ====================================================
\clearpage

\section{Additional metrics on early adopter networks}
\label{sec:app_centralized_left_right}

We measured additional metrics on early-adopter networks, the comparative findings are summarized in \Cref{table:measures}.

\header{\textit{Network density}} is the fraction of existing edges among all possible pair-wise edges. It quantifies the connectivity of the follower network. A higher value indicates the early adopters are more cohesively followed among themselves.

\header{\textit{Maximum indegree}} is the maximum indegree value in the network. It provides information about the most influential user in the network. 

\header{\textit{Global efficiency}} is computed as the number of closed triplets over the total number of triplets. It measures the average efficiency over all pairs of distinct users. Higher global efficiency reflects higher capacity to diffuse information over the network.

\header{\textit{Gini coefficient of betweenness centrality}} measures the dispersion in fraction of all shortest paths in the network that pass through a given user. Higher coefficient implies that a few early adopters are on the shortest path for most node-pairs, but the rest of the early adopters are not.

% ====================================================
\section{Intersecting relative engagement with view and tweet count}
\label{sec:app_intersect}

\begin{figure}[!htb]
  %%%% TWITTER FOLLOWERS %%%%%
  \centering
    \begin{subfigure}[b]{0.32\linewidth}
    \includegraphics[width=\linewidth]{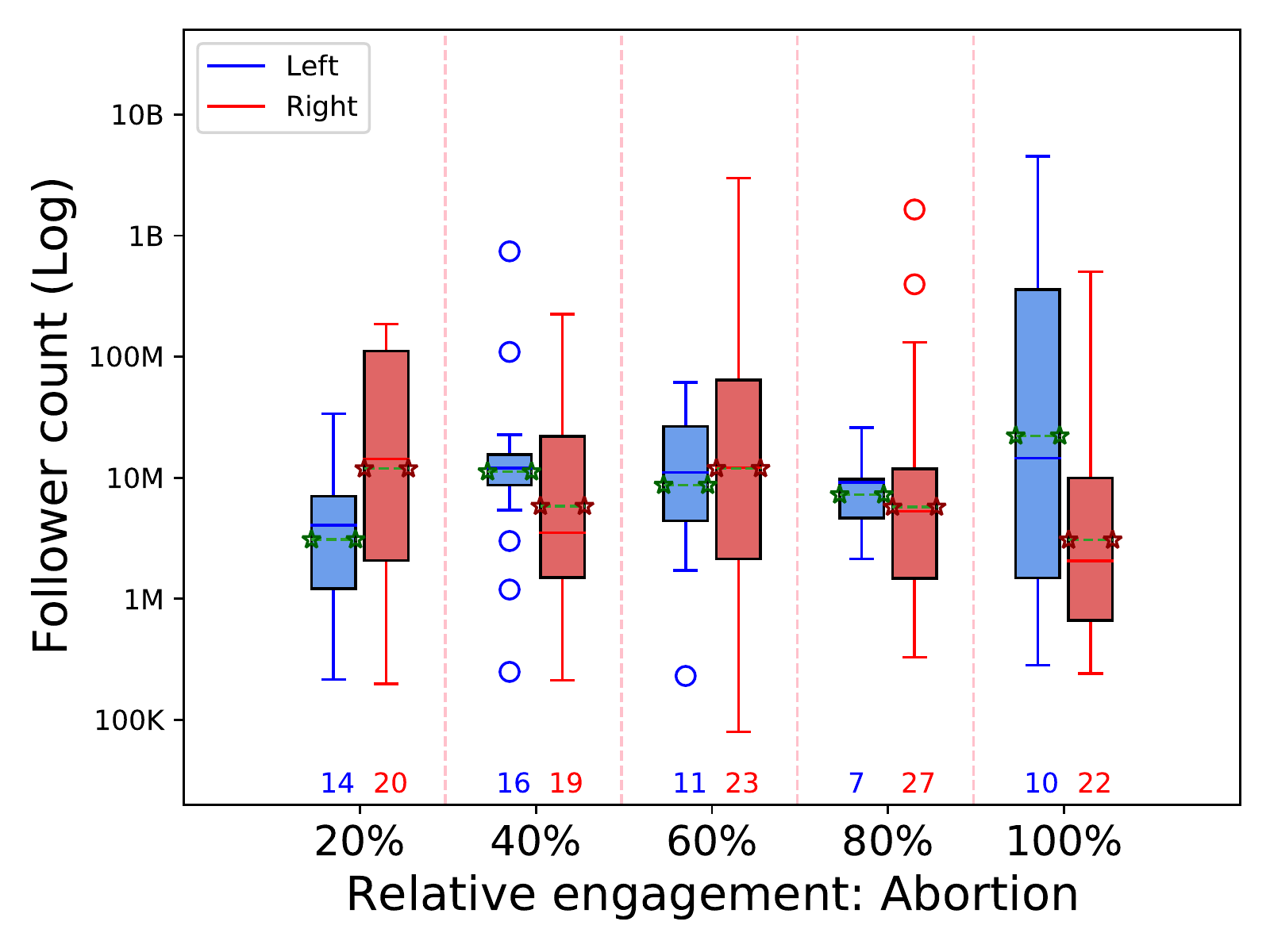}
    %\caption{\abortion}
  \end{subfigure}
  \begin{subfigure}[b]{0.32\linewidth}
    \includegraphics[width=\linewidth]{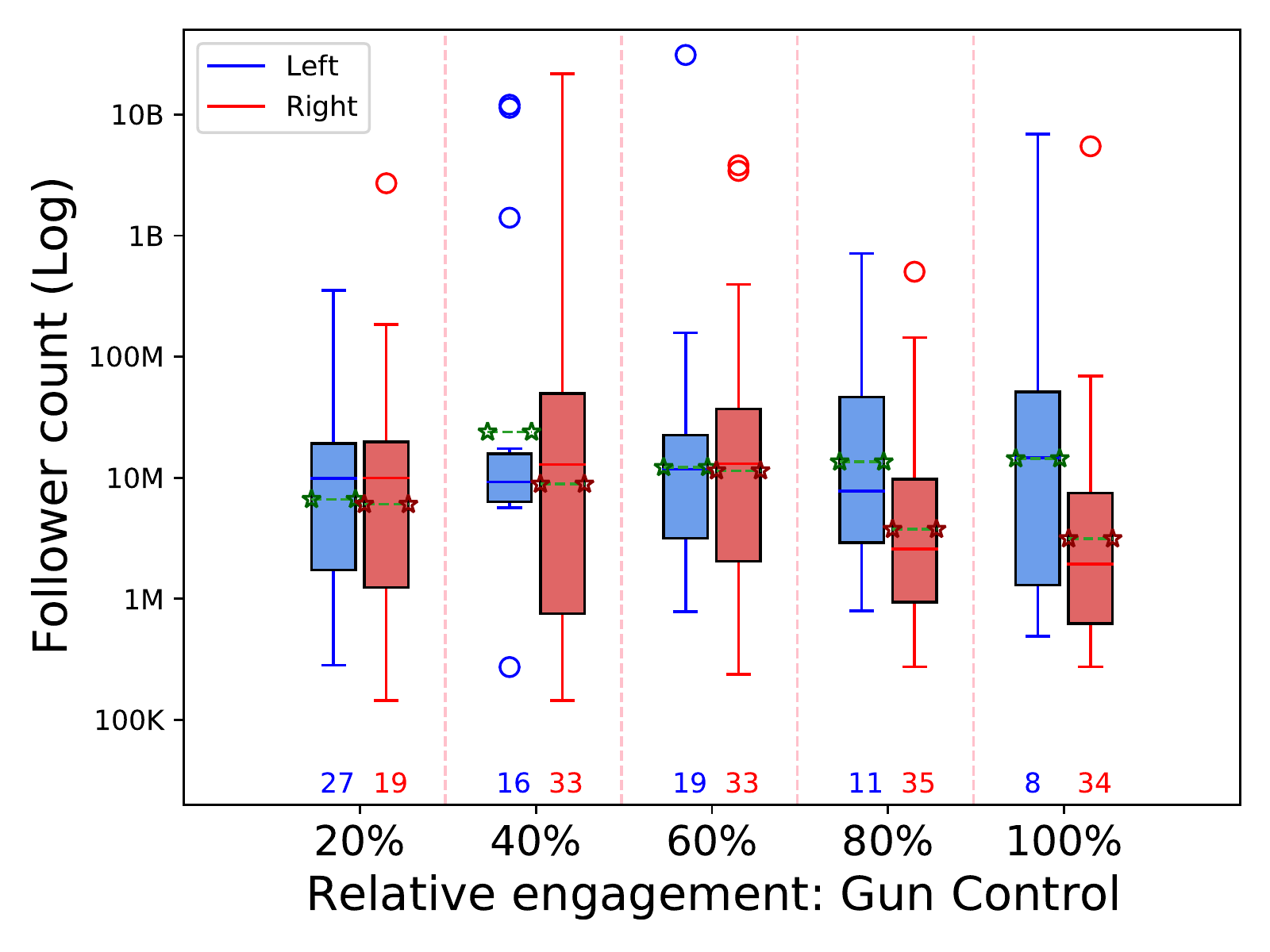}
    %\caption{\guncontrol}
  \end{subfigure}
  \begin{subfigure}[b]{0.32\linewidth}
  \includegraphics[width=\linewidth]{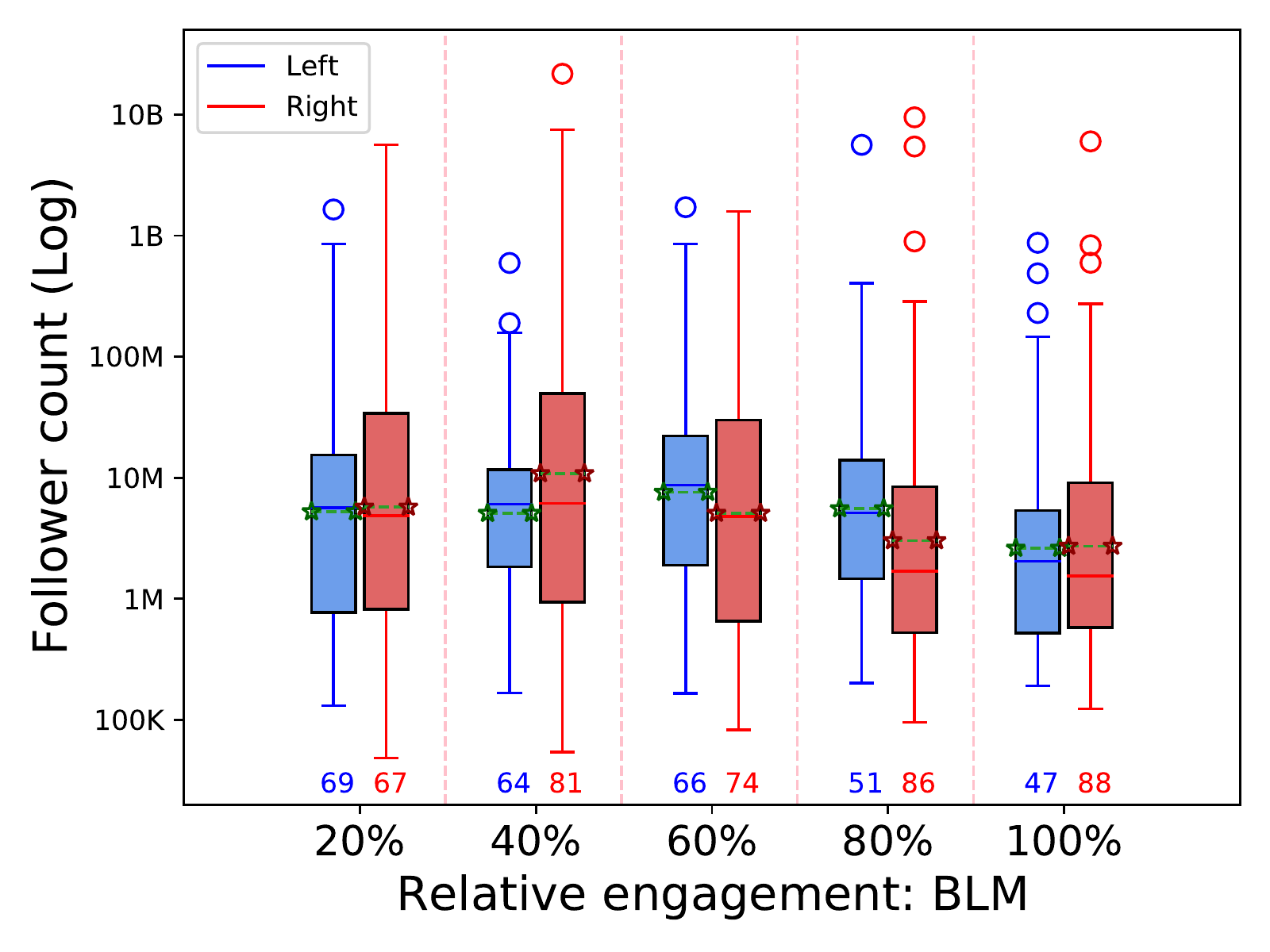}
  %\caption{\blm}
  \end{subfigure}
  %%% TWEETS @ 120 %%%%
  \begin{subfigure}[b]{0.32\linewidth}
    \includegraphics[width=\linewidth]{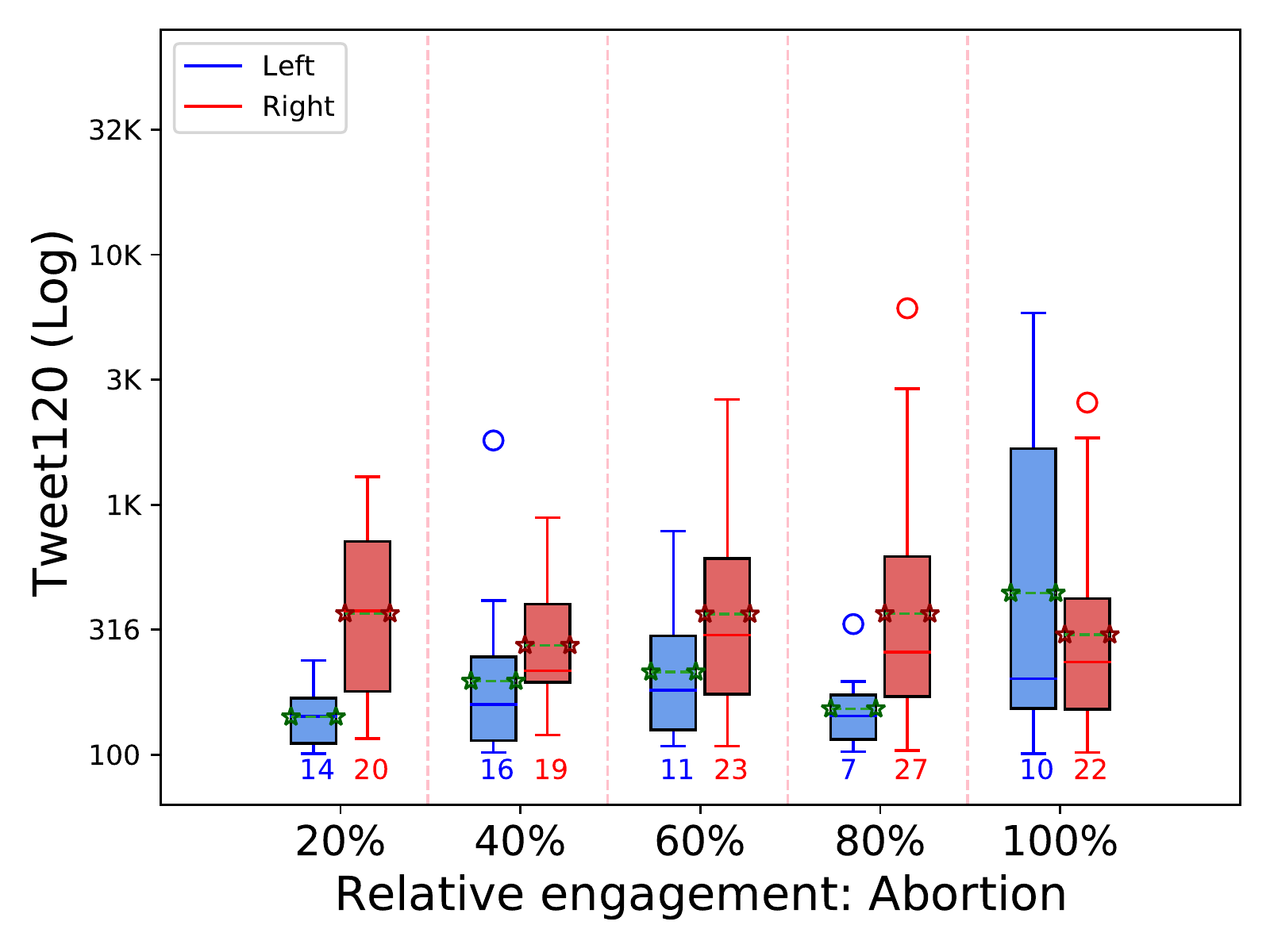}
    %\caption{\abortion}
  \end{subfigure}
  \begin{subfigure}[b]{0.32\linewidth}
    \includegraphics[width=\linewidth]{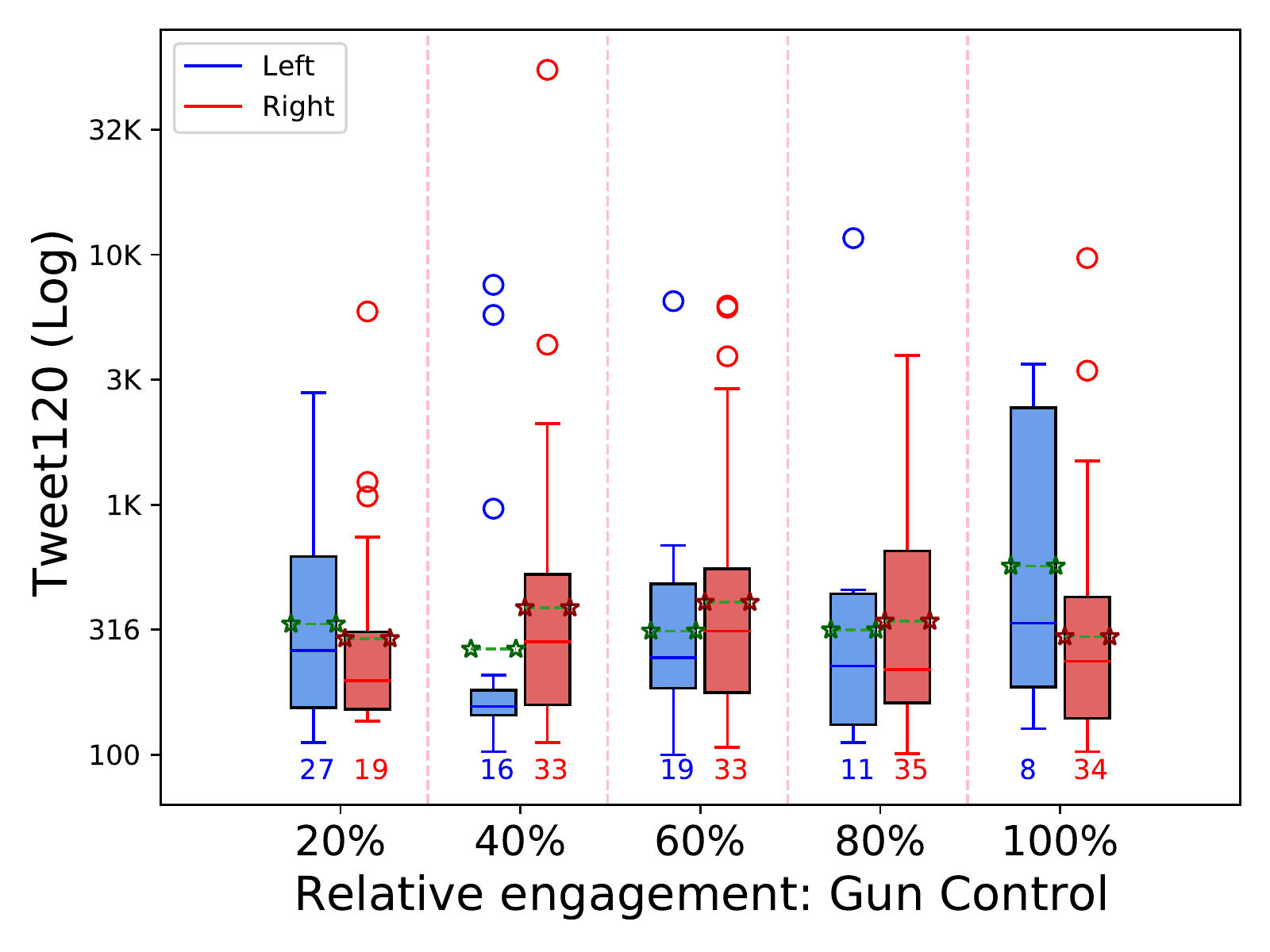}
    %\caption{\guncontrol}
  \end{subfigure}
  \begin{subfigure}[b]{0.32\linewidth}
  \includegraphics[width=\linewidth]{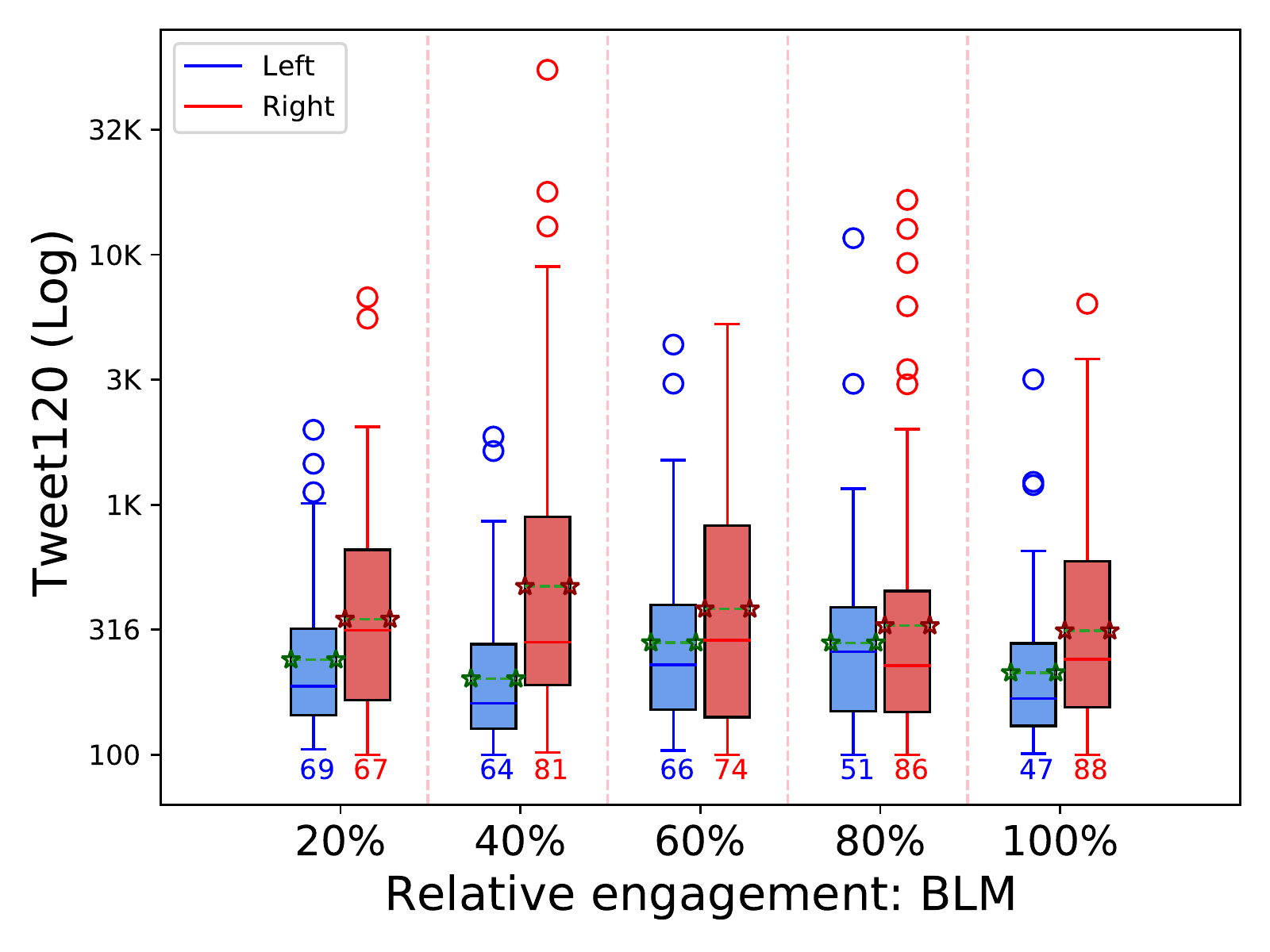}
  %\caption{\blm}
  \end{subfigure}
  %%%% VIEWS @ 120 %%%%
  \begin{subfigure}[b]{0.32\linewidth}
    \includegraphics[width=\linewidth]{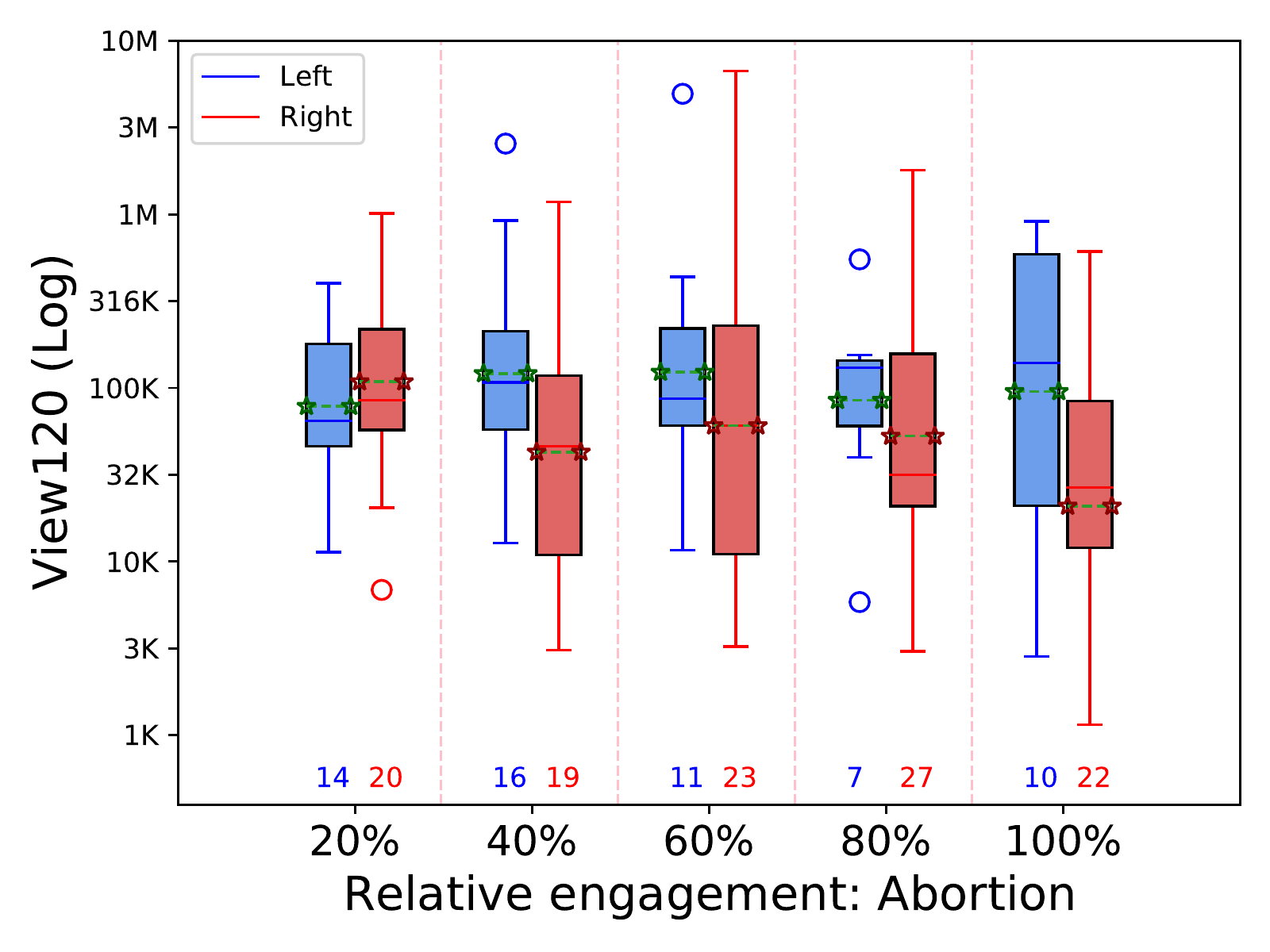}
    %\caption{\abortion}
  \end{subfigure}
  \begin{subfigure}[b]{0.32\linewidth}
    \includegraphics[width=\linewidth]{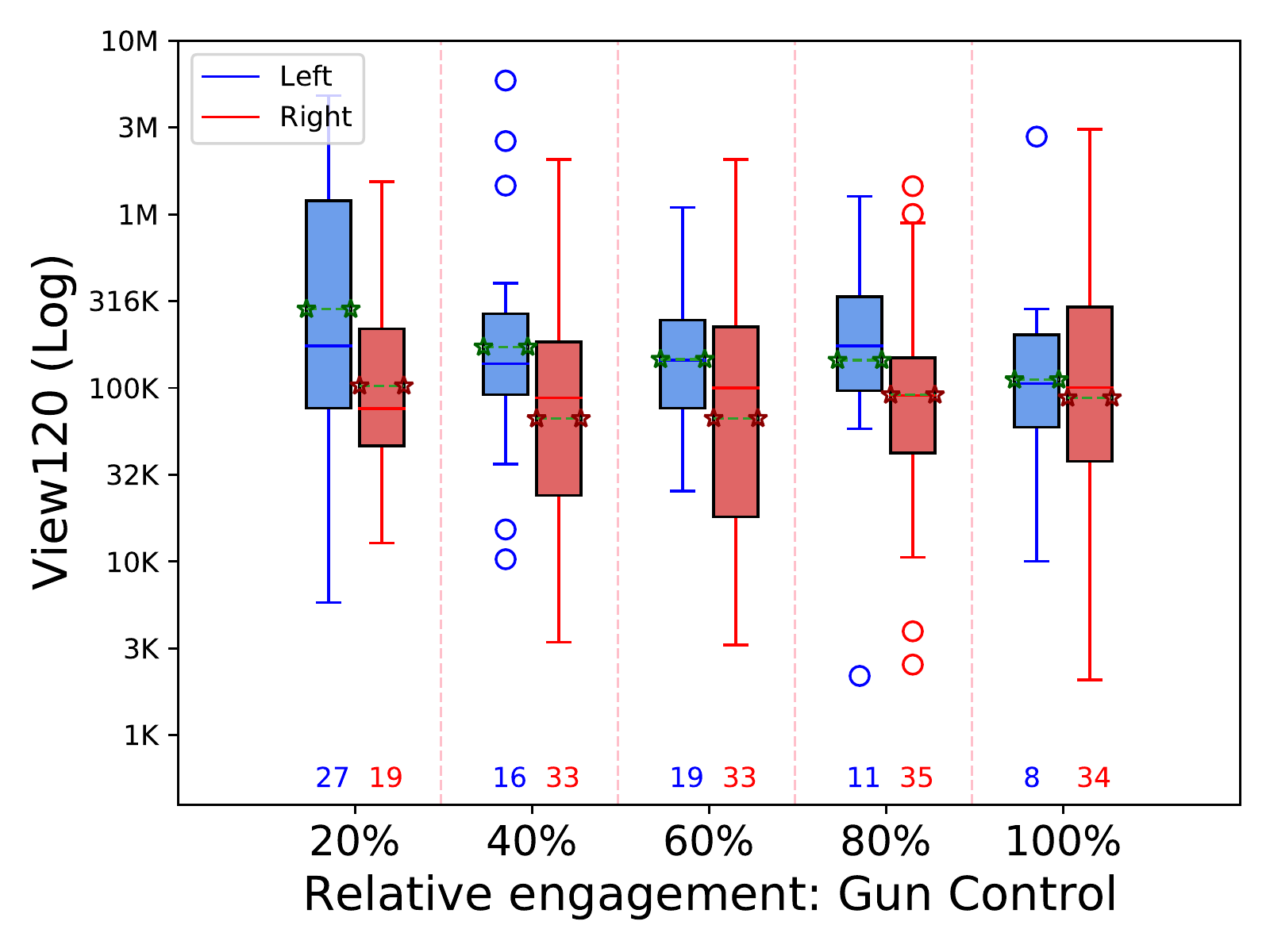}
    %\caption{\guncontrol}
  \end{subfigure}
  \begin{subfigure}[b]{0.32\linewidth}
  \includegraphics[width=\linewidth]{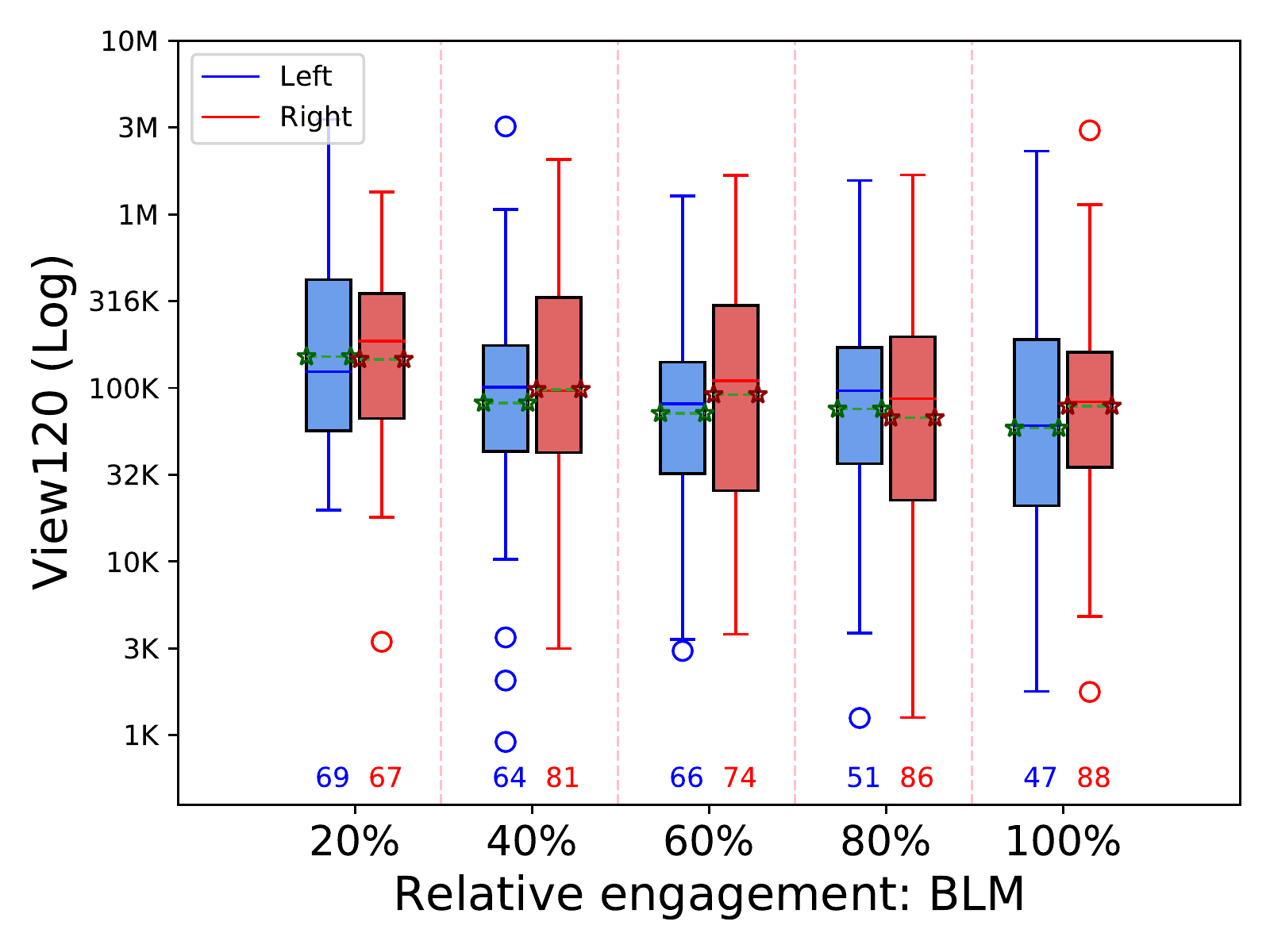}
  %\caption{\blm}
  \end{subfigure}
  %%%% 
  \caption{{\bf (Rows)} Boxplots of three attention metrics  -- number of Twitter followers, the number of tweets, and the number of video views at 120 days -- broken down by the relative engagement of each video in five groups of 20-percent each, {\bf(Columns)} for each of the three topics - \abortion, \guncontrol and \blm. 
  Population means are denoted by a dashed green line ending with stars.
  Left-leaning videos are promoted by less tweets overall (except the most engaging group of \guncontrol, and the least engaging groups for \abortion and \guncontrol), but on average attracts more views (except the 40, 60 and 100 percentile group for \blm).}
  \label{fig:app_intersecting}
\end{figure}

We link the measure of relative engagement of YouTube videos to the number of tweets, the number of followers, and the number of views. The number of tweets and views are accumulated for the first 120 days after upload.
Rather than looking at the overall correlation, which is dominated by large variations, we choose to group videos according to relative engagement. This allows us to interpret the change in Twitter user base, tweeting behavior, and YouTube attention across videos of different engagement levels.

We compute relative engagement score for each video and rank the videos in descending order of relative engagement score. Then, the videos are divided into 5 groups, each including the top 20\% videos, top 20\%\textasciitilde{}40\% videos, top 40\%\textasciitilde{}60\% videos, top 60\%\textasciitilde{}80\% videos and top 80\%\textasciitilde{}100\% videos. 
For each group, left and right-leaning videos are shown as boxplots of \textit{Tweet120}, \textit{Follower count}, and \textit{View120} (\Cref{fig:app_intersecting}).

As seen earlier, left-leaning videos have less number of tweets on average which can be seen with {\it Tweet120}. Especially, in the case of \abortion, the middle buckets (20\%, 40\% and 60\%) show that left-leaning videos have significantly less tweets but have more views on average. The same pattern holds for \guncontrol. This is consistent with the previous finding that left-leaning videos are more effectively promoted (higher viral potential) with fewer number of tweets.

We can observe that higher numbers of followers does not imply higher numbers of tweets on average as witnessed by most of middle buckets. However, we note that the top three outliers in left-leaning videos in 40\% bucket of {\it Follower count} boxplot for \guncontrol corresponds to the top three outliers of left-leaning videos in the same bucket of {\it Tweet120}.

In terms of {\it View120}, the average trends support findings in \Cref{ssec:aggregate_obs} in that left-leaning videos in \abortion and \guncontrol have higher views whereas left-leaning videos in \blm have lower views. Although view count is considered as an important measure of video popularity in general, no noticeable correlation is shown between relative engagement and {\it View120}. This is expected as engagement metric and popularity metric capture very different perspective of online attention~\cite{wu2019estimating}.

\nociteAP{*}
\bibliographystyleAP{aaai22}
\bibliographyAP{xplatform-ap-ref}
% ====================================================

\end{document}